\documentclass[twocolumn]{aastex63}

\usepackage{multirow}
\hypersetup{linkcolor=blue,citecolor=blue,filecolor=magenta,urlcolor=cyan}

\shorttitle{Neptune's resonances: 5:3 and 7:4}
\shortauthors{Thirouin \& Sheppard}

\begin{document}

\title{Rotational Study of 5:3 and 7:4 Resonant Objects within the Main Classical Trans-Neptunian Belt}

\correspondingauthor{Audrey Thirouin}
\email{thirouin@lowell.edu}

\author[0000-0002-1506-4248]{Audrey Thirouin}
\affiliation{Lowell Observatory, 1400 W Mars Hill Road, Flagstaff, Arizona, 86001, USA}

\author[0000-0003-3145-8682]{Scott S. Sheppard}
\affiliation{Earth and Planets Laboratory, Carnegie Institution for Science, \\ 5241 Broad Branch Rd. NW, Washington, District of Columbia, 20015, USA.}



\begin{abstract}
The 5:3 and 7:4 mean motion resonances of Neptune are at 42.3 and 43.7~au, respectively, and overlap with objects in the Classical trans-neptunian belt (Kuiper belt). We report the complete/partial lightcurves of 13 and 14 trans-Neptunian objects (TNOs) in the 5:3 and 7:4 resonances, respectively. We report a most likely contact binary in the 7:4 resonance, 2013~FR$_{28}$, with a periodicity of 13.97$\pm$0.04~h and a lightcurve amplitude of 0.94$\pm$0.02~mag. With a V-/U-shaped lightcurve, 2013~FR$_{28}$ has one of the largest well-sampled TNO amplitude observed with ground-based observations, comparable to the well-determined contact binary 2001~QG$_{298}$. 2013~FR$_{28}$ has a mass ratio q$\sim$1 with a density $\rho$$\sim$1~g cm$^{-3}$. We find several objects with large amplitudes and classify 2004~SC$_{60}$, 2006~CJ$_{69}$, and 2013~BN$_{82}$ as likely contact binaries, and 2001~QF$_{331}$, 2003~YW$_{179}$, and 2015~FP$_{345}$ as likely elongated objects. We observe the 17:9 resonant or classical object 2003~SP$_{317}$ that we classify as a likely contact binary. A lower estimate of 10-50~\% and 20-55~\% for the fraction of (nearly) equal-sized contact binaries is calculated in the 5:3 and 7:4 resonances, respectively. Surface colors of 2004~SC$_{60}$, 2013~BN$_{82}$, 2014~OL$_{394}$, and 2015~FP$_{345}$ have been obtained. Including these colors with ones from the literature reveals that elongated objects and contact binaries share the same ultra-red surface color, except Manw\"e-Thorondor and 2004~SC$_{60}$. Not only are the colors of the 7:4 and 5:3 TNOs similar to the Cold Classicals, but we demonstrate that the rotational properties of the 5:3 and 7:4 resonants are similar to those of the Cold Classicals, inferring a clear link between these sub-populations.

\end{abstract}



\section{Introduction} 
\label{sec:intro}

Our Solar System's outer regions are home to the Trans-Neptunian objects (TNOs) which are small icy leftovers from the era of planet formation. Some of these planetesimals get caught in resonances with Neptune as this planet migrated through the Solar System \citep{Nesvorny2022, Nesvorny2021, Pirani2021, Volk2019, Lawler2019, Gladman2012, Levison2008, Malhotra1995}. A small body is in Neptune's resonance if its orbital period is in a specific ratio to Neptune's orbital period. As an example, a TNO in the 5:3 resonance will orbit 3 times around the Sun while Neptune will make 5 revolutions around the Sun in the same amount of time.

  \begin{figure*}
 \includegraphics[width=9.5cm, angle=0]{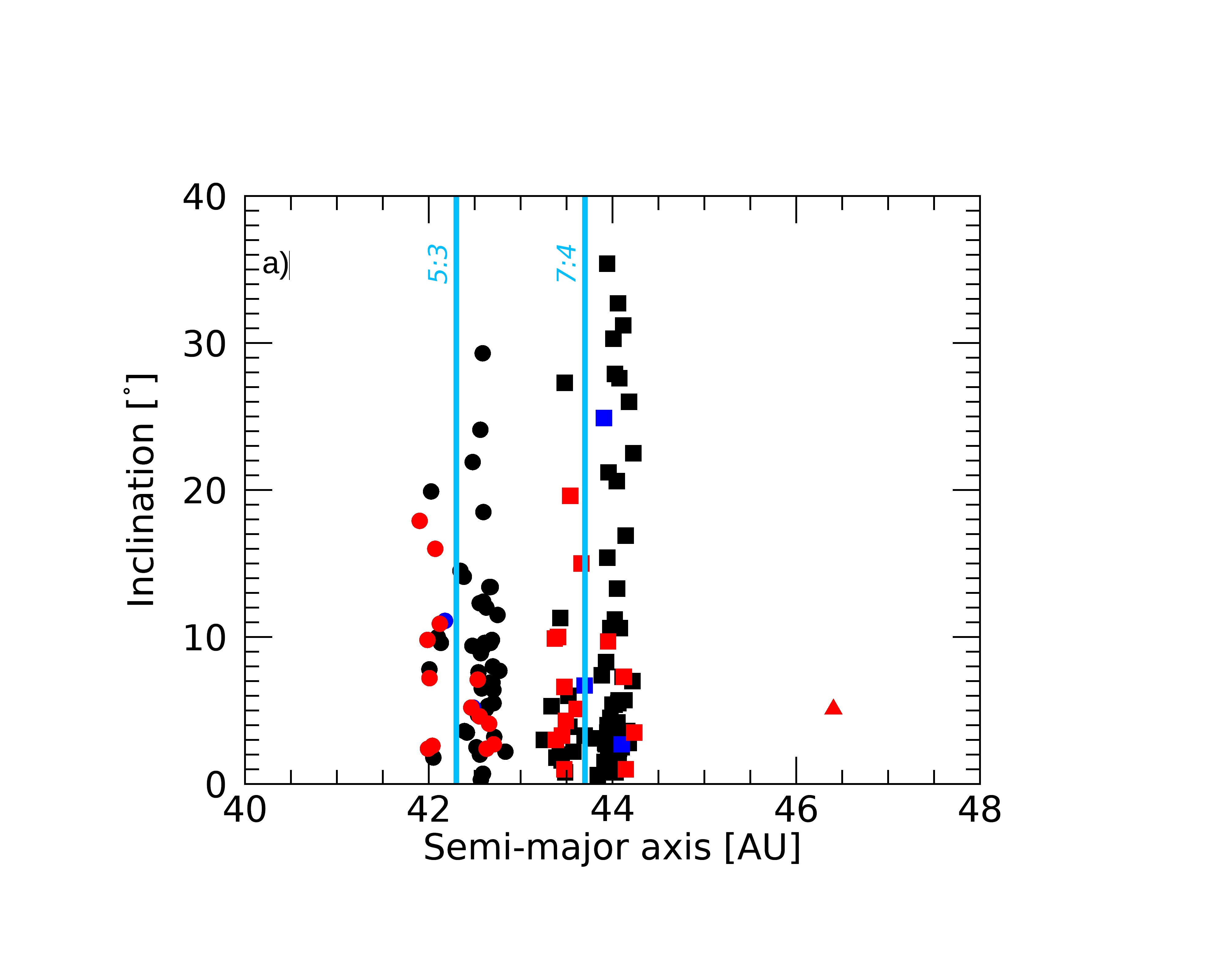}
 \includegraphics[width=9.5cm, angle=0]{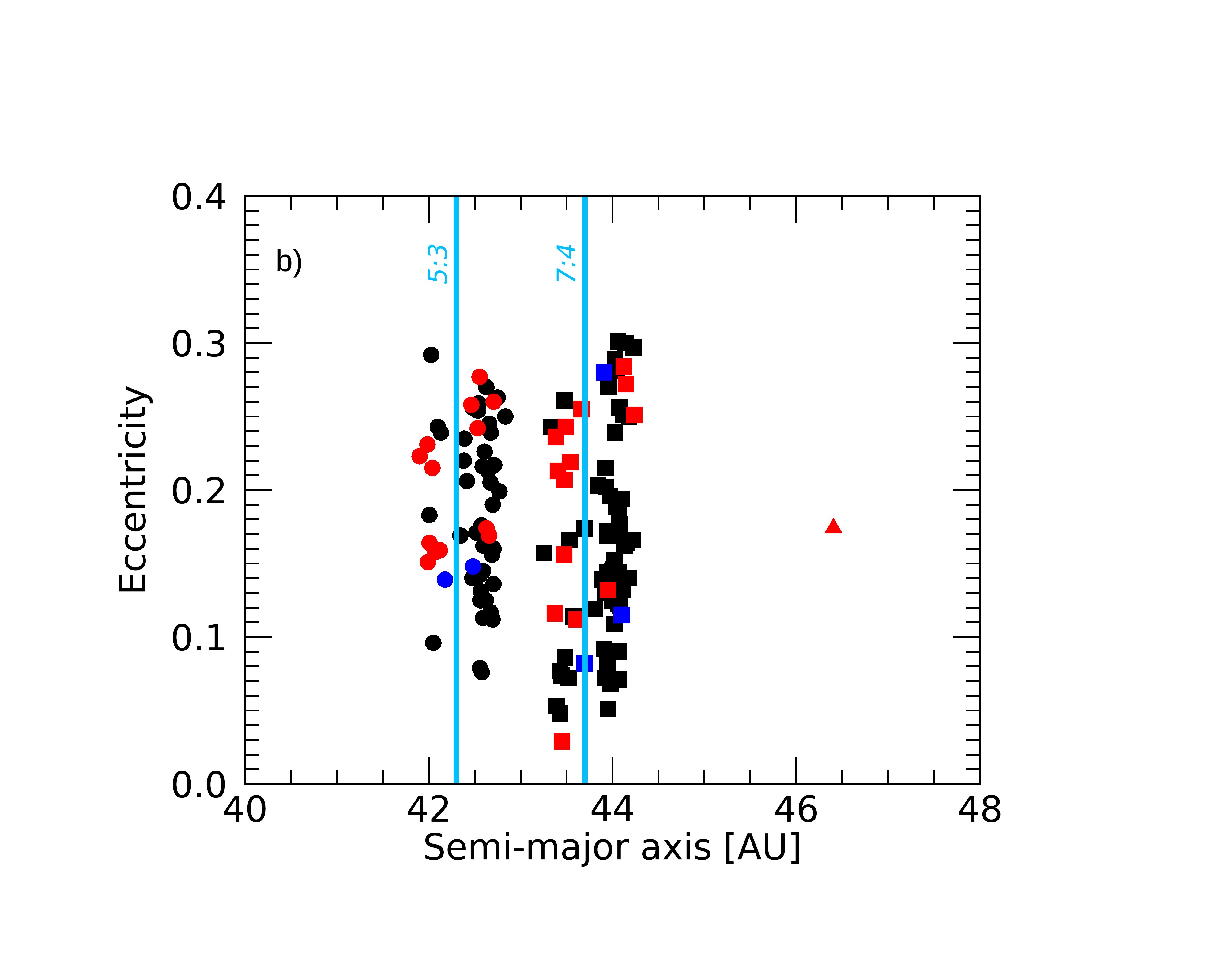}
 \includegraphics[width=9.5cm, angle=0]{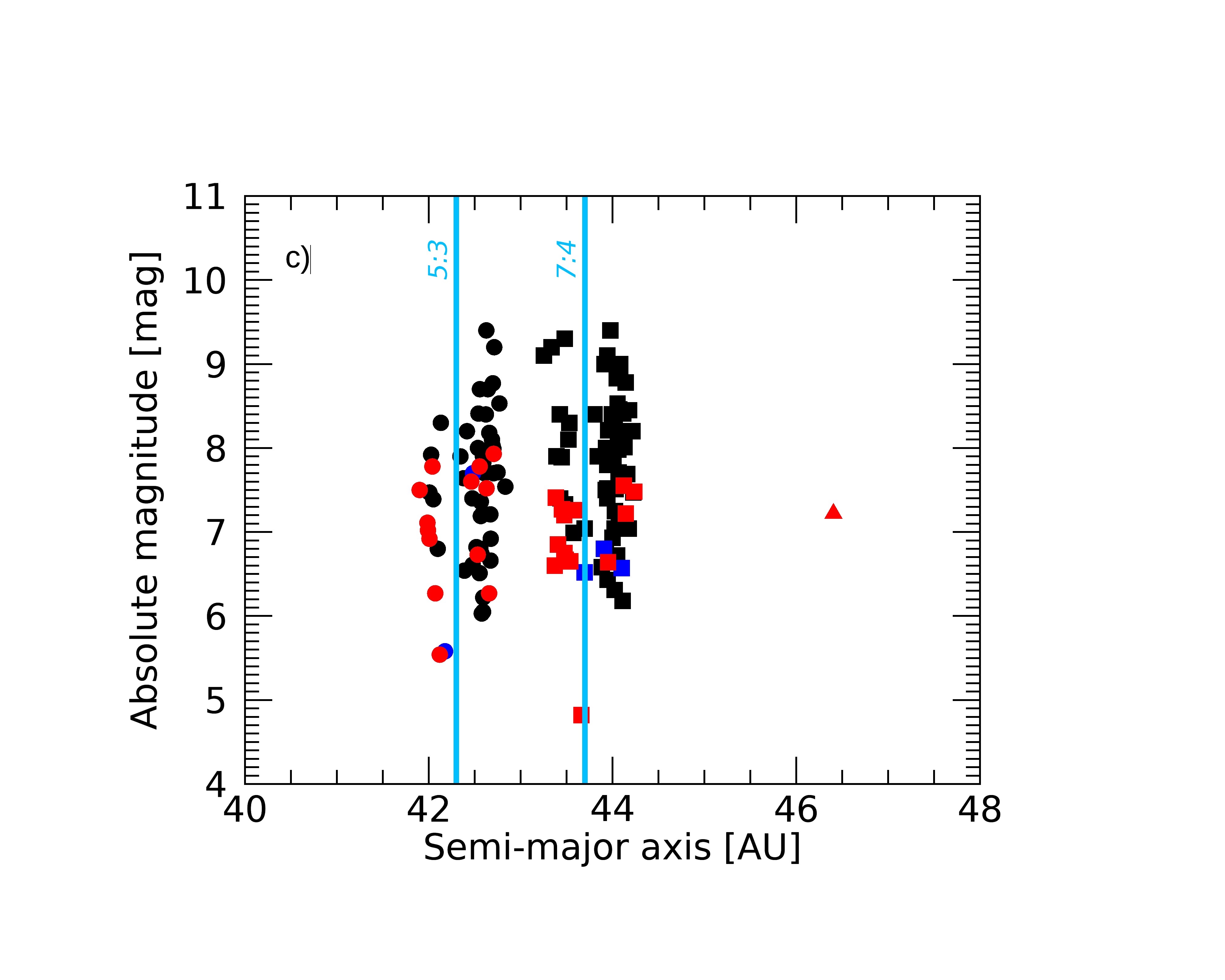}
 \includegraphics[width=9.5cm, angle=0]{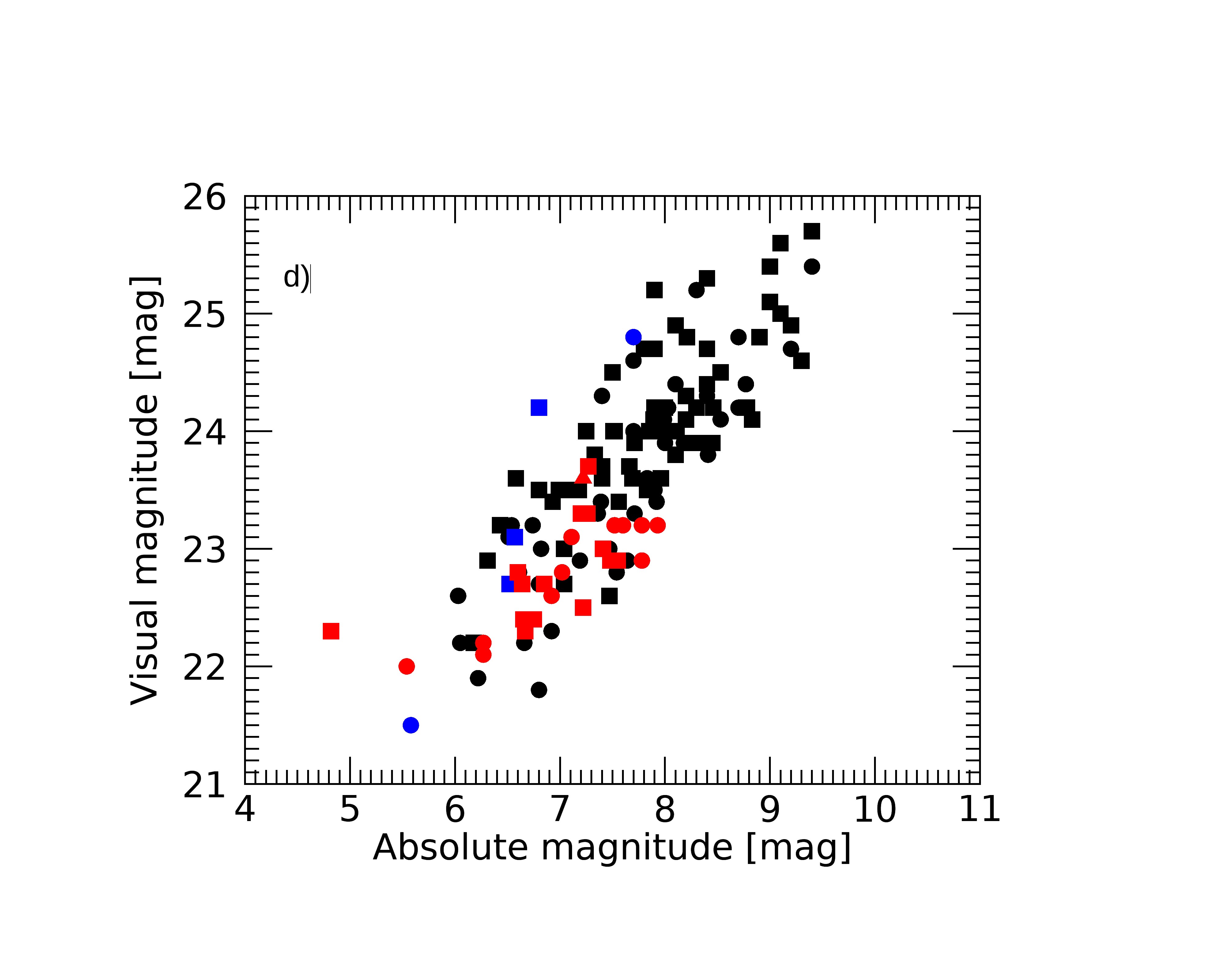}
\caption{Trans-neptunian objects trapped in the 5:3 (circles) and 7:4 (squares) resonances with Neptune based on the \textit{Deep Ecliptic Survey} classification are plotted with the following legend: black symbols for TNOs never observed for lightcurve studies, blue symbols for TNOs with some photometric information in the literature (Table~\ref{literature}), and red symbols for TNOs observed during our survey (Table~\ref{Summary_photo}). The red triangle corresponds to the object 2003~SP$_{317}$ whose dynamical classification is 17:9 resonant or classical trans-neptunian object. \underline{\textit{Note:}} Our survey targeted 2008~CS$_{190}$ and 2001~XP$_{254}$ (plotted with red symbols) which   have been studied as well by \citet{Kecskemethy2023}.  }
\label{fig:Orb}
\end{figure*}

Several resonances are located within the Classical trans-neptunian belt which is between $\sim$40~AU and $\sim$47~AU. The main resonances in the Classical population are the 5:3 and 7:4 (the main focus of this work), but there are also several higher-order resonances; the 8:5, 9:5, 17:9, 15:8, among others \citep{Gladman2008, Bannister2018}. The \textit{Deep Ecliptic Survey (DES$\footnote{\url{https://www.boulder.swri.edu/~buie/kbo/desclass.html}}$)} lists 59 and 76 TNOs in the 5:3 and 7:4 resonances, respectively, whereas higher-order resonances have a handful of detected objects so far (Figure~\ref{fig:Orb}). During Neptune's migration, objects formed at different heliocentric distances got pushed to the Solar System's outer edge and some of them ended up stuck into resonances, but it is also important to point out that due to the 5:3 and 7:4 locations, the Classical belt plays a significant role by supplying objects into these resonances.

\citet{MurrayClay2011} proposed the migration-induced capture scenario in which TNOs can be captured into resonances. This scenario suggested that the 2:1 resonant TNOs at low inclinations should have a higher fraction of binary compared to the 2:1 resonant TNOs at high inclinations whereas the TNOs at low inclination in the 3:2 resonance should have a low fraction of binaries as this resonance did not pass over the Cold Classical population. Both predictions were confirmed observationally and are discussed in the review paper by \citet{Noll2020}. \citet{MurrayClay2011} also predicted that the 3:2 and 2:1 resonances should have a low inclination Cold Classical component. \citet{Sheppard2012} demonstrated that up to an inclination (i) of 10$^\circ$, resonant TNOs are mainly from the dynamically Cold Classical population whereas at i$>$10$^\circ$ the objects' main source is the dynamically Hot Classical population. However, \citet{Sheppard2012} conclusions are based only on color differences between the high/low inclination resonant TNOs and color alikeness with the Cold Classicals. One may wonder if the differences/similarities between the resonant and Cold classical TNOs extend to other physical and rotational properties. 

To complement our global picture of the resonant TNOs within the Classical belt, we conduct a photometric survey to obtain lightcurves, contact binary fractions, as well as amplitude and period distributions of the 5:3 and 7:4 resonances to infer the differences and similitudes of these resonances with the dynamically Cold Classical population. 

\section{Our survey: Telescopes, Instruments, and Strategy} 
\label{sec:obs}

We report a photometric study carried out from September 2016 to June 2023 with the 6.5~m \textit{Magellan-Baade telescope} at the Las Campanas Observatory, Chile, and the 4.3~m \textit{Lowell Discovery Telescope} (LDT) next to Happy Jack in Arizona, USA. This combination of medium to large facilities allows us to observe objects in the Southern and Northern hemispheres down to a visual magnitude of about 23-24~mag. At both sites, we use their imager instruments; the wide-field imager called Inamori-Magellan Areal Camera and Spectrograph with a 27.4$\arcmin$ diameter field for a 0.20$\arcsec$/pixel scale at the \textit{Magellan-Baade Telescope} and the Large Monolithic Imager with 12.5$\arcmin$$\times$12.5$\arcmin$ field of view for a 0.12$\arcsec$/pixel scale at the \textit{LDT}. For lightcurve observations, we select broad-band filters (VR-filter at \textit{LDT} and WB4800–7800 filter at \textit{Magellan-Baade}) aiming to increase the small body's signal-to-noise ratio. For surface color studies we choose the g'r'i' Sloan filters. Exposure times are adapted to the facility, filter, weather/seeing conditions, and the target's brightness. 

We select 5:3 and 7:4 resonant TNOs with a visual magnitude V$\lesssim$23-23.5~mag encompassing a variety of inclinations, eccentricities, and absolute magnitudes (i.e., sizes). All objects targeted by our survey$\footnote{For this work, all orbital elements, visual and absolute magnitudes have been extracted in February 2023 from the Minor Planet Center (MPC) webpages: \url{https://minorplanetcenter.net/iau/lists/t_tnos.html} and \url{https://minorplanetcenter.net/iau/lists/t_centaurs.html}.}$, 13 in the 5:3, 14 in the 7:4, are highlighted in Figure~\ref{fig:Orb} and summarized in Table~\ref{Summary_photo}. The TNO (385458) 2003~SP$_{317}$ is also in our target list, but its dynamical classification is questionable. With a semi-major axis of 46.405~AU, an inclination of 5.1$^{\circ}$, and an eccentricity of 0.174, it is a classical TNO for the \textit{Deep Ecliptic Survey}$\footnote{\url{https://www.boulder.swri.edu/~buie/kbo/astrom/385458.html}. For more details about the \textit{Deep Ecliptic Survey}, see \citet{Elliot2005}}.$, but \citet{Alexandersen2019} classify it as a 17:9 resonant. Because of this classification issue, we do not plot the other classical nor the other 17:9 TNOs on Figure~\ref{fig:Orb} and 2003~SP$_{317}$ will remain as an isolated object for this work.  

During each observing night, we obtain dome flats and/or sky flats, and biases to calibrate our science images. To optimize our observing time, we observe 3 to 5 objects alternatively on the same night to obtain their sparse lightcurves. With one image per object every 40-45~minutes or so for several hours, there is a good enough sampling to evaluate the lower limits of the object's period and variability. Ideally (weather and observing schedule dependent), we reschedule the objects on at least one different night to confirm the first sparse lightcurve results. If an object displays a moderate/large variability ($\Delta m$$\gtrsim $0.3~mag), it will be re-observed over several nights to derive its full lightcurve. All our science images are reduced and analyzed with the standard steps described in \citet{Thirouin2010, Thirouin2014}. Basically, we obtain the aperture photometry of our targets followed by a periodicity search such as the Lomb periodogram \citep{Lomb1976}. The Lomb periodogram's highest peak gives the strongest periodicity detected in the dataset, but the proper rotational period can be a multiple of the detected period. Assuming a single-peaked lightcurve, the true period of the object will be the one favored by the Lomb periodogram, but if the lightcurve is double-peaked, then the true rotational period is twice the one favored by the periodogram. \citet{Thirouin2014} showed that a spheroidal object has a low amplitude single-peaked lightcurve whereas an elongated object as well as contact binary have moderate to high amplitude double-peaked lightcurves (more details in the next Section).    
   

\section{Lightcurves interpretations}
\label{sec:interpretation}

Based on \citet{Leone1984}, \citet{SheppardJewitt2004} proposed a ``lightcurve classification" using rotational period (P) and lightcurve amplitude ($\Delta m$) to infer if a lightcurve is due to albedo difference(s) on the body's surface, or/and due to a non-spherical shaped object and/or if the object is, in fact, a binary system. Three regions were identified (see Figure~5 in \citet{SheppardJewitt2004}); in Region A, a small lightcurve amplitude ($\Delta m$$<$0.25~mag) could be caused by an elongated object, albedo, or binarity; in Region B, a moderate to large lightcurve amplitude ($\Delta m$$>$0.25~mag) and a fast rotation is most likely due to a rotational elongated object; and in Region C: a moderate to large lightcurve amplitude with an average or slow rotation is most likely due to a nearly equal-sized binary. \citet{SheppardJewitt2004} adopted a cut-off at $\Delta m$=0.25~mag, but this limit was revaluated at about 0.15~mag by \citet{Thirouin2010}. Below, we will use a threshold at 0.20~mag which is in between both estimates. 

Additional criteria like the morphology of the lightcurve can be used to complement the \citet{SheppardJewitt2004} classification. \citet{LacerdaJewitt2007, Lacerda2008} showed that an elongated triaxial (Jacobi ellipsoid) object has a sinusoidal$\footnote{A second-order Fourier series can fit a sinusoidal lightcurve, but it will not fit a contact binary lightcurve. The formula for a second-order Fourier series fit is: \begin{equation}  Fit_{Fourier} = a + b  \cos(2\pi \phi) + c \sin(2\pi \phi)  
+ d  \cos(4 \pi \phi) + e \sin(4 \pi \phi) \end{equation} with $\phi$ as the rotational phase and a, b, c, d, and e as constants}$ lightcurve with a moderate variability (typically, between 0.2 and 0.4~mag), but a spherical object (McLaurin spheroid) with/without albedo disparity on its surface has a low/flat amplitude (typically, lower than 0.2~mag) whereas a nearly equal-sized contact binary lightcurve has a large amplitude with an inverted U-shape at the maximum of brightness and a V-shape at the minimum of brightness (i.e., non-sinusoidal lightcurve) from shadowing effects between the components. However, the U-/V-shapes can be less obvious if the contact binary is not observed equator-on. The modeling of the contact binary 2001~QG$_{298}$ by \citet{Lacerda2011} illustrates how the system's geometry affects the morphology and amplitude of a contact binary's lightcurve.   

According to \citet{Jeans1919, Weidenschilling1980, Leone1984, Chandrasekhar1987}, a triaxial object$\footnote{Triaxial object with axes: a$>$b$>$c, rotating along the c-axis.}$ with an axis ratio a/b$=$2.3 is unstable due to rotational fission, meaning that this object is so ``stretched" that as it rotates it will break into two components creating a close/contact binary system. Considering an equatorial view ($\xi$ = 90$^{\circ}$), lightcurve amplitude and axis ratio are related: 
\begin{equation}
\Delta{m} = 2.5~log\left(\frac{a}{b}\right) 
\end{equation}
and so, an axis ratio of 2.3 corresponds to an amplitude of 0.9~mag. Therefore, we consider an object whose lightcurve has a $\Delta m$$\geq$0.9~mag are \textit{most likely} contact binaries. However, if the viewing geometry is not equatorial  ($\xi$ $\neq$ 90$^{\circ}$):
\begin{equation}
\Delta{m} = 2.5~log\left( \frac{a}{b}\right)  - 1.25~log\left(
\frac{a^{2}\cos^{2}\xi + c^{2}\sin^{2}\xi}{b^{2}\cos^{2}\xi +
c^{2}\sin^{2}\xi}\right)
\end{equation} 
then an object with a/b=2.3 will have an amplitude smaller than 0.9~mag. Therefore, a lightcurve with an indication of V-/U-shape (non-sinusoidal lightcurve) and with a large amplitude can also be caused by a contact binary, even if the threshold at 0.9~mag is not reached. An object with a non-sinusoidal large amplitude lightcurve and with a Fourier series chi-square ($\chi$$^{2}$) value above 1 is a \textit{likely} contact binary when the 0.9~mag limit is not met. In the case of a likely contact binary, only a lightcurve at a significantly different epoch of at least four or more years (depending on the obliquity of the system) and system modeling will confirm the system's characteristics \citep{Lacerda2014, Lacerda2011}. 

In some instances, only a partial lightcurve with lower limits for the periodicity and object's variability can be available, and thus the presence of a V-/U-shape can be difficult/impossible to identify. If the object displays a large variability, it is a \textit{potential} contact binary, and additional observations are required to confirm this conclusion. 

In this paper, we consider that a close/contact binary can be a small body with a bi-lobed shape or two objects touching in one point, as well as two objects with a small separation of less than a few hundred kilometers. Following \citet{Nesvorny2019} definition, a contact binary/close binary has an a$_{B}$/R$_{B}$$<$10 where a$_{B}$ is the binary semimajor axis and R$^{3}_{B}$=R$^{3}_{primary}$+R$^{3}_{secondary}$ with R$_{primary}$ and R$_{secondary}$ being the radii of the system's components.

Following, we will interpret all sinusoidal lightcurves to be caused by elongated small bodies. But, we point out that for these bodies, a lightcurve at a different epoch is warranted to discard a future contact binary-shaped lightcurve. A nearly equal-sized contact binary not imaged nearly edge-on may not display signs of U-/V-shapes, and so only a new lightcurve at a different epoch can determine the object's shape.

\section{Lightcurves Results} 
\label{sec:res}

Below, the rotational and physical properties of the 28 TNOs observed during our survey will be discussed (Table~\ref{Summary_photo} and Table~\ref{modeling}). The entire photometry and partial/flat lightcurves are in Appendix~\ref{sec:appA} and \ref{sec:appB}. For each complete lightcurve study, we report the Lomb periodogram (plot a) on the upcoming figures) described in Section~\ref{sec:obs} and the lightcurve with a rotational phase between 0 and 1.2 on plot b).  

  \subsection{Objects with a $\Delta m$$\geq$0.9~mag}

\begin{figure}
  \includegraphics[width=9cm, angle=0]{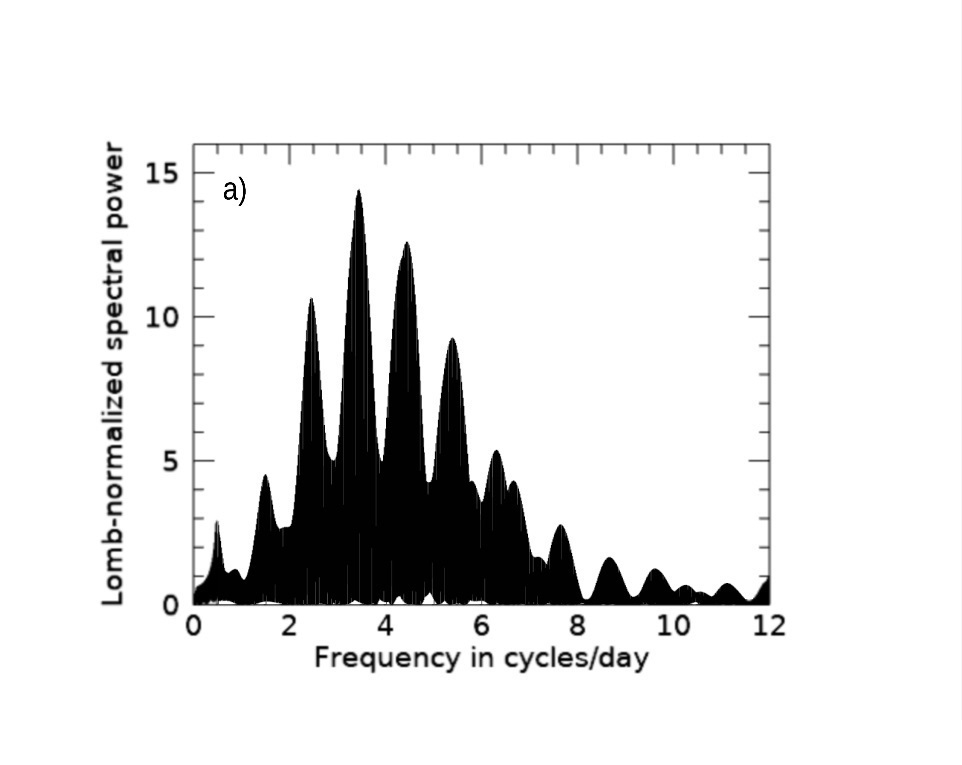}
  \includegraphics[width=9cm, angle=0]{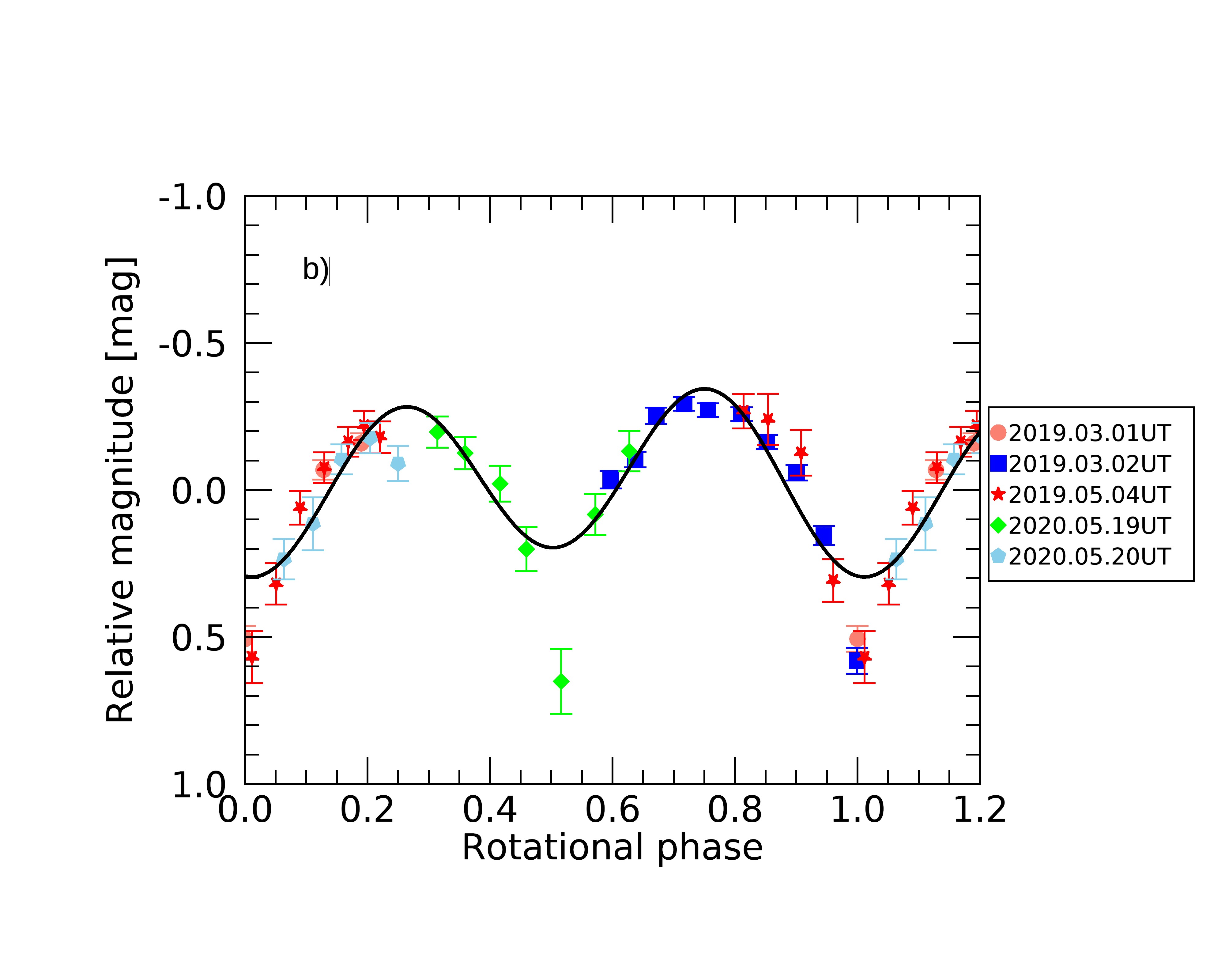}
 \caption{The main peak of the Lomb periodogram (Plot a)) is located at 3.44~cycles day$^{-1}$. The double-peaked lightcurve of 2013~FR$_{28}$ with P$^{double}_{rotational}$=13.97$\pm$0.04~h has an amplitude larger than the cut-off at 0.9~mag and has the V-/U-morphology of a most likely contact binary (Plot b)). The Fourier series fit (black continuous line) cannot match the observations with a $\chi$$^2$=4.90}.  
 \label{fig:FR28}
 \end{figure}

 
  \paragraph{(532039) 2013~FR$_{28}$} With the \textit{Magellan-Baade} telescope and the \textit{LDT}, we observed this TNO in 2019 and 2020. The periodogram prefers a frequency of f=3.44~cycles day$^{-1}$ equivalent to a period of P=6.99~h. Due to the lightcurve morphology, the object's rotation is double, so P$^{double}_{rotational}$=13.97$\pm$0.04~h. Our study concludes that 2013~FR$_{28}$ has a non-sinusoidal lightcurve with a $\Delta m$=0.94$\pm$0.02~mag (Figure~\ref{fig:FR28}). The Fourier series does not match our data because of the distinctive U- and V-shapes. Because of the morphology and amplitude above the 0.9~mag limit, 2013~FR$_{28}$ is a most likely contact binary.

\citet{Leone1984} studied the figures of equilibrium for binary systems which we can use for crude modeling to derive some basic information. Following their procedure, we derive that 2013~FR$_{28}$ has a mass ratio$\footnote{Because the error bars for the period and amplitude are not taken into account, we prefer to use approximate values for the density and mass ratio.}$ q$\sim$1 with $\rho$$\sim$1~g cm$^{-3}$. The properties extracted from this modeling are available in Table~\ref{modeling}. We can also infer based on this lightcurve that 2013~FR$_{28}$ is nearly equator-on \citep{Lacerda2011}.

In the trans-neptunian belt, such a well-sampled and extreme variability is only surpassed by the 3:2 resonant contact binary 2001~QG$_{298}$ whose variability was 1.14$\pm$0.04~mag in 2002-2003 \citep{SheppardJewitt2004}. With the addition of 2013~FR$_{28}$, there are only two most likely trans-neptunian contact binaries detected based on their peculiar lightcurve morphology and large amplitude over 0.9~mag. Further observations will allow its pole orientation and true maximum amplitude to be determined.
 
  \subsection{Objects with a $\Delta m$ between 0.5 and 0.9~mag}

\begin{figure}
  \includegraphics[width=9cm, angle=0]{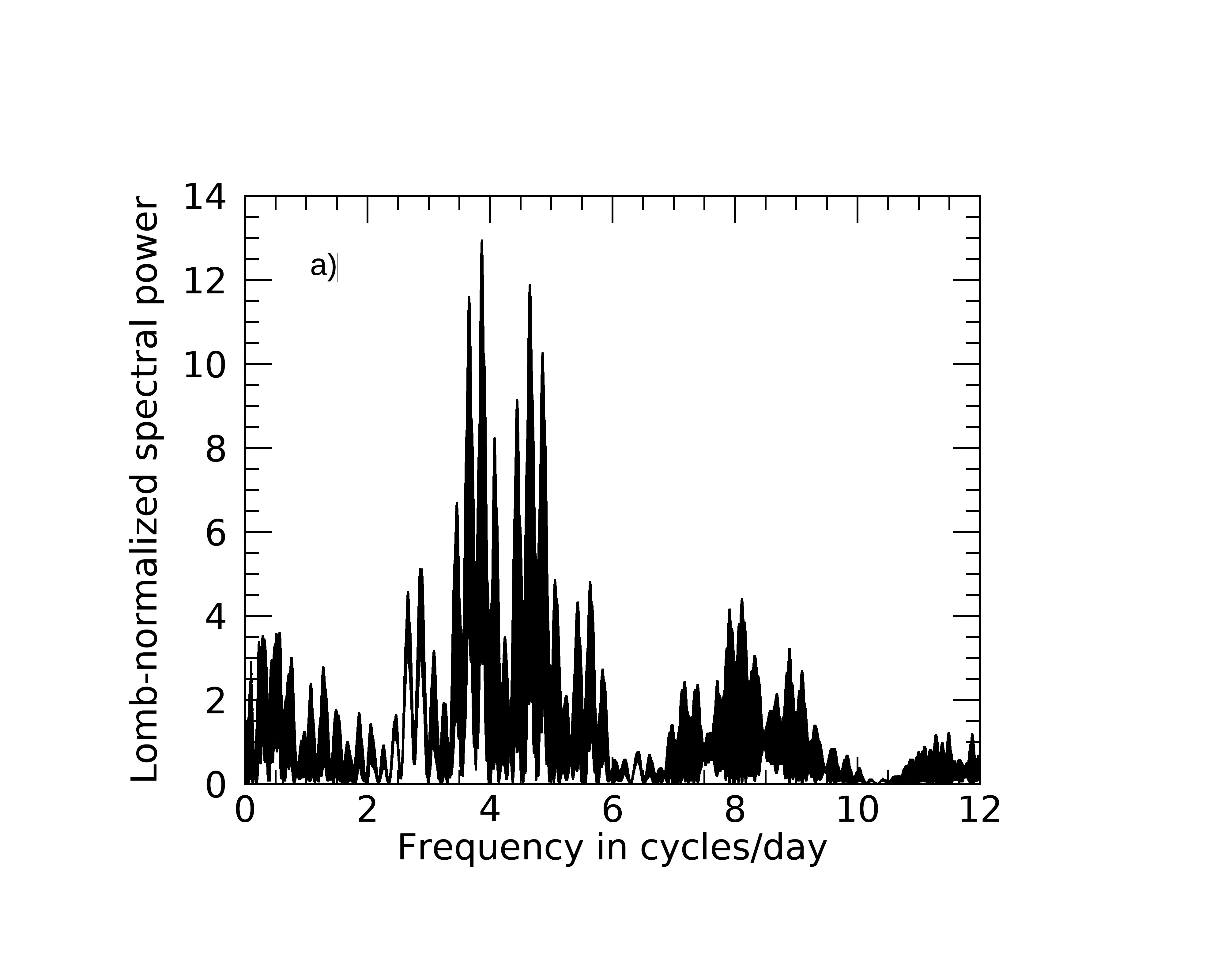}
  \includegraphics[width=9cm, angle=0]{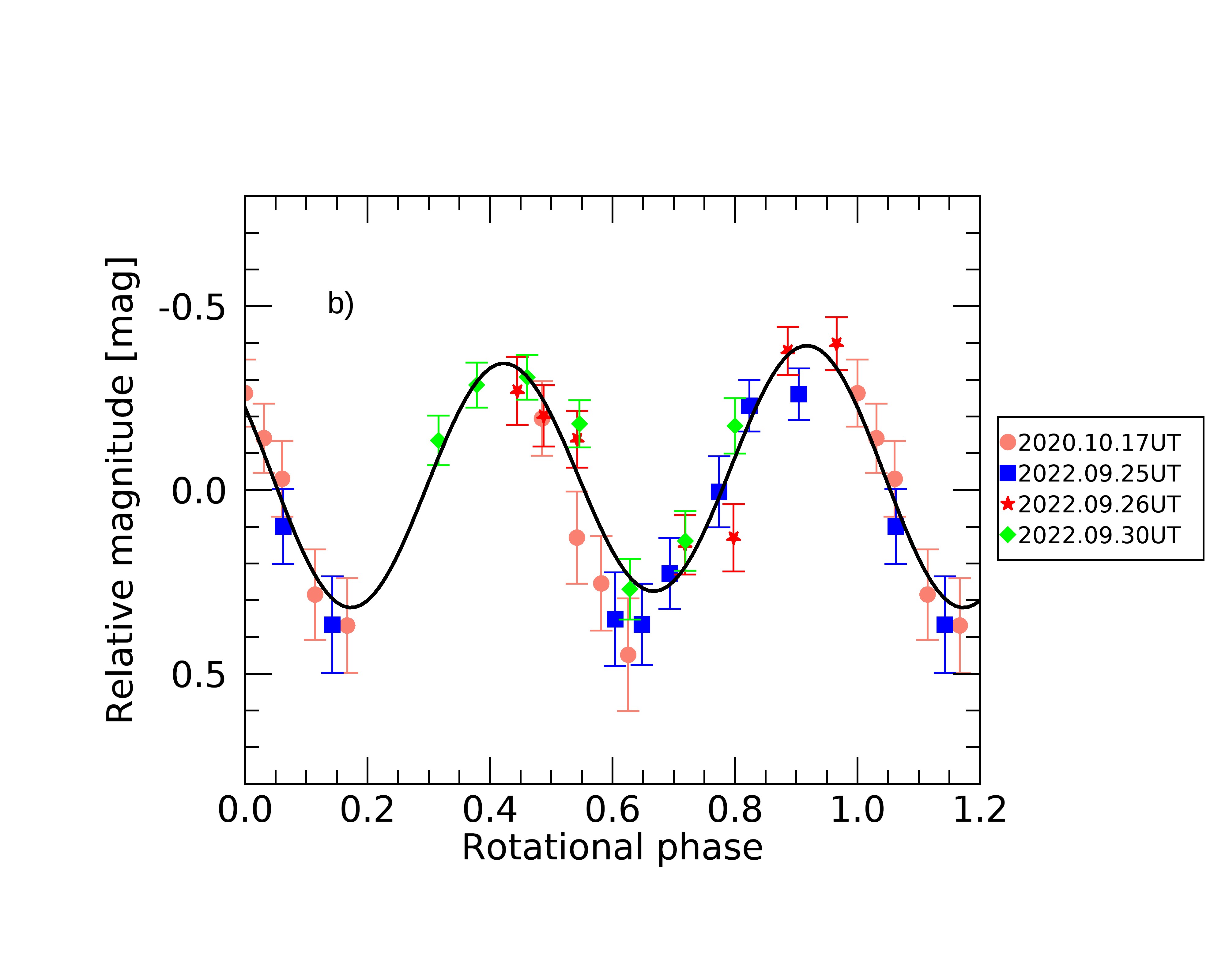}
 \caption{The lightcurve of 2003~SP$_{317}$ is asymmetric and double-peaked with a period P$^{double}_{rotational}$=12.39$\pm$0.03~h and $\Delta m$=0.85$\pm$0.03~mag. We classify this body as a likely contact binary due to its high variability, its V-shape at the minima, and because the Fourier series fit is unable to reproduce the lightcurve ($\chi$$^2$=1.02). }
 \label{fig:SP317}
 \end{figure}

  \paragraph{(385458) 2003~SP$_{317}$} We obtained images of 2003~SP$_{317}$ in 2020 and 2022 over four nights with the \textit{LDT}. The periodogram's highest peak is at f=3.87~cycles day$^{-1}$ (P=6.20~h), but based on the lightcurve asymmetry and variability, a double-peaked period of 12.39$\pm$0.03~h is more appropriate (Figure~\ref{fig:SP317}). 2003~SP$_{317}$ has a $\Delta m$=0.85$\pm$0.03~mag. 

Using the HSC (Hyper Suprime-Cam) on the \textit{Subaru} telescope, \citet{Alexandersen2019} observed 2003~SP$_{317}$ for short-term variability. They report an incomplete lightcurve with a periodicity of 12.45~h which is similar to the periodicity derived from our observations. They only sample the maximum of the curve over their two observing nights for an amplitude limit of 0.56~mag. 

In Figure~\ref{fig:SP317}, we overplot a Fourier series fit to the lightcurve, but it is unable to reproduce the second minimum showing that the lightcurve of 2003~SP$_{317}$ is not sinusoidal. The lightcurve presents the classical V-shape of a contact binary, but not the U-shape. Because $\Delta m$ is just below the 0.9~mag cut-off, we conclude that 2003~SP$_{317}$ is a likely contact binary.  

Following \citet{Leone1984}, we estimate that 2003~SP$_{317}$ has a mass ratio between q$_{min}$$\sim$0.75 with $\rho_{min}$$\sim$1~g cm$^{-3}$ and q$_{max}$$\sim$1 with $\rho_{max}$$\sim$1.25~g cm$^{-3}$ (Table~\ref{modeling}). 

\begin{figure}
  \includegraphics[width=9cm, angle=0]{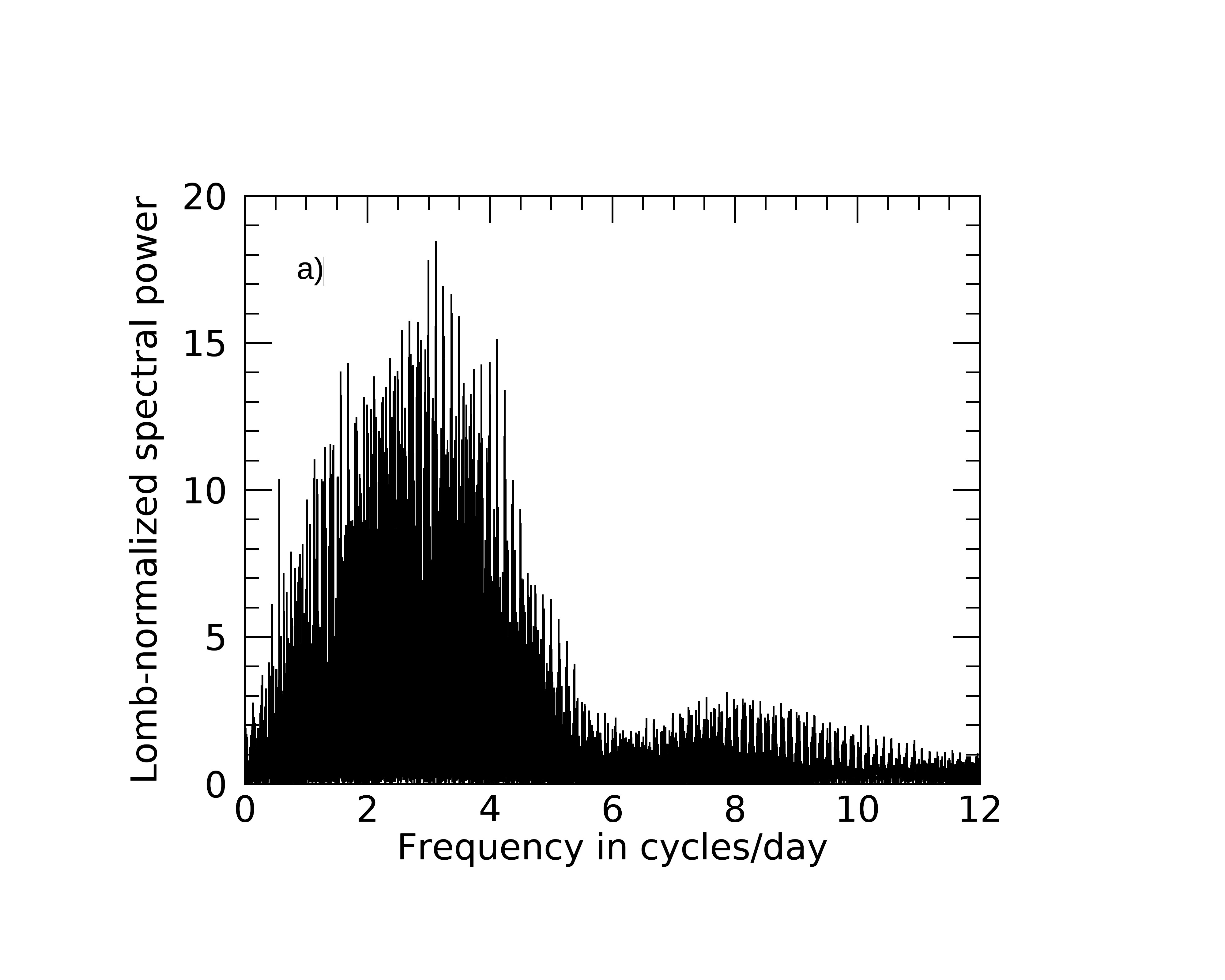}
  \includegraphics[width=9cm, angle=0]{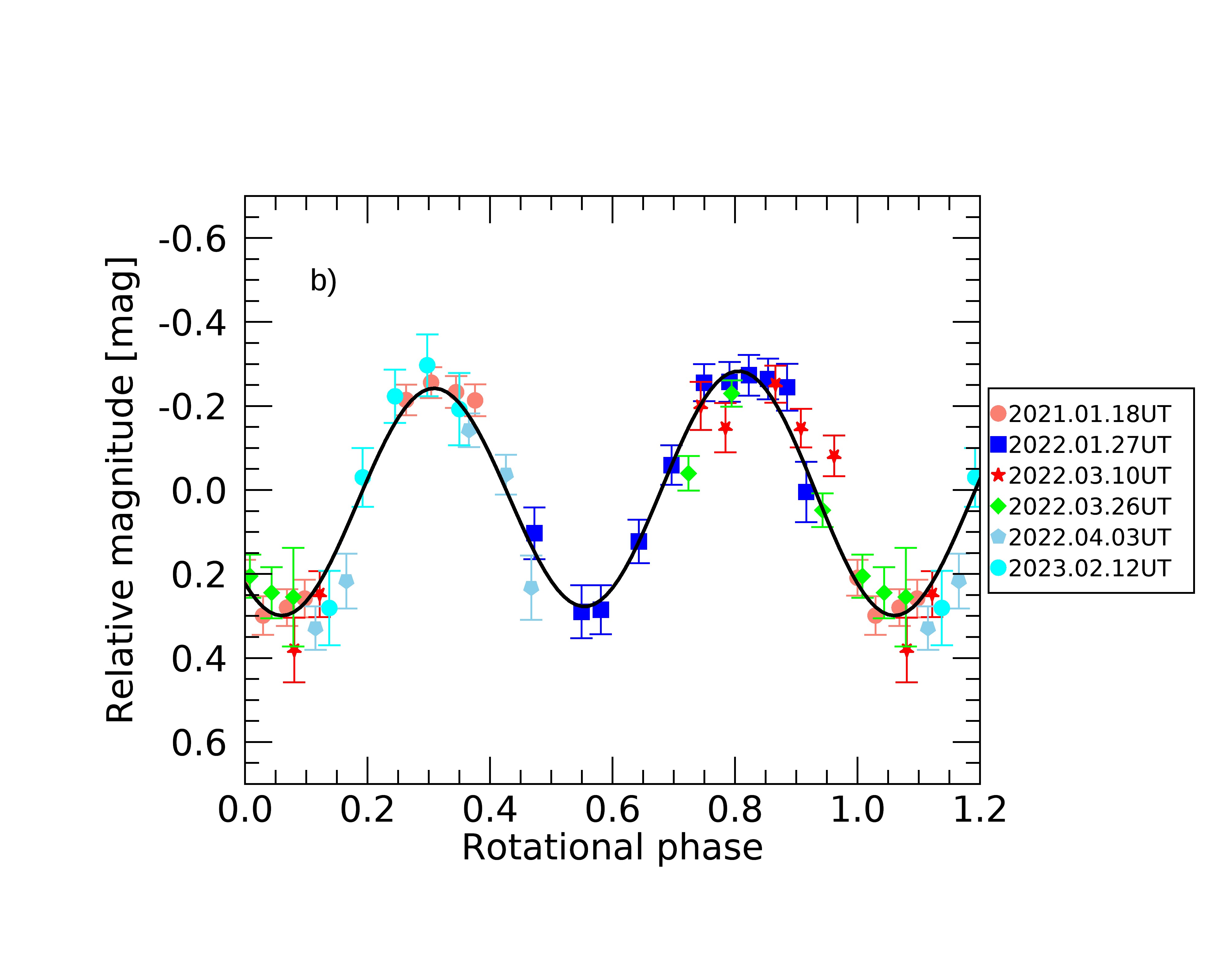}
 \caption{The periodogram's main peak is at 3.12~cycles/day (Plot a)). Because of the large photometric variability of 2003~YW$_{179}$, the double-peaked rotational period of 15.41~h is best suited. A second-order Fourier fit (black curve) is a good match to our datasets inferring that this object is elongated (Plot b)).  }
 \label{fig:YW179}
 \end{figure}

  \paragraph{(469584) 2003~YW$_{179}$} We observed this object in six instances from 2021 to 2023. The periodogram suggests a period of 3.12~cycles/day. As the symmetric lightcurve of 2003~YW$_{179}$ has a large variability, the best rotational period is 15.41$\pm$0.03~h (Figure~\ref{fig:YW179}). We derive an amplitude of 0.58$\pm$0.02~mag from the fit which is in good agreement with the data. We conclude that 2003~YW$_{179}$ is a highly elongated object with axis ratios such as a/b=1.71 and c/a=0.43 with a density $\rho$$>$0.18~g cm$^{-3}$ for a view of $\xi$=90$^\circ$ (Table~\ref{modeling}). However, as for 2001~QF$_{331}$, additional lightcurves with amplitude changes are warranted to secure our interpretation.

\begin{figure}
  \includegraphics[width=9cm, angle=0]{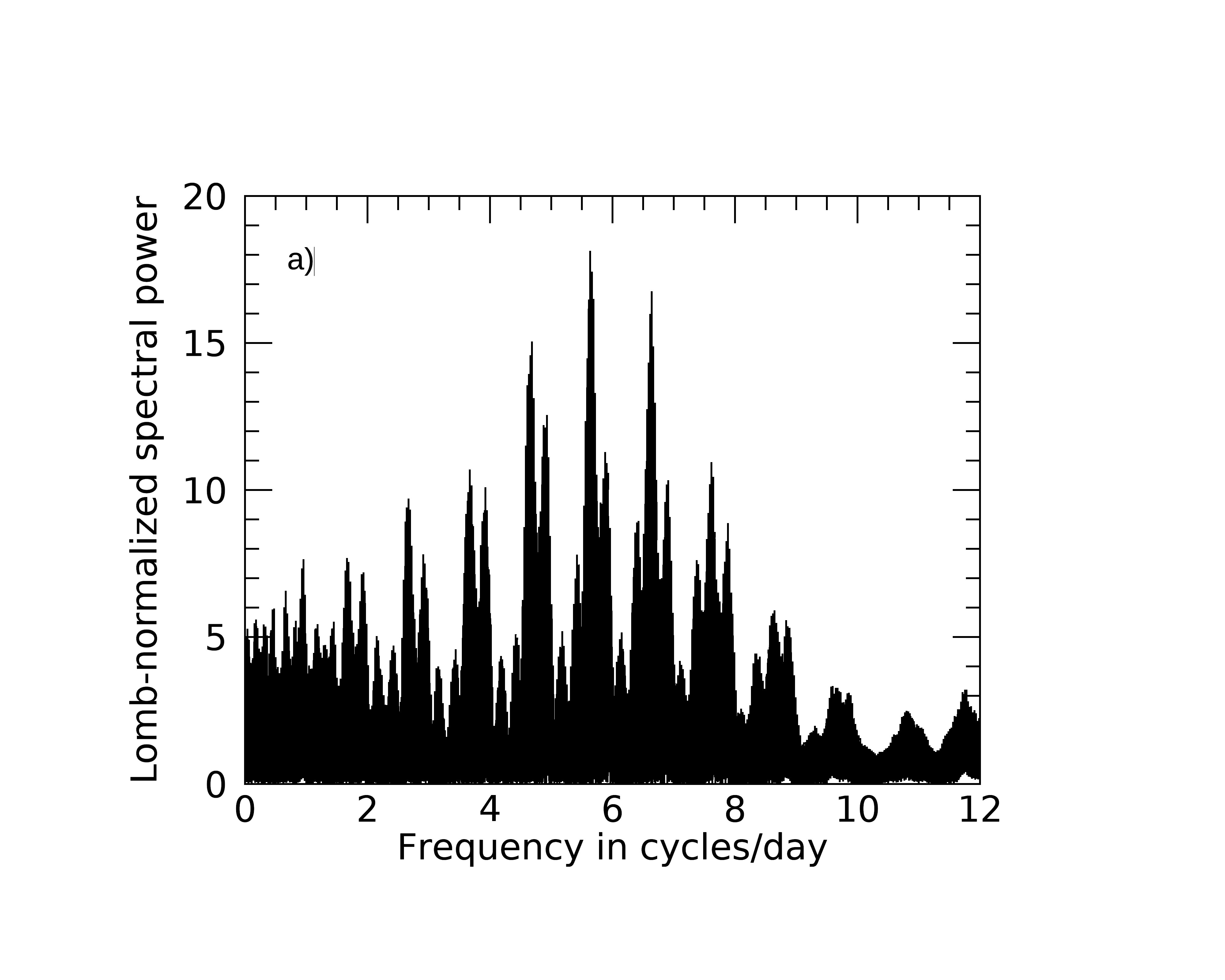}
          \includegraphics[width=9cm, angle=0]{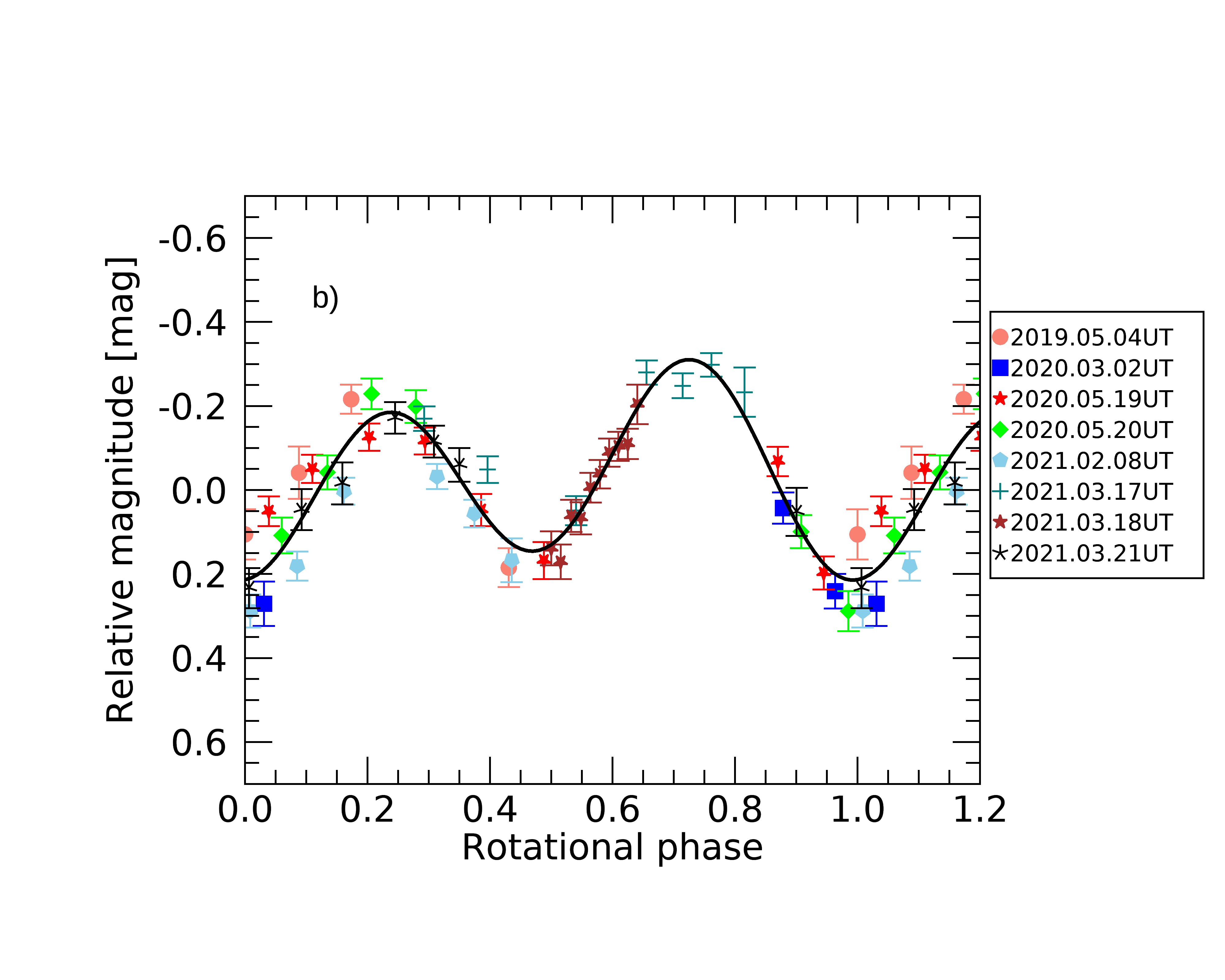}
\caption{We infer that the lightcurve of 2015~FP$_{345}$ is double-peaked with P$^{double}_{rotational}$=8.47$\pm$0.02~h. This lightcurve is fitted by a Fourier series demonstrating that this object is elongated and likely has spot(s) on its surface.  }
\label{fig:FP345}
\end{figure}

    \paragraph{(536922) 2015~FP$_{345}$} From 2019 to 2021, we imaged 2015~FP$_{345}$ on eight occasions for short-term variability and once in 2022 for colors. In Figure~\ref{fig:FP345}, the period determination favors a 5.66~cycles/day, as the lightcurve of 2015~FP$_{345}$ is asymmetric, its true rotation is P$^{double}_{rotational}$=8.47$\pm$0.02~h. A second-order Fourier series nicely fits the lightcurve, and thus we conclude that 2015~FP$_{345}$ is an elongated triaxial object with albedo variegation(s). An asymmetric lightcurve can be due to albedo marking(s) or an irregular shape. In the case of Haumea, \citet{Lacerda2008} demonstrated that a dark red spot is causing an asymmetric lightcurve. A topographic feature on 2003~AZ$_{84}$ was discovered by \citet{DiasOliveira2017}, but its lightcurve is symmetric. Therefore, we prefer the option of albedo spot(s) to explain an asymmetric lightcurve. \citet{Thirouin2010, Lacerda2008, JewittSheppard2002, Degewij1979} indicate that the asymmetry is due to albedo spot(s) which are typically between 4 and 10~\% on the TNO surfaces. The lightcurve amplitude is 0.52$\pm$0.04~mag.   

Considering an equatorial viewing geometry ($\xi$=90$^\circ$), the axis ratios of 2015~FP$_{345}$ are a/b=1.61, c/a=0.44 for a density $\rho$$>$0.59~g cm$^{-3}$ (Table~\ref{modeling}).

On May 2022, we imaged 2015~FP$_{345}$ with three Sloan filters for suface colors which are g'-i'=1.34$\pm$0.07~mag and g'-r'=0.88$\pm$0.07~mag making it an ultra-red object (colors are discussed in Section~\ref{sec:colors}). 
   
  \paragraph{(118378) 1999~HT$_{11}$} In April 2023, we obtained two observing blocks of 3~h and 3.5~h for this object with the \textit{LDT}. The photometry of 1999~HT$_{11}$ presents a higher uncertainty than the rest because it is one of the faintest targets in our sample and because of its high variability of 0.52~mag and 0.44~mag over the respective observing blocks. We do not have enough data to provide its complete lightcurve but we consider 1999~HT$_{11}$ a potential contact binary due to its amplitude.

  \subsection{Objects with a $\Delta m$ between 0.3 and 0.5~mag}

\begin{figure}
  \includegraphics[width=9cm, angle=0]{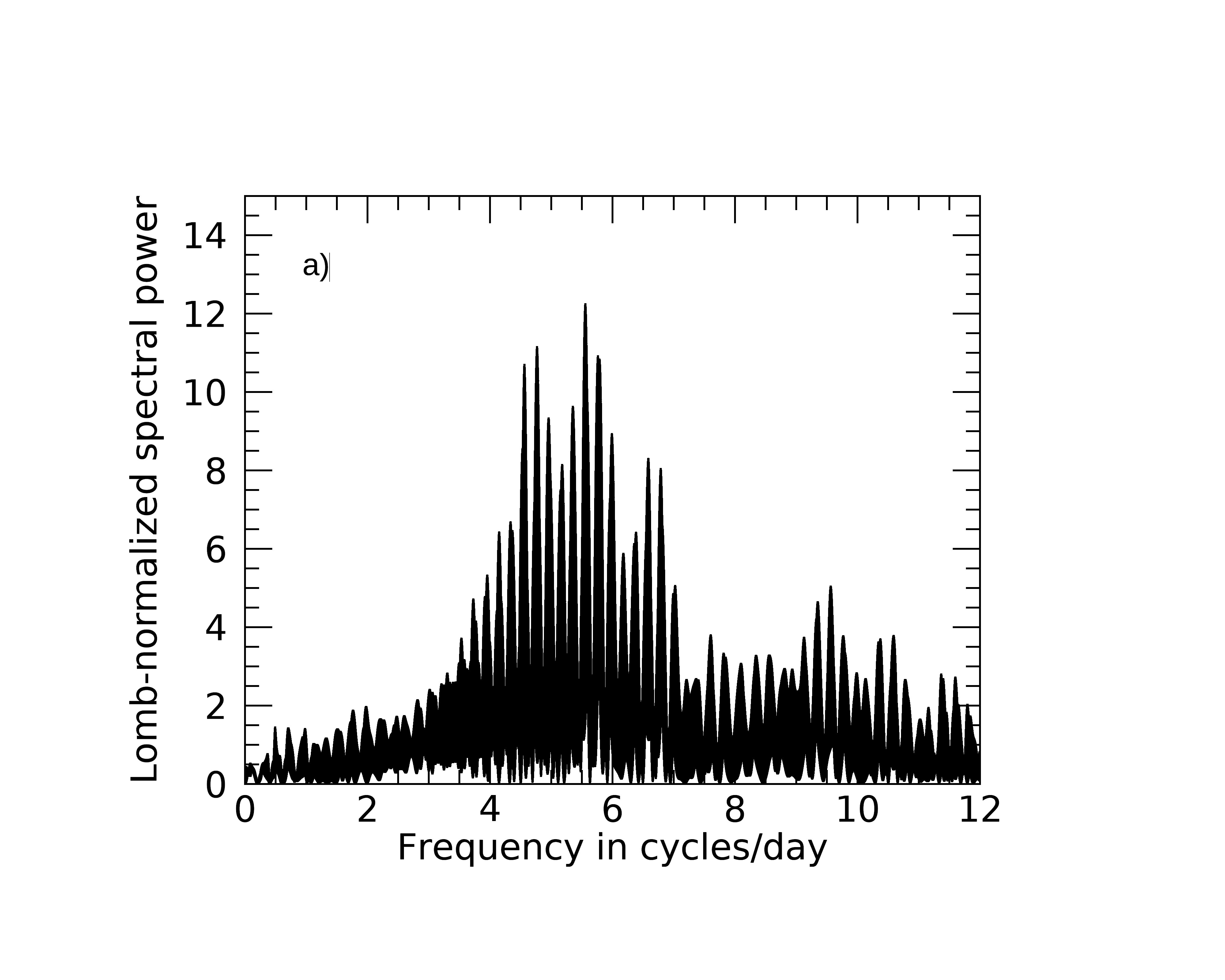}
 \includegraphics[width=9cm, angle=0]{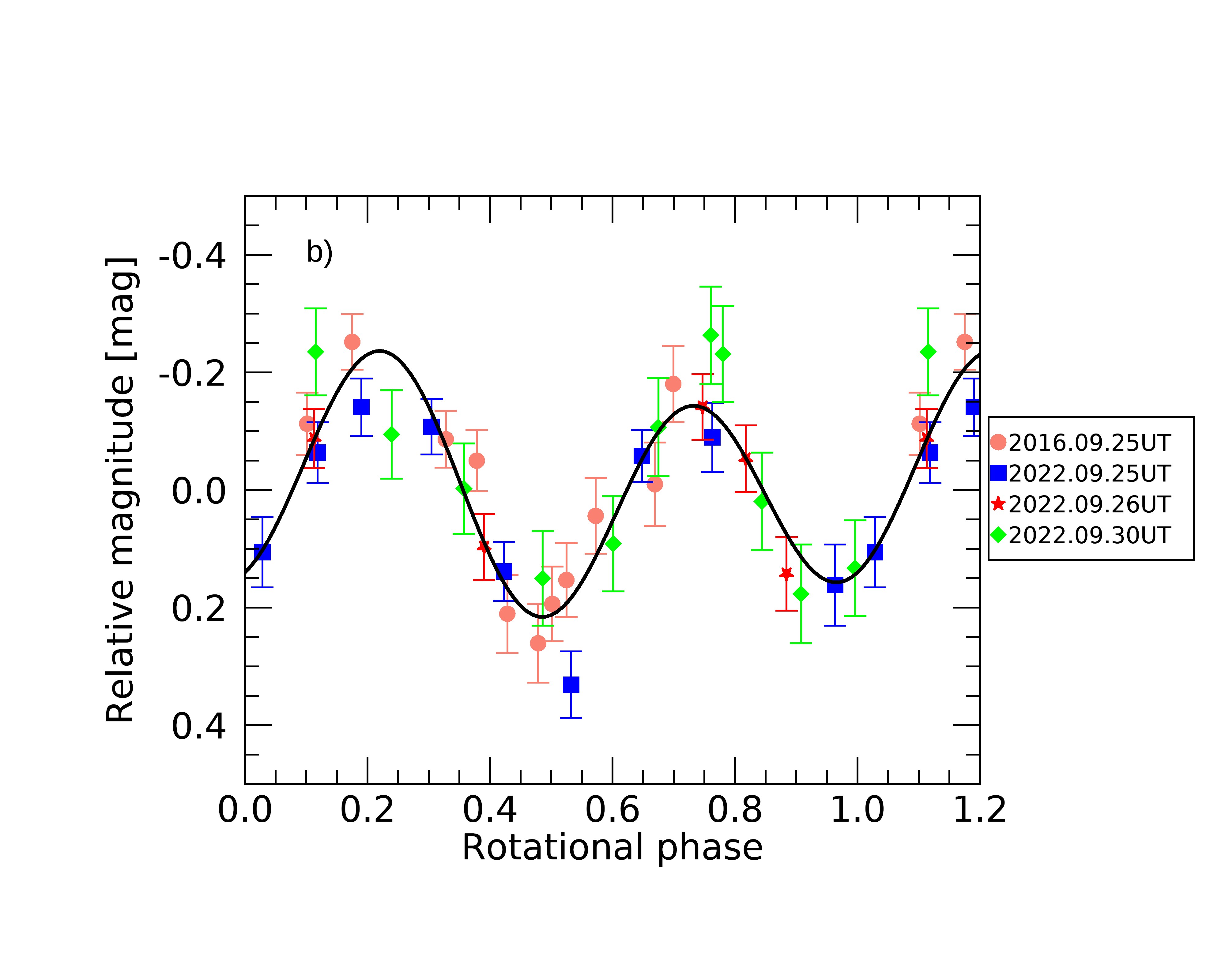}
\caption{The periodogram's tallest peak is at 5.56~cycles/day, but as the lightcurve of 2001~QF$_{331}$ is asymmetric and has a large amplitude, the double-peaked period of 8.63~h is the adequate solution. The lightcurve is fitted with a second-order Fourier series (black curve) suggesting that this object is elongated.  }
\label{fig:QF331}
\end{figure}

  \paragraph{(503883) 2001~QF$_{331}$} We observed 2001~QF$_{331}$ over four nights with the \textit{LDT} in 2016 and 2022. The periodicity search favors a frequency of 5.56~cycles/day. However, due to the object's large variability and also due to an asymmetric$\footnote{A lightcurve is considered asymmetric if its two maxima (and/or both minima) do not reach the same magnitude. In the case of 2001~QF$_{331}$, the two maxima are not reaching the same magnitude and its two minima are also not at the same magnitude.}$ lightcurve, the double-peaked solution is more appropriate for this object. We conclude that the period of 2001~QF$_{331}$ is 8.63$\pm$0.04~h (Figure~\ref{fig:QF331}). The Fourier series fit the lightcurve and infers that the amplitude from the tallest maximum to the deepest minimum is 0.45$\pm$0.03~mag. The lightcurve interpretation suggests that 2001~QF$_{331}$ is not a nearly equal-sized contact binary imaged (nearly) edge-on, but rather a single elongated object with some albedo variations on its surface. Only future observations to determine its pole orientation from amplitude changes can distinguish between an elongated object and a nearly equal-sized contact binary.

Assuming that small bodies have a fluid-like behavior and are in hydrostatic equilibrium, \citet{Chandrasekhar1987} studied the figures of equilibrium for triaxial objects (among other object configurations) and using his work, we can derive lower limits to the object's density and its axis ratios for a certain viewing angle ($\xi$): if $\xi$=90$^\circ$, a/b=1.51, c/a=0.46 and $\rho$$>$0.56~g cm$^{-3}$. Values for $\xi$\footnote{\citet{SheppardJewitt2004, Sheppard2004} showed that an average viewing angle is $\xi$=60$^{\circ}$}=60$^{\circ}$ are in Table~\ref{modeling}.

\clearpage
\startlongtable
\begin{deluxetable*}{lcccccc|cc|cccc|c}
\tabletypesize{\scriptsize}
\tablecaption{\label{Summary_photo} Circumstances of observing runs as well as rotational period and lightcurve amplitude estimates. \\  }
\tablewidth{0pt}
\tablehead{Object  & Date  &   r$_h$ &  $\Delta$ & $\alpha$   & Filter & Telescope & P$_{rot}$ & $\Delta$m & H$_{MPC}$ & a &e & i& Wide  \\
        &     UT       &  [AU]  &  [AU]  &  [$^{\circ}$]   & &  & [h] & [mag] & [mag]& [AU] & & [$^\circ$]&Binary?$^{*}$ }
\startdata
\multicolumn{14}{c}{\textit{Mean motion resonance: 5:3}}\\                                                                 
(612086) 1999~CX$_{131}$ & 02/13/2021      & 37.138&  38.086  &0.4 & VR &   LDT& ... & $\sim$0.1 & 7.11 &41.986&0.231&9.8 & no \\
(503883) 2001~QF$_{331}$ & 09/25/2016         &32.338& 31.376 &0.5 & VR &   LDT& 8.63$\pm$0.04 & 0.45$\pm$0.03 & 7.93 &42.707 &0.260 &2.7 & ?\\
					& 09/25/2022 &     30.939 & 31.806     &  0.9 &VR& LDT  &  ...& ... & ... & ... & ...	&... &  ...  \\
					& 09/26/2022 &      30.930  &31.805   &  0.9 &VR& LDT  &  ...& ... & ... & ... & ...	&... &  ...  \\
					& 09/30/2022 &      30.899 & 31.805       &  0.8 &VR& LDT  &  ...& ... & ... & ... & ...	&... &  ...  \\
 (469420) 2001~XP$_{254}$ & 01/18/2021   & 32.251 & 33.171 & 0.6       & VR  & LDT  &  $>$6 & $>$0.12  &  7.78 & 42.040 & 0.215	& 2.6 & yes \\
 (149349)  2002~VA$_{131}$ & 02/08/2021  & 35.697 & 35.814  & 1.6       & VR  & LDT  &  $>$6 & $>$0.05 &  6.73 &42.533 & 0.242	& 7.1 & no \\
    					   & 01/27/2022   &35.262&  35.615   & 1.5    &VR& LDT  &  ...& ... & ... & ... & ...	&... &  ...  \\
 2002~VV$_{130}$ & 02/08/2021    &   35.398 & 35.444   &  1.6 & VR  & LDT  &  $>$3& $>$0.14  & 7.52 & 42.629 &  0.174 &  2.4  & no \\
& 11/27/2021 &     34.438 & 35.407     & 0.3  &VR& LDT  & $>$8& $>$0.14 & ... & ... & ...	&... &  ...  \\
& 12/19/2022 &  34.521  &35.363      & 0.8   &VR& LDT  & $>$3.5 & $>$0.12 & ... & ... & ...	&... &  ...  \\
(469584) 2003~YW$_{179}$  & 01/18/2021   & 35.311 & 36.205  & 0.7      & VR  & LDT  & 15.41$\pm$0.03 & 0.58$\pm$0.02 &  7.02 &41.992 & 0.151 & 2.4 & no \\
 		& 01/27/2022  & 35.337 & 36.276 & 0.5    & VR  & LDT  &  ...& ... & ... & ... & ...& ...& ... \\
 		& 03/10/2022   &   35.385&  36.283   &  0.7 & VR  & LDT   &  ...& ... & ... & ... & ...& ...& ... \\
		& 03/26/2022    &   35.536 & 36.286  &  1.0 & VR  & LDT   &  ...& ... & ... & ... & ...& ...& ... \\
		& 04/03/2022    &  35.635 & 36.288    &  1.2 & VR  & LDT  &  ...& ... & ... & ... & ...& ...& ... \\
		& 02/12/2023    &  35.368 & 36.353  &  0.1& VR  & LDT  &  ...& ... & ... & ... & ...& ...& ... \\
 2004~VE$_{131}$ & 01/18/2021  &34.901 & 35.512 & 1.3       & VR  & LDT  &  $>$5.5  & $>$0.17  & 7.6 &  	42.463 & 0.258 & 5.2 & no \\
      & 02/08/2021   &35.209 & 35.499 & 1.5  &VR  & LDT    & $>$3.5  & $>$0.28   & ... & ... & ...	&... &  ...  \\
        & 11/27/2021  &  34.333  &35.317 & 0.1 &VR  & LDT    &$>$7  & $>$0.34  & ... & ... & ...	&... &  ...  \\   
        & 12/03/2021  &  34.329 & 35.313& 0.1 &VR  & LDT    & $>$8  & $>$0.40   & ... & ... & ...	&... &  ...  \\   
        & 12/05/2021  &34.331 & 35.312 & 0.1 &VR  & LDT   & $>$7  & $>$0.40    & ... & ... & ...	&... &  ...  \\   
 (434709) 2006~CJ$_{69}$ & 02/08/2021      &  32.263 & 33.152  &  0.7& VR &   LDT& 23.39$\pm$0.07  & 0.35$\pm$0.04 & 7.50 &41.901 &0.223&17.9  & ? \\
        &  02/13/2021  &   32.229  & 33.151 &  0.6& VR  & LDT  &  ...  &  ... &  ... & ...&  ...	& ... &  ...  \\
        &  04/13/2021     & 32.386&  33.135   & 1.2  & VR  & LDT  &  ...  &  ... &  ... & ...&  ...	& ... &  ...  \\
        &  01/27/2022   &  32.299 & 33.060  & 1.1  & VR  & LDT  &  ...  &  ... &  ... & ...&  ...	& ... &  ...  \\
         &  04/01/2022   &  32.165 & 33.044  & 0.8  & VR  & LDT  &  ...  &  ... &  ... & ...&  ...	& ... &  ...  \\
        &  04/03/2022   &   32.179  &33.044  & 0.9  & VR  & LDT  &  ...  &  ... &  ... & ...&  ...	& ... &  ...  \\
   (470523) 2008~CS$_{190}$ &  02/28/2019       & 36.252 & 37.233 &  0.2  & WB & Magellan& - & $\sim$0.1  & 6.27 &42.071 &0.158  & 16.0   & no\\
 	                            & 03/01/2019           & 36.250  & 37.234 &  0.2 & WB & Magellan&  ... & ...& ...  &...&...&... & ...\\
 	                             &    03/02/2019      &  36.249 & 37.234 &  0.2  & WB & Magellan&  ... & ...& ...   &...&...&...& ...\\
    2012~BY$_{154}$ &  02/28/2019        & 34.684 & 35.655 &  0.3  &WB & Magellan&-  & $\sim$0.1 & 6.92 & 42.008 & 0.164 & 7.2  & ?\\
 	                             &    03/02/2019      &  34.677 & 35.655 &  0.3  &WB & Magellan&  ... & ...& ...   &...&...&...& ...\\
  (523688) 2014~DK$_{143}$ & 05/14/2021     & 42.730 & 43.714 & 0.3      &VR  & LDT  &  8.99$\pm$0.03& 0.21$\pm$0.03  &  5.54 &42.120 & 0.159	& 10.9 & ?\\
 	                             &    04/21/2023      &   43.161 & 43.990    &  0.7  &VR& LDT&  ... & ...& ...   &...&...&...& ...\\
 	                             &    04/22/2023      &  43.152 & 43.990  & 0.7   &VR& LDT&  ... & ...& ...   &...&...&...& ...\\
 	                             &    04/25/2023      &   43.127 & 43.992   & 0.7   &VR& LDT&  ... & ...& ...   &...&...&...& ...\\
 	                             &    04/27/2023      &   43.112 & 43.992   &  0.6  &VR& LDT&  ... & ...& ...   &...&...&...& ...\\
 	                             &    06/18/2023      &  43.097 & 44.012  &  0.6  &g'r'i'& LDT&  ... & ...& ...   &...&...&...& ...\\
(523731) 2014~OK$_{394}$&    02/12/2023      & 36.846 & 37.007 & 1.5 & VR & LDT &$>$4   & $>$0.05 &6.27   &42.587& 0.169  & 4.1 & ?\\
 (543734) 2014~OL$_{394}$ &  09/25/2016    & 30.775 & 29.825   & 0.6 & VR & LDT & $>$4& $>$0.28 &7.78 &42.554 &0.277 &4.6  &?\\
 	           &    10/28/2017     &   30.785  & 29.811 &  0.4 & VR & LDT& $>$4 &$>$0.41 & ...   &...&...&...& ... \\
	 & 10/16/2023 &   30.120 & 31.081 & 0.5   &g'r'i'& LDT  &  ...& ... & ... & ... & ...	&... &  ...  \\
 \hline \hline 
\multicolumn{14}{c}{\textit{Mean motion resonance: 7:4}}\\                                                                 
1999~HG$_{12}$ &    04/22/2023     & 38.727 & 39.631 & 0.6 & WB & Magellan&   $>$5 &$>$0.18&7.2 &43.968 & 0.160 & 1.0 & no \\
 (129772) 1999~HR$_{11}$ &    05/16/2018     & 42.177   &41.224 & 0.5  & WB & Magellan&   $>$7 &$>$0.17&7.27 &43.450 & 0.029 & 3.3&? \\
 	                                  &    05/17/2018     & 42.177   &41.229 & 0.5  & WB& Magellan&  ... & ...& ...   &...&...&...& ...\\
(118378) 1999~HT$_{11}$ &    04/25/2023     & 38.664 & 39.566& 0.7 &VR & LDT&   $>$3  &$>$0.52 &7.26 &43.773 & 0.115 & 5.1 & no \\
 	                                  &    04/27/2023     & 38.649  &39.566& 0.6  & VR& LDT&  $>$3.5  &$>$0.44 & ...   &...&...&...& ...\\
 (60620) 2000~FD$_{8}$$^{a}$ &  04/04/2021      &     35.308&  36.214  & 0.7 &  VR & LDT & $>$2 &  $>$0.09  &  6.65  & 43.690  & 0.222 & 19.5 & no \\
&     05/14/2021       &    35.245  &36.196  & 0.5 &  VR & LDT & $>$2 &   $>$0.03  & ...  & ...  & ... & ... & \\
 (385527) 2004~OK$_{14}$ & 07/03/2017    &  34.304  & 33.540 & 1.1  & VR &  LDT & - & $\sim$0.1 &7.48 &  44.236 &0.251& 3.5 &no\\
  2004~OQ$_{15}$$^{a}$ & 07/02/2017     &     38.379 & 39.278 & 0.7 &  VR & LDT &   $>$2  & $>$0.28   &  6.64 & 43.952 & 0.132 &9.7 & ? \\
 2004~SC$_{60}$  &  10/03/2019 &  31.702&  32.669& 0.5& VR  & LDT  &  58.09$\pm$0.08 & 0.44$\pm$0.04 &  7.22 & 44.145 &  0.272	&  1.0&  ? \\
  &  10/06/2019    &   31.690 & 32.669 & 0.4  & VR  & LDT  &  ...  &  ... &  ... &... &  ...& ...&  ... \\
         &  12/01/2019  &   31.949 & 32.653  & 1.2  & WB  & Magellan  &  ...  & ...  &  ... &  ... &  ...& ...& ... \\
          &  12/02/2019  & 31.961 & 32.653  & 1.2  & WB  & Magellan  &  ...  & ...  &  ... &  ... &  ...& ...& ... \\
  &  09/24/2020    &  31.669 & 32.576& 0.8  & VR  & LDT  &  ...  &  ... &  ... &... &  ...& ...&  ... \\
 &  09/13/2021   &  31.704 & 32.492   & 1.1  & VR  & LDT  &  ...  &  ... &  ... &... &  ...& ...&  ... \\
				& 09/30/2022 &  31.497 & 32.411     & 0.7   &VR& LDT  &  ...& ... & ... & ... & ...	&... &  ...  \\
				& 10/16/2023 &     31.359 & 32.339   & 0.3   &g'r'i'& LDT  &  ...& ... & ... & ... & ...	&... &  ...  \\
 (531917) 2013~BN$_{82}$  &  02/08/2021   &  34.114 & 35.095 & 0.2  & VR  & LDT  &   18.22$\pm$0.04  &  0.40$\pm$0.04 &  6.73 &43.478 &  0.207	& 6.6 &   ? \\
        &  02/13/2021  &   34.122 & 35.097 & 0.3  & VR  & LDT  &  ...  &  ... &  ... & ...&  ...	& ... &  ...  \\
        &  03/08/2021  &     34.251 & 35.102   & 0.8  & VR  & LDT  &  ...  &  ... &  ... & ...&  ...	& ... &  ...  \\
        &  03/10/2022  &    34.341  & 35.194  &  0.8 & VR  & LDT  &  ...  &  ... &  ... & ...&  ...	& ... &  ...  \\
        &  03/26/2022  &   34.512&  35.198  & 1.2  & VR  & LDT  &  ...  &  ... &  ... & ...&  ...	& ... &  ...  \\
        &  03/27/2023  &  34.596 & 35.297& 1.2  & g'r'i'  & LDT  &  ...  &  ... &  ... & ...&  ...	& ... &  ...  \\
(532039) 2013~FR$_{28}$ & 03/01/2019     & 33.588  & 34.252 &  1.2 & WB & Magellan & 13.97$\pm$0.04 & 0.94$\pm$0.02 & 7.41 &43.384 &0.236 &3.0 & ?\\ 
	           &    03/02/2019      &  33.575 & 34.253 & 1.2  & WB& Magellan&  ... & ...& ...  &...&...&... & ...\\
	         &    05/04/2019    &  33.304 &  34.275 & 0.5 & VR & LDT&  ... & ...& ...   &...&...&...& ...\\
 	       &  05/19/2020      & 33.535 & 34.412    & 0.9  & VR &  LDT &  ... & ...& ...   &...&...&...& ...\\
	       &  05/20/2020    & 33.544 & 34.413   & 0.9  & VR &  LDT &  ... & ...& ...   &...&...&...& ...\\
   2013~SJ$_{102}$ &  09/24/2020  &  31.579 & 32.526 & 0.6  & VR  & LDT  &  -  & $\sim$0.1 &  7.55 &	44.124 &  0.284	& 7.3 & ?   \\
 (533028) 2014~AL$_{55}$ & 03/18/2017     &  35.432  & 34.781& 1.2  &  VR & LDT & $>$3  & $>$0.11 &6.67 &43.492 &0.243&4.3& ?\\  
  &    02/08/2021   &  33.752 & 34.736 &  0.1 &  VR & LDT &  ...  &... &...  & ... &  ... & ... &  ...\\
(523742) 2014~TZ$_{85}$   &    02/12/2023   &  53.419 & 54.308 &  0.5 &  VR & LDT &  $>$4   & $>$0.08 & 4.82  &43.662 &  0.255 & 15.0 &  ?\\
(559179) 2015~BR$_{518}$  &  03/02/2020   & 39.440 & 40.364  &0.5  &  VR & LDT & $>$3.5 & $>$0.06 & 6.60  &43.379&  0.114 & 9.9   & ?\\
&  02/13/2021   &    39.683 &40.452   & 0.9  &  VR & LDT & $>$5 & $>$0.14 & ... &...&  ... &...  & ...\\
(536922) 2015~FP$_{345}$ & 05/04/2019    & 35.228 & 36.196 & 0.5 &  VR & LDT &  8.47$\pm$0.02 & 0.52$\pm$0.04 &6.85 &43.407 &0.213 & 10.0 & ?\\
 &  03/02/2020    &   35.702 &  36.323 &  1.2 & VR  & LDT  & ...   &...   &...   &... & ... 	& ... &  ...  \\
   &  05/19/2020   &  35.464 & 36.356  & 0.8  & VR  & LDT  & ...   &...   &...   &... & ... 	& ... &  ...  \\
  &  05/20/2020   &35.472 & 36.356   & 0.8& VR  & LDT  & ...   &...   &...   &... & ... 	& ... &  ...  \\
         &  03/17/2021    &  35.711 & 36.486    & 1.0 & VR  & LDT  & ...   &...   &...   &... & ... 	& ... &  ...  \\
       &  03/18/2021  &   35.700 & 36.486   & 1.0 & WB  & Magellan  & ...   &...   &...   &... & ... 	& ... &  ...  \\ 
       &  03/21/2021 &  35.671 & 36.487  & 0.9 & WB & Magellan  & ...   &...   &...   &... & ... 	& ... &  ...  \\ 
      &  05/25/2022   &  35.799 & 36.677    & 0.8  & g'r'i'  & LDT  & ...   &...   &...   &... & ... 	& ... &  ...  \\
 \hline \hline 
\multicolumn{14}{c}{\textit{Mean motion resonance: 17:9 (or Classical TNO)}}\\                                                                 
(385458) 2003~SP$_{317}$  &  10/17/2020   &  41.260 & 42.239     & 0.3  & VR  & LDT  &  12.39$\pm$0.03  & 0.85$\pm$0.03   & 7.22   &	46.405 & 0.174  	& 5.1 &  no  \\
      &  09/25/2022  &   40.983  & 41.953   &  0.4 & VR  & LDT  & ...   &...   &...   &... & ... 	& ... &  ...  \\
      &  09/26/2022  &  40.978  & 41.953   & 0.3  & VR  & LDT  & ...   &...   &...   &... & ... 	& ... &  ...  \\
      &  09/30/2022  &   40.964 & 41.951  & 0.2  & VR  & LDT  & ...   &...   &...   &... & ... 	& ... &  ...  \\
 \hline
\hline
\enddata
\tablenotetext{}{\textit{\underline{Notes}}: \\
$^{a}$: Both objects are probably 7:4 resonants, but other dynamical classifications are possible: \url{https://www.boulder.swri.edu/~buie/kbo/astrom/04OQ15.html} and \url{https://www.boulder.swri.edu/~buie/kbo/astrom/60620.html}.    \\
$^{*}$: Objects were imaged by the \textit{HST} programs: 9060, 10514, 10800, 11113, and 12468 (K.S. Noll et al.), 11644 (M.E. Brown et al.), and 13664 (S.D. Benecchi et al.).}
\end{deluxetable*}

\startlongtable
\begin{deluxetable*}{cccccccc||ccccccccc}
\tablecaption{\label{modeling} Properties derived from the lightcurve of the contact binaries and elongated objects (only objects with a full lightcurve). For the contact binaries: mass ratio (q), density ($\rho$), axis ratios of the primary and secondary (abc and a'b'c', respectively), D =(a+a')/d where d is the orbital separation as defined by \citet{Leone1984} with D=1 if the two objects are in contact, albedo (Alb), and diameter ($\varnothing$). For the elongated objects: lower limit to the density, and axis ratios derived with two different viewing angles ($\xi$). Object's albedos are unknown, thus we use two by-default values of 0.04 and 0.20 \citep{Lacerda2014Albedo}.  }
\tablewidth{0pt}
\tablehead{ Object & q & $\rho$ & \(\frac{b}{a}\) &  \(\frac{c}{a}\)  &  \(\frac{b'}{a'}\)  &  \(\frac{c'}{a'}\)  & D & Alb.& $\varnothing$ &a&b&c&a'&b'&c'&d\\
      &  & [g cm$^{-3}$] &&  &   &  & & &[km]&[km]&[km]&[km]&[km]&[km]&[km]&[km]}
\startdata
\multicolumn{16}{c}{\textit{Most likely and likely contact binaries}}\\                                                                 
2003~SP$_{317}$    & 0.75 & 1 & 0.85& 0.78 & 0.79 & 0.73& 0.77 &0.04&239&74&63&57&70&56&51&187\\ 
& ... & ... & ...& ... & ... & ...& ... &0.20&107&33&28&26&32&25&23&84\\ 
& 1& 1.25 & 0.86& 0.81 & 0.79 & 0.75& 0.69 &0.04&239&69&59&56&73&58&55&205\\ 
& ... & ... & ...& ... & ... & ...& ... &0.20&107&31&27&25&33&26&24&92\\ 
2004~SC$_{60}$    &  0.4 & 1  & 1 & 1  & 1  & 1 & 0.25  &0.04& 239 &67 &67 &67 &53 &53 &53 &478 \\ 
                            & ...  & ...  & ... & ...  &...   &...  &  ... &0.20&107  &30 &30 & 30&24 &24 &24 &214 \\  
2006~CJ$_{69}$    & 0.2  & 1  & 0.99 & 0.96  &  0.84 & 0.83 & 0.50  &0.04&  210& 68& 67&65 &44 &37 & 37&224 \\ 
		     & ...  & ...  & ... & ...  &...   &...  &  ... &0.20& 94 & 30& 30& 29& 20& 17&16 &100 \\ 
2013~BN$_{82}$  &  0.3  & 1  & 0.97	&0.93&0.90&0.86  &  0.51  &0.04&  300& 94& 91& 87& 67& 60& 57&315 \\ 
 		& ...  & ...  & ... & ...  &...   &...  &  ...  &0.20&  134& 42& 41&39 &30 &27 &26 &141 \\ 
2013~FR$_{28}$    & 1 & 1 & 0.87& 0.81 & 0.80 & 0.75& 0.68 &0.04&219&63&54&51&67&53&50&190\\ 
& ... & ... & ...& ... & ... & ...& ... &0.20&98&28&24&23&30&24&22&85\\ 
\hline
\multicolumn{16}{c}{\textit{Elongated objects, $\xi$=90$^{\circ}$}}\\                                                                 
2001~QF$_{331}$    & - & $>$0.56 & 0.66 & 0.46 & - & - & - &0.04&172 &139&92&42&-&-&-&-\\ 
& - & ... & ... & ... & - & - & - &0.20&77 &62&41&19&-&-&-&-\\ 
2003~YW$_{179}$    & - & $>$0.18 & 0.59 & 0.43 & - & - & - &0.04&262&224&132&57&-&-&-&-\\ 
& - & ... & ... &... & - & - & - &0.20&117&100&59&25&-&-&-&-\\ 
2014~DK$_{143}$    & - & $>$0.49   & 0.82 & 0.53 & - & - & - &0.04 & 518 & 373&	306	&162  &-&-&-&-\\ 
 & - & ...  & ...  & ... & - & - & - &0.20& 232& 167&	137	&73&-&-&-&-\\ 
2015~FP$_{345}$    & - & $>$0.59  & 0.62  & 0.44 & - & - & - &0.04&283 & 237& 147& 65 &-&-&-&-\\ 
 & - & ...  & ...  & ... & - & - & - &0.20&127 & 106 & 66 & 29 &-&-&-&-\\ 
\multicolumn{16}{c}{\textit{Elongated objects, $\xi$=60$^{\circ}$}}\\                                                                 
2001~QF$_{331}$    & - & $>$0.58 & 0.59 & 0.43 & - & - & - &0.04&172 &147&87&37&-&-&-&-\\ 
& - & ... & ... & ... & - & - & - &0.20&77 &66&39&17&-&-&-&-\\ 
2003~YW$_{179}$    & - & $>$0.19 & 0.52 & 0.40 & - & - & - &0.04&262&238&124&49&-&-&-&-\\ 
& - & ... & ... &... & - & - & - &0.20&117&106&55&22&-&-&-&-\\ 
2014~DK$_{143}$    & - & $>$0.50   &0.74   &0.49  & - & - & - & 0.04& 518 & 397&	294	&144  &-&-&-&-\\ 
 & - & ...  & ...  & ... & - & - & - &0.20&232 & 178	&131	&64 &-&-&-&-\\ 
2015~FP$_{345}$    & - & $>$0.62  & 0.55  & 0.41 & - & - & - &0.04&283 & 251& 138& 57 &-&-&-&-\\ 
 & - & ...  & ...  & ... & - & - & - &0.20&127 & 113 & 62 & 25 &-&-&-&-\\ 
\hline
\hline
\enddata
\end{deluxetable*}

\begin{figure}
  \includegraphics[width=9cm, angle=0]{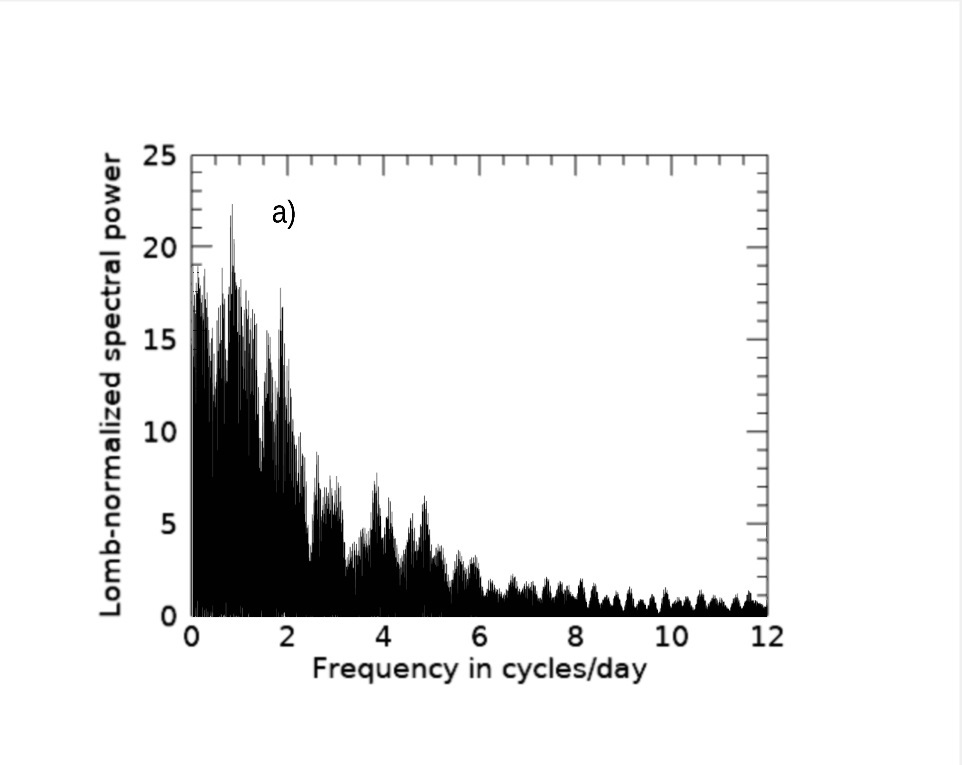}
  \includegraphics[width=9cm, angle=0]{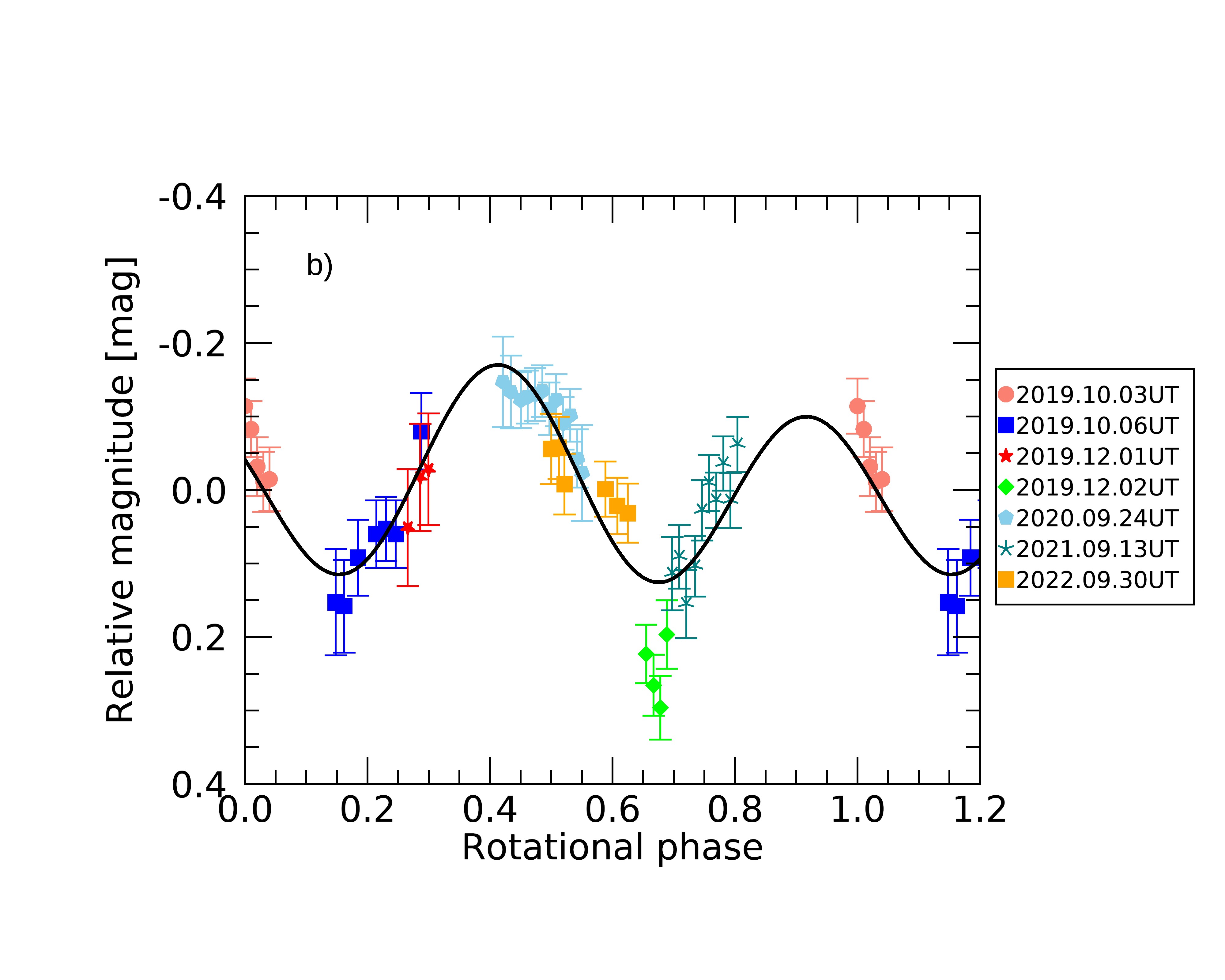}
 \caption{We suggest that 2004~SC$_{60}$ has a double-peaked rotation of 58.09$\pm$0.08~h and a $\Delta m$=0.44$\pm$0.04~mag. The lightcurve morphology is best interpreted if 2004~SC$_{60}$ is a likely contact binary. The $\chi$$^2$ of the Fourier series fit is 1.65.}
 \label{fig:SC60}
 \end{figure}

  \paragraph{2004~SC$_{60}$} We observed this object with both telescopes from 2019 to 2022. By merging all our datasets of 2004~SC$_{60}$, a long periodicity of 0.083~cycles day$^{-1}$ is found. In Figure~\ref{fig:SC60}, we report a fragmented lightcurve with P$^{double}_{rotational}$=58.09$\pm$0.08~h with $\Delta m$=0.44$\pm$0.04~mag (second maximum of the curve is incomplete thus the amplitude can be a bit larger than the one reported here). Once again, the Fourier series fit is unable to reproduce this lightcurve. Due to the fragmentary state of the lightcurve, the presence of the V-/U-shapes is difficult to evaluate, but the second minimum appears to be sharp. We infer that 2004~SC$_{60}$ is a likely contact binary with q$_{min}$$\sim$0.4 and $\rho$$_{min}$$\sim$1~g cm$^{-3}$. As the amplitude is likely larger than the one reported here, the mass ratio is probably underestimated and the density is overestimated.

On October 2023, we estimate that the surface colors of 2004~SC$_{60}$ are very-red with g'-i'=1.15$\pm$0.05~mag and g'-r'=0.83$\pm$0.05~mag.
 
\begin{figure}
  \includegraphics[width=9cm, angle=0]{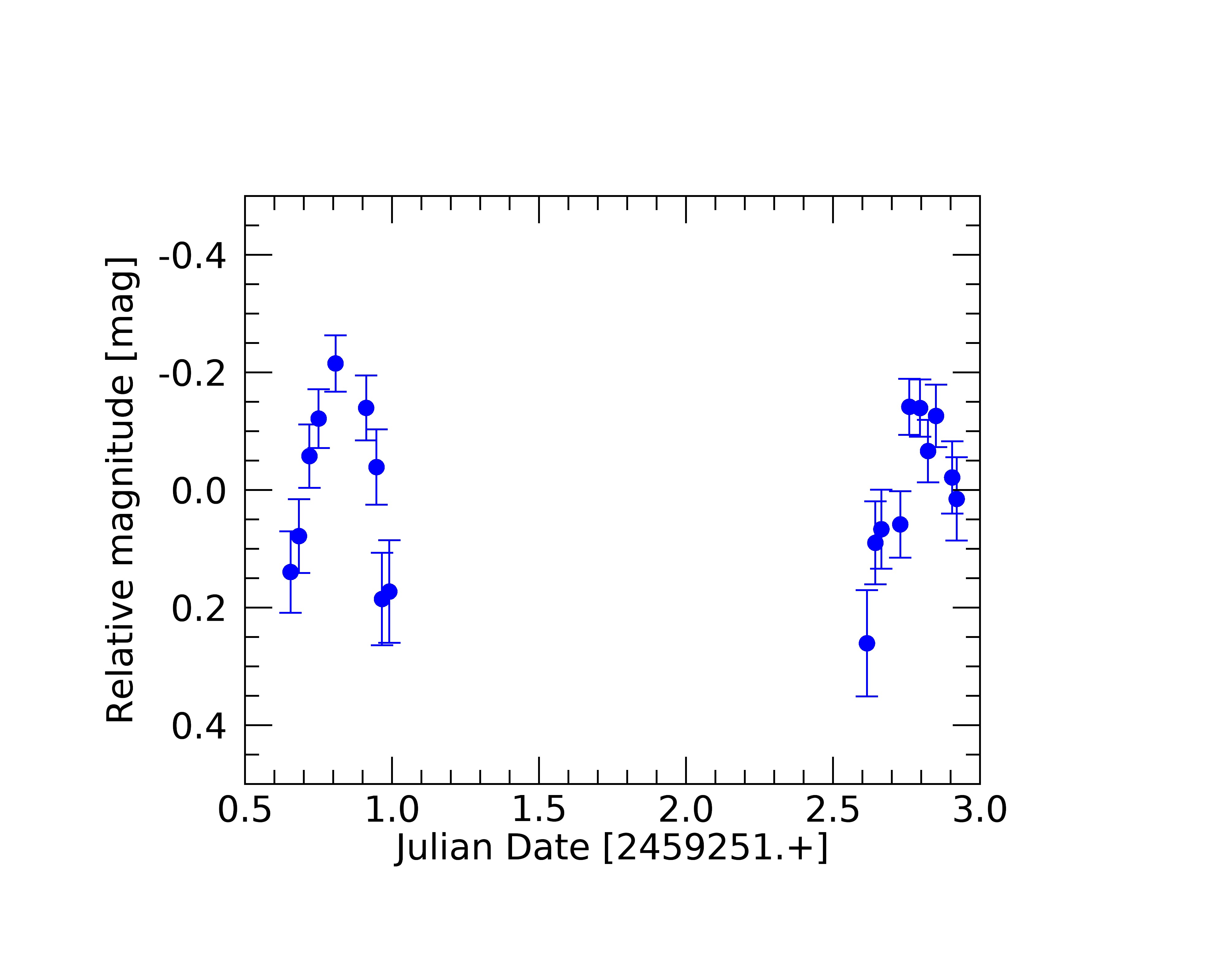}
 \caption{The partial lightcurve of 2004~VE$_{131}$ using only the December 2021 data from the \textit{LDT} display a large amplitude and gives us a hint that the lightcurve of this object is asymmetric. 2004~VE$_{131}$ is a potential contact binary. }
 \label{fig:VE131}
 \end{figure}

  \paragraph{2004~VE$_{131}$} This object was observed five times over 2021. The shortest observing block was 3.5~h whereas the longest was about 8~h, and over most blocks, 2004~VE$_{131}$ displays a large amplitude from $\sim$0.3 to 0.4~mag (Appendix~\ref{sec:appA} and Figure~\ref{fig:VE131}). None of our datasets shows a consecutive maximum and minimum over one observing night and thus we assume that this object rotates very to extremely slowly (probably several days for its period). The fourth and fifth partial lightcurves sample one (or both) of the lightcurve maxima and they look like U-shaped maxima. Also, the third partial lightcurve is a sharp minimum which can be a V-shape. Therefore, even without enough data to produce the full lightcurve of 2004~VE$_{131}$, we have some hints that this object is of interest and we propose that it is a potential contact binary. Due to the lack of a full lightcurve, we cannot extract information about the system. No other lightcurve study has been published for this object, so we cannot reaffirm our results. We call for more data to characterize this object/system. 

\begin{figure}
  \includegraphics[width=9cm, angle=0]{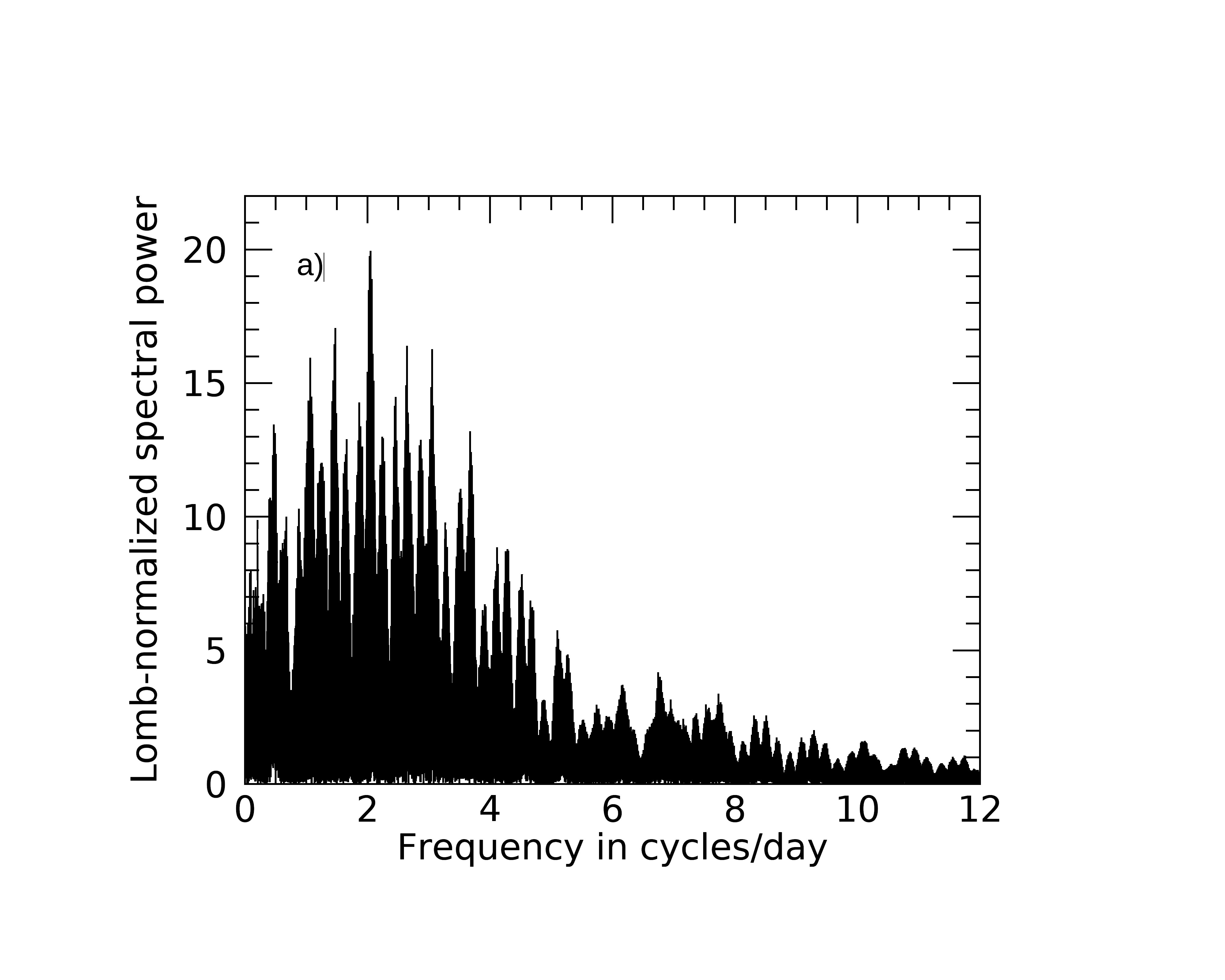}
  \includegraphics[width=9cm, angle=0]{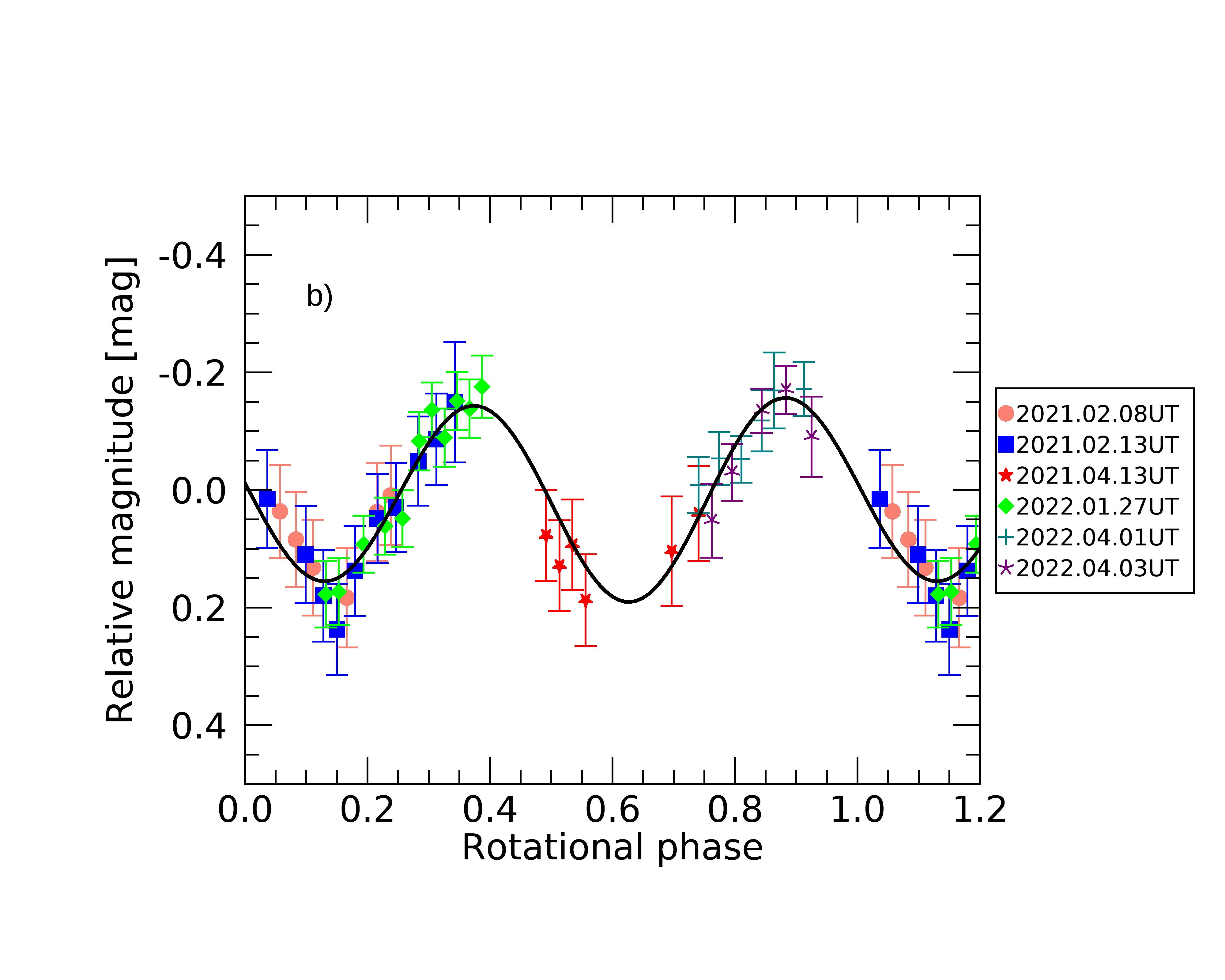}
 \caption{The periodicity of 2006~CJ$_{69}$ is P$^{double}_{rotational}$=23.39$\pm$0.07~h (with the periodogram tallest peak at f=2.05~cyles/day, plot a)). 2006~CJ$_{69}$ is a likely contact binary as shown by its variability and non-sinusoidal lightcurve. The $\chi$$^2$ of the Fourier series fit is 1.19.}
 \label{fig:CJ69}
 \end{figure}

  \paragraph{(434709) 2006~CJ$_{69}$} From 2021 to 2022, we observed 2006~CJ$_{69}$ with the \textit{LDT}. The tallest peak in Figure~\ref{fig:CJ69} is at  2.05~cycles/day. Because of the large lightcurve amplitude, the double-peaked periodicity with P$^{double}_{rotational}$=23.39$\pm$0.07~h seems adequate. The lightcurve amplitude is 0.35$\pm$0.04~mag. The overplotted fit shows that the 2006~CJ$_{69}$ lightcurve is not sinusoidal and the minimum is V-shaped. Based on this work, we interpret that 2006~CJ$_{69}$ is a likely contact binary with a mass ratio q$_{min}$$\sim$q$_{max}$$\sim$0.2 for $\rho$$_{min}$=1~g cm$^{-3}$ or $\rho$$_{max}$=5~g cm$^{-3}$. Because $\rho$$_{max}$ is unlikely for a TNO, $\rho$$_{min}$ is used for our modeling in Table~\ref{modeling}.    
 
\begin{figure}
  \includegraphics[width=9cm, angle=0]{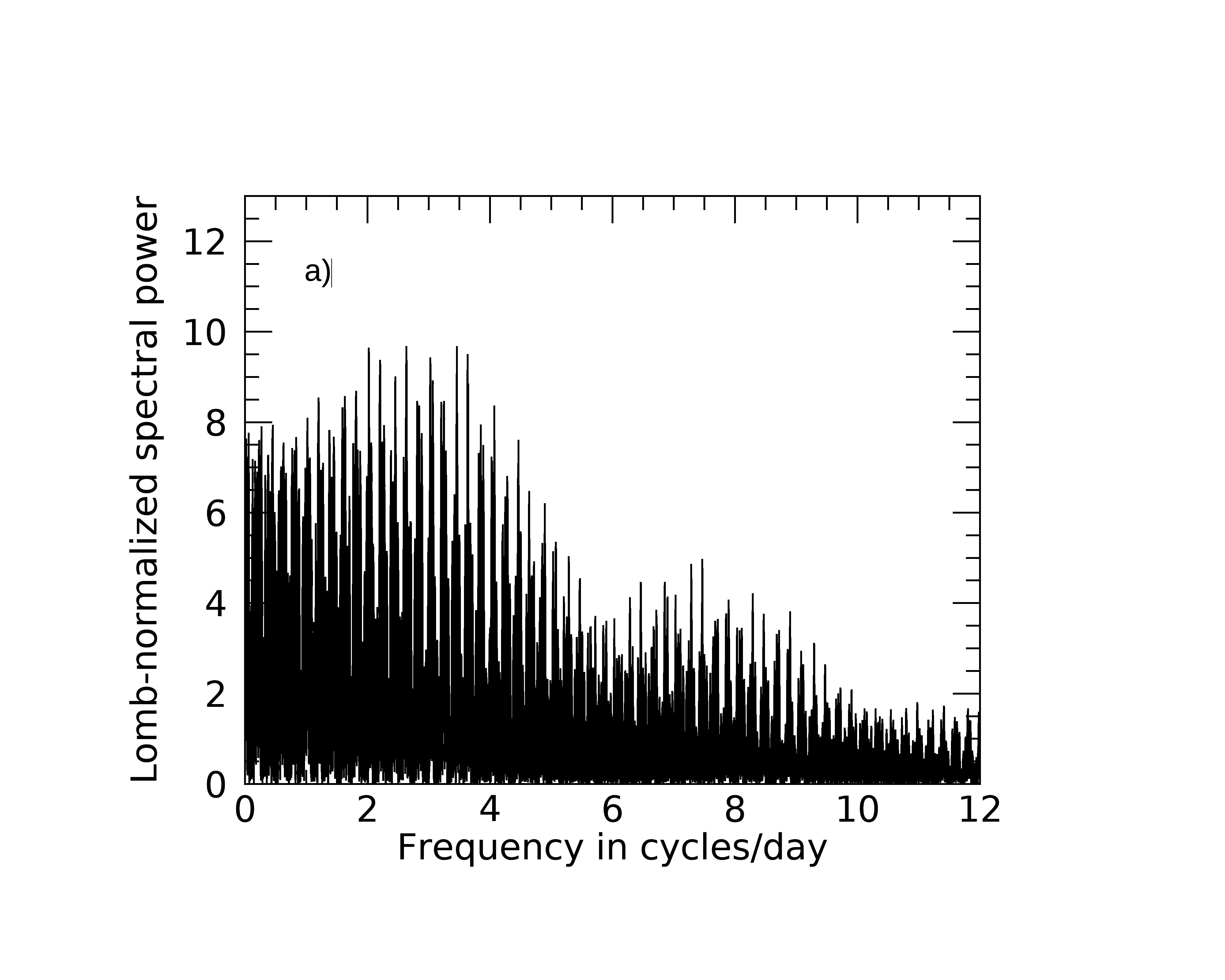}
  \includegraphics[width=9cm, angle=0]{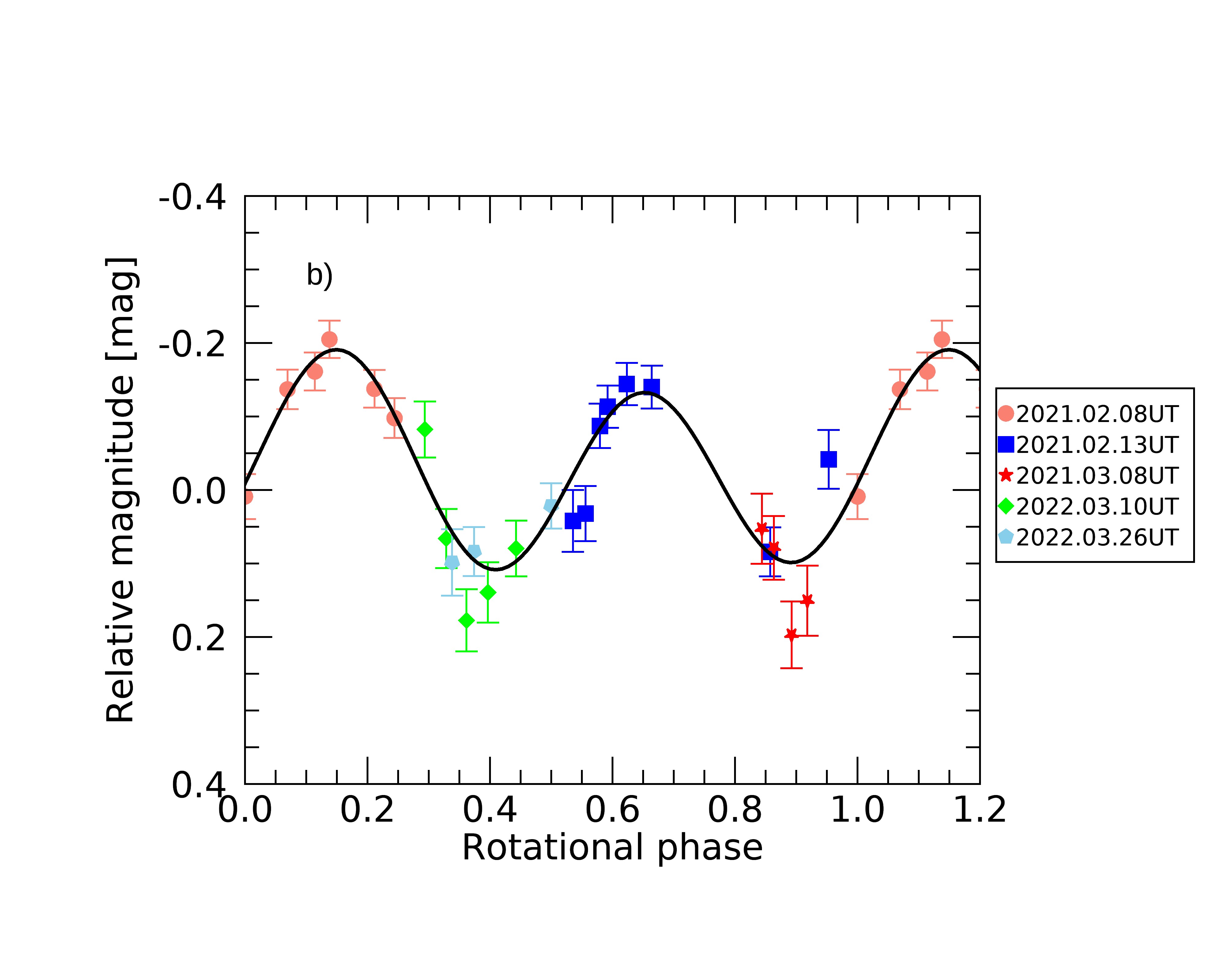}
\caption{A double-peaked lightcurve with of P$^{double}_{rotational}$=18.22$\pm$0.04~h and $\Delta m$=0.40$\pm$0.04~mag is favored for 2013~BN$_{82}$. The fit is not perfect for our data ($\chi$$^2$=1.53.) inferring that 2013~BN$_{82}$ is a likely contact binary. }
\label{fig:BN82}
\end{figure}

    \paragraph{(531917) 2013~BN$_{82}$} We imaged this object over two years with the \textit{LDT}. The strongest identified periodicity is 2.63~cycles/day, but the double-peaked lightcurve with P$^{double}_{rotational}$=18.22$\pm$0.04~h is a better match with a $\Delta m$=0.40$\pm$0.04~mag (figure~\ref{fig:BN82}). The Fourier series fit is not a perfect match, but we must point out that the data points forming the lightcurve are a bit sparse. We suggest that 2013~BN$_{82}$ is a likely contact binary, but we caution the reader that more data are required to secure this conclusion. 

We model 2013~BN$_{82}$ with a mass ratio of q$_{min}$$\sim$0.25 to q$_{max}$$\sim$0.3 with a density of $\rho$$_{min}$=1~g cm$^{-3}$ to $\rho$$_{max}$=5~g cm$^{-3}$. In Table~\ref{modeling}, we use a conservative q=0.3 and $\rho$=1~g cm$^{-3}$ to derive some basic information about this binary. 

Some g', r', i' images infer that 2013~BN$_{82}$ is an ultra-red TNO with g'-i'=1.31$\pm$0.05~mag and g'-r'=0.98$\pm$0.05~mag (Section~\ref{sec:colors}).  
   
\paragraph{(543734) 2014~OL$_{394}$} In 2016 and 2017, we obtained some images of 2014~OL$_{394}$ over approximately 4~h in both cases. Its variability is about 0.28~mag in 2016 and about 0.41~mag in 2017. We conclude that 2014~OL$_{394}$ rotates in more than 4~h with an amplitude higher than 0.41~mag. More observations are warranted to complete this lightcurve because based on its amplitude, 2014~OL$_{394}$ is a potential contact binary. 

2014~OL$_{394}$ is an ultra-red object with g'-i'=1.34$\pm$0.06~mag and g'-r'=0.87$\pm$0.06~mag.
 
  \subsection{Objects with a $\Delta m$ between 0.2 and 0.3~mag}

\begin{figure}
  \includegraphics[width=9cm, angle=0]{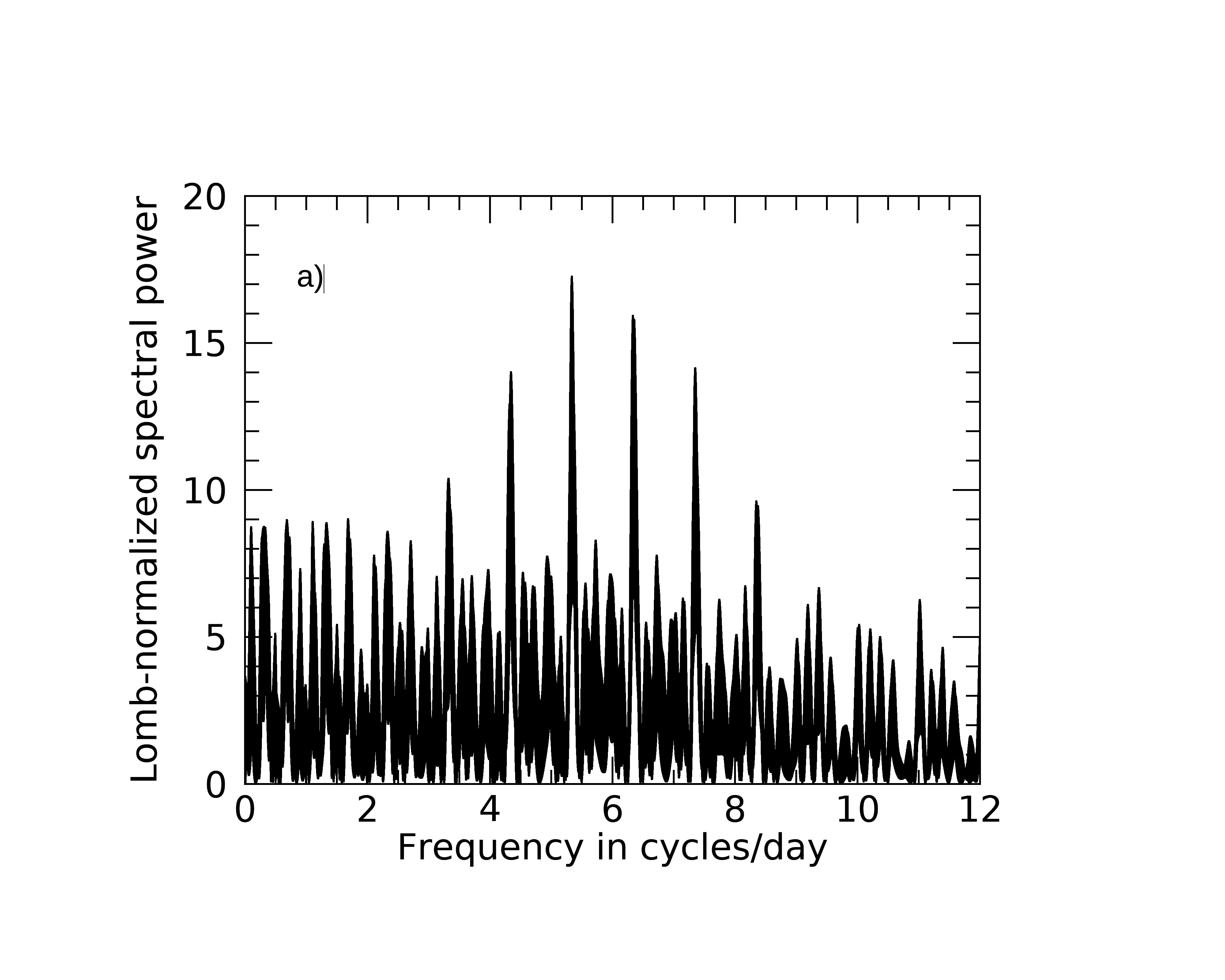}
          \includegraphics[width=9cm, angle=0]{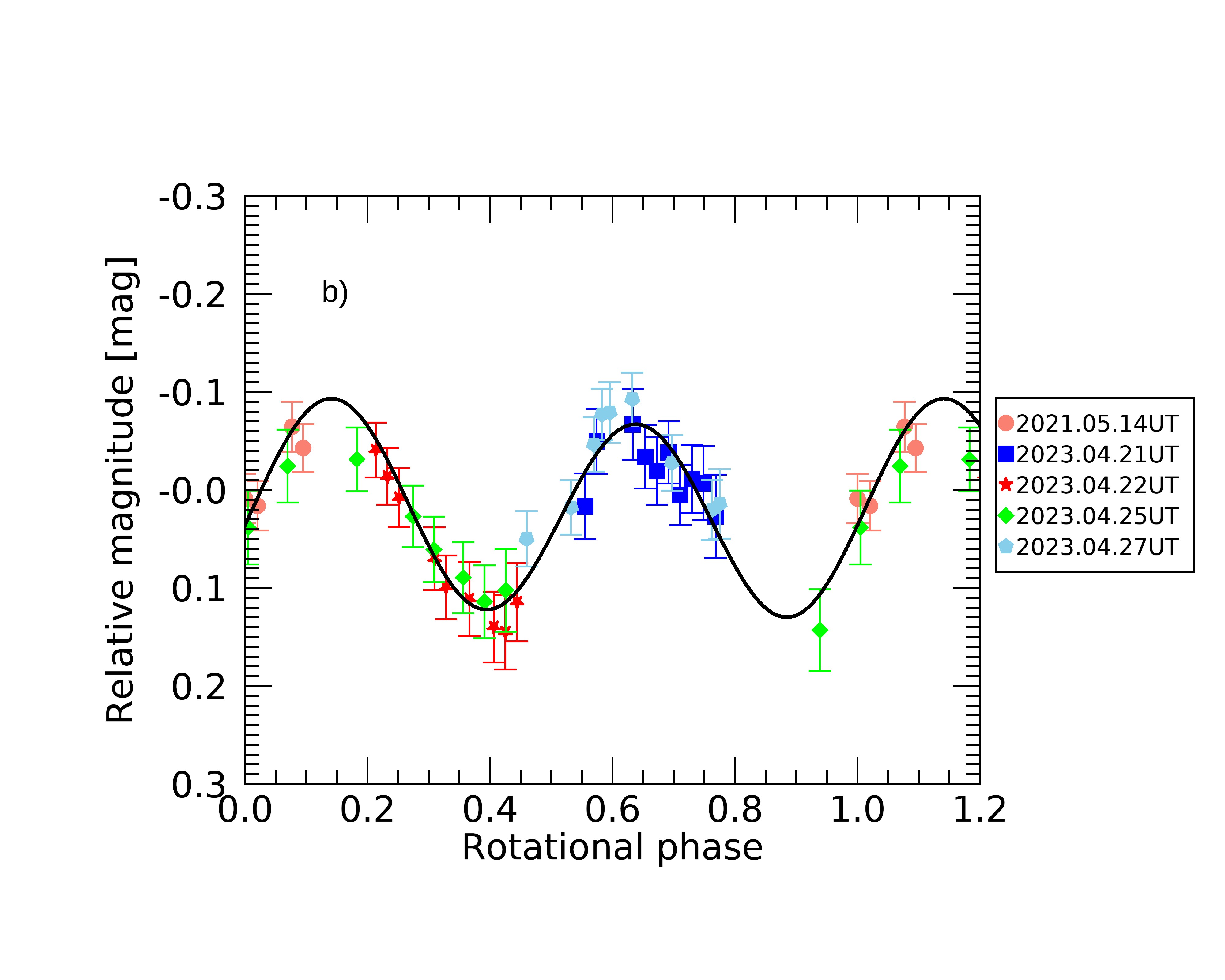}
\caption{The rotation of 2014~DK$_{143}$ is P$^{double}_{rotational}$=8.99$\pm$0.03~h. The asymmetric lightcurve with $\Delta m$=0.21$\pm$0.03~mag is well-fitted by a Fourier series demonstrating that this object is elongated and has spot(s) on its surface.  }
\label{fig:DK143}
\end{figure}

\paragraph{(523688) 2014~DK$_{143}$} We derive the complete lightcurve of 2014~DK$_{143}$ with one observing night in May 2021 and four in April 2023. The tallest peak of the periodogram in Figure~\ref{fig:DK143} is at 5.34~cycles/day. The lightcurve is asymmetric stating that the double-peaked period of 8.99$\pm$0.03~h is satisfactory with an amplitude of 0.21$\pm$0.03~mag. The lightcurve's sinusoidal nature advises that 2014~DK$_{143}$ is a moderately elongated object with a/b=1.21, c/a=0.53 and $\rho$$>$0.49 g/cm$^3$ if $\xi$=90$^\circ$ (Table~\ref{modeling}). 

In June 2023, we imaged 2014~DK$_{143}$ with the g', r', i' Sloan filters deriving its colors: g'-r'=0.80$\pm$0.03~mag and g'-i'=1.12$\pm$0.03~mag. These colors indicate that this object is one of the few moderately red objects trapped in the 5:3 resonance (Section~\ref{sec:colors}). 

\paragraph{2014~OQ$_{15}$} The partial lightcurve has a $\Delta m$$\sim$0.28~mag over about 2~h. 

  \subsection{Objects with a $\Delta m$$<$0.2~mag}

\paragraph{1999~HG$_{12}$} With one night of observing, the amplitude of 1999~HG$_{12}$ varies by about 0.18~mag in 5~h. 
\paragraph{(129772) 1999~HR$_{11}$} Over two consecutive nights, 1999~HR$_{11}$ has an amplitude of $\sim$0.17~mag. We constrain its rotational period to be more than 7~h. 
\paragraph{(60620) 2000~FD$_{8}$} Two observing instances of about 2~h each show a very low varaibility of 0.03 to 0.09~mag.  
\paragraph{(469420) 2001~XP$_{254}$} Over approximately 6~h of observations, the brightness of this TNO decreased by $\sim$0.12~mag. 2001~XP$_{254}$ was also observed by \citet{Kecskemethy2023} with the \textit{K2} space telescope. They identified two long periodicities (46.719~h and 117.157~h) that we cannot probe with our 6~h observing block with an amplitude of about 0.28~mag. 
\paragraph{(149349) 2002~VA$_{131}$} Over approximately 6~h, this object's variability is low at about 0.05~mag.  
\paragraph{2002~VV$_{130}$} From three observing blocks of 3~h, 8~h, and 3.5~h, we are unable to retrieve a periodicity and can only conclude that 2002~VV$_{130}$ rotates in more than 8~h with a $\Delta m$$\sim$0.14~mag. 
\paragraph{(533028) 2014~AL$_{55}$} We observed twice this object and its variability is $\sim$0.11~mag in 3~h. 
\paragraph{(523731) 2014~OK$_{394}$} After 4~h, we conclude that this object has a $\sim$0.05~mag variability. 
\paragraph{(523742) 2014~TZ$_{85}$} Based on an observing run carried out in February 2023, the variability of 2014~TZ$_{85}$ is only 0.08~mag in $\sim$4~h. 
\paragraph{(559179) 2015~BR$_{518}$} With images from 2020 over 3.5~h and about 5~h in 2021, we suggest that this small body rotates in more than 5~h with a variability larger than 0.14~mag. 

\paragraph{Flat lightcurves} Five TNOs -- (612086) 1999~CX$_{131}$, (385527)  2004~OK$_{14}$, (470523) 2008~CS$_{190}$, 2012~BY$_{154}$, and 2013~SJ$_{102}$ -- have such an extremely low variability that their lightcurves are flat over the observing time per target. The causes of a flat lightcurve are a spheroidal small body with a homogeneous surface, and/or a slow rotator with a period longer than the time spanned imaging the object, and/or a pole-on orientation. Based on \textit{Kepler 2} data obtained over $\sim$38~days, \citet{Kecskemethy2023} also conclude that the variability of 2008~CS$_{190}$ is extremely low with a $\Delta m$$<$0.022~mag, and no retrievable period.



\section{Discussion} 
\label{sec:dis}

 \subsection{Our survey and the literature}
 
Prior to our survey, only four 5:3 and three 7:4 resonant TNOs had some published short-term variability information (Figure~\ref{fig:Orb} and Table~\ref{literature}). Our survey increases by more than 300~\% the number of short-term studies of resonant TNOs. Alongside targeting TNOs brighter than $\sim$24~mag, selected TNOs have a mixture of absolute magnitudes and orbital elements to probe differences (if any) within the populations.

The histograms in Figure~\ref{fig:Histo} present the amplitude distributions of both resonances (and 2003~SP$_{317}$) by mixing our survey and the literature. Lightcurve amplitude ranges from flat to about 1~mag, but only $\sim$21~\% have no noticeable variability. Overall, both resonances have objects with variability up to 0.6~mag, with the exception of 2013~FR$_{28}$ and 2003~SP$_{317}$ whose amplitudes are 0.85 and 0.94~mag. The entire sample average amplitude is 0.47~mag for full lightcurve. Complete lightcurves in the 5:3/7:4 resonances have an average of 0.34/0.52~mag while the partial lightcurves have an average of 0.20/0.19~mag. The average amplitude in the 5:3 is consistent with the one in the Cold Classical population \citep{ThirouinSheppard2019a}. Bodies in the 7:4 have larger variabilities than the Cold Classicals whose high variability was demonstrated by \citet{ThirouinSheppard2019a}.

\startlongtable
\begin{deluxetable*}{cccccc|c}
\tablecaption{\label{literature} Published photometric studies of 5:3 and 7:4 resonant TNOs.   }
\tablewidth{0pt}
\tablehead{  Object & P$^{single}_{rotational}$  & P$^{double}_{rotational}$& $\Delta m$  & H$_{MPC}$ & Reference$^{a}$ & Wide Binary$^{b}$   \\
      & [h] & [h] &[mag]& [mag] &   & yes/no/?  }
\startdata
\multicolumn{7}{c}{\textit{Mean motion resonance: 5:3}}\\                                                                 
(126154) 2001~YH$_{140}$$^{c}$    & 6.22/8.45$\pm$0.05/12.99 &- & 0.19$\pm$0.14& 5.58 & O06 & no\\
                                                & 13.25$\pm$0.2 &- & 0.21$\pm$0.04&  ... & S07& ...\\   
                                                               			                  & 13.19& - & 0.13$\pm$0.05&  ... & T10, T13&...\\
&13.705$\pm$0.039& 27.397$\pm$0.172 & 0.095/0.229$\pm$0.019&  ... & K23&...\\
 (469420) 2001~XP$_{254}$   & 46.719$\pm$0.422  & - & 0.282$\pm$0.056 &  7.78 & K23 & yes\\
    & 117.157$\pm$2.328  & - & 0.275$\pm$0.056 &  ... & K23 & ... \\
 (470523) 2008~CS$_{190}$   &   &  & $<$0.022  &  6.27 & K23 & no\\
  2015~RJ$_{278}$ & -  & - &  $<$0.10 & 7.7 & A19&? \\
\hline
\hline
\multicolumn{7}{c}{\textit{Mean motion resonance: 7:4}}\\                                                                 
(119066) 2001~KJ$_{76}$$^{d}$    & 3.38$\pm$0.39 & - & 0.34$\pm$0.06 & 6.52  & K06& ? \\
(385446) 2003~QW$_{111}$      & -  & 11.92678$\pm$0.0007 & 0.48$\pm$0.01 & 6.57  & R20 & yes\\
 Manw\"e-Thorondor$^{e}$    &   &  &  &   & &\\
 2013~UK$_{17}$ &  -  & -  &$<$0.10 & 6.8  & A19 &?\\
\hline
\enddata
 \tablenotetext{a}{References are: 
 K06: \citet{Kern2006}, O06: \citet{Ortiz2006}, S07: \citet{Sheppard2007}, T10: \citet{Thirouin2010}, T13: \citet{Thirouin2013}, A19: \citet{Alexandersen2019}, R20: \citet{Rabinowitz2020}, K23: \citet{Kecskemethy2023}.} 
 \tablenotetext{b}{Known resolved binaries are listed with a ''yes", single objects without a satellite detected by the \textit{Hubble Space Telescope} (HST) have a "no", and objects never observed by \textit{HST} (and so with an unknown resolved binary nature) have a "?". The same definition has been applied in Table~\ref{Summary_photo}. Objects in this table were imaged by the \textit{HST} programs: 11113 and 12468 (K.S. Noll et al.), 11178 and 13404 (W.M. Grundy et al.).}
\tablenotetext{c}{Based on a dataset from December 2004, \citet{Ortiz2006} favored a single-peaked of 8.45~h for 2001~YH$_{140}$, but also discussed two potential aliases at 6.22~h and 12.99~h. \citet{Thirouin2010} reanalyzed the \citet{Ortiz2006} dataset and favored a rotation of 13.19~h which is in agreement with the periodicity obtained by \citet{Sheppard2007} based on an independent dataset from December 2003. By merging the datasets from \citet{Sheppard2007} and \citet{Thirouin2010}, \citet{Thirouin2013} favored a rotational period of 13.19~h. \citet{Kecskemethy2023} found a similar single-peaked periodicity of about 13~h, but they favored the double-peaked solution. We favor the single-peaked lightcurve for this paper.  }  
\tablenotetext{d}{The single-peaked lightcurve of 2001~KJ$_{76}$ reported by \citet{Kern2006} has a rotational period of 3.38~h. Due to the large amplitude and the TNO spin barrier at $\sim$4~h, the double-peaked periodicity (i.e., 6.76~h) seems more appropriate and will be used for this paper \citep{Thirouin2010, Thirouin2014}. However, \citet{Kern2006} study is based on only six images, therefore we emphasize that the periodicity and the amplitude are highly uncertain. }
\tablenotetext{e}{(385446) 2003~QW$_{111}$ is a resolved binary system whose primary is Manw\"e and its companion is Thorondor \citep{Noll2006}. \citet{Grundy2014} predicted that the system would have a mutual event season in 2015-2017. Based on \citet{Rabinowitz2020}, the system's mutual event season is explainable if Manw\"e is a contact binary and Thorondor is highly elongated. }  
\underline{\textit{Notes}}: \\
- The 7:4 resonant/classical (\url{https://www.boulder.swri.edu/~buie/kbo/astrom/160147.html}) TNO named (160147) 2001~KN$_{76}$ was observed by \citet{Kern2006}. Unfortunately, the lightcurve (and the photometry of each image) is unavailable, so we cannot assess the lightcurve quality. For the purpose of this work, we exclude the results from \citet{Kern2006} about 2001~KN$_{76}$. \\
- 2015~RE$_{278}$ is classified as a 7:4 resonant TNO by \citet{Alexandersen2019}, but as a Classical TNO according to the DES (\url{https://www.boulder.swri.edu/~buie/kbo/astrom/15RE278.html}). Because of an ambiguous classification, this TNO has been excluded from our analysis. 
\end{deluxetable*}

\begin{figure}
 \includegraphics[width=9.0cm, angle=0]{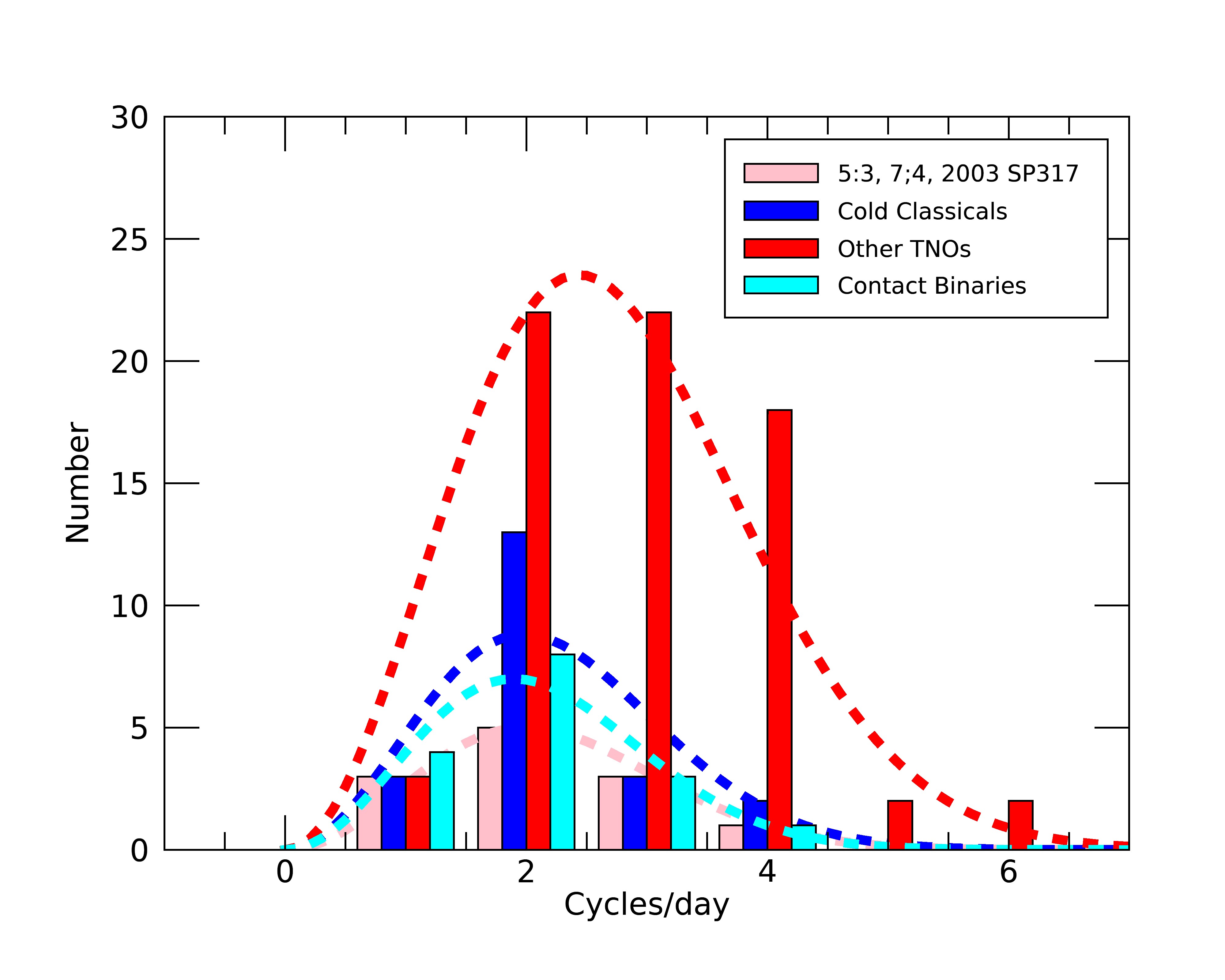}
 \caption{We overplot a Maxwellian distribution to fit the rotational frequency distribution of several sub-populations such as the Cold Classicals, the contact binaries across the trans-neptunian belt, the other TNOs including all the objects except for the Cold Classicals and the resolved binaries, and the TNOs discussed in this paper. All the sub-populations follow a Maxwellian distribution and we infer that the mean rotational period of the Cold Classicals and TNOs from this paper are similar. } 
 \label{fig:Maxwellian}
 \end{figure}

\begin{figure*}
  \includegraphics[width=9cm, angle=0]{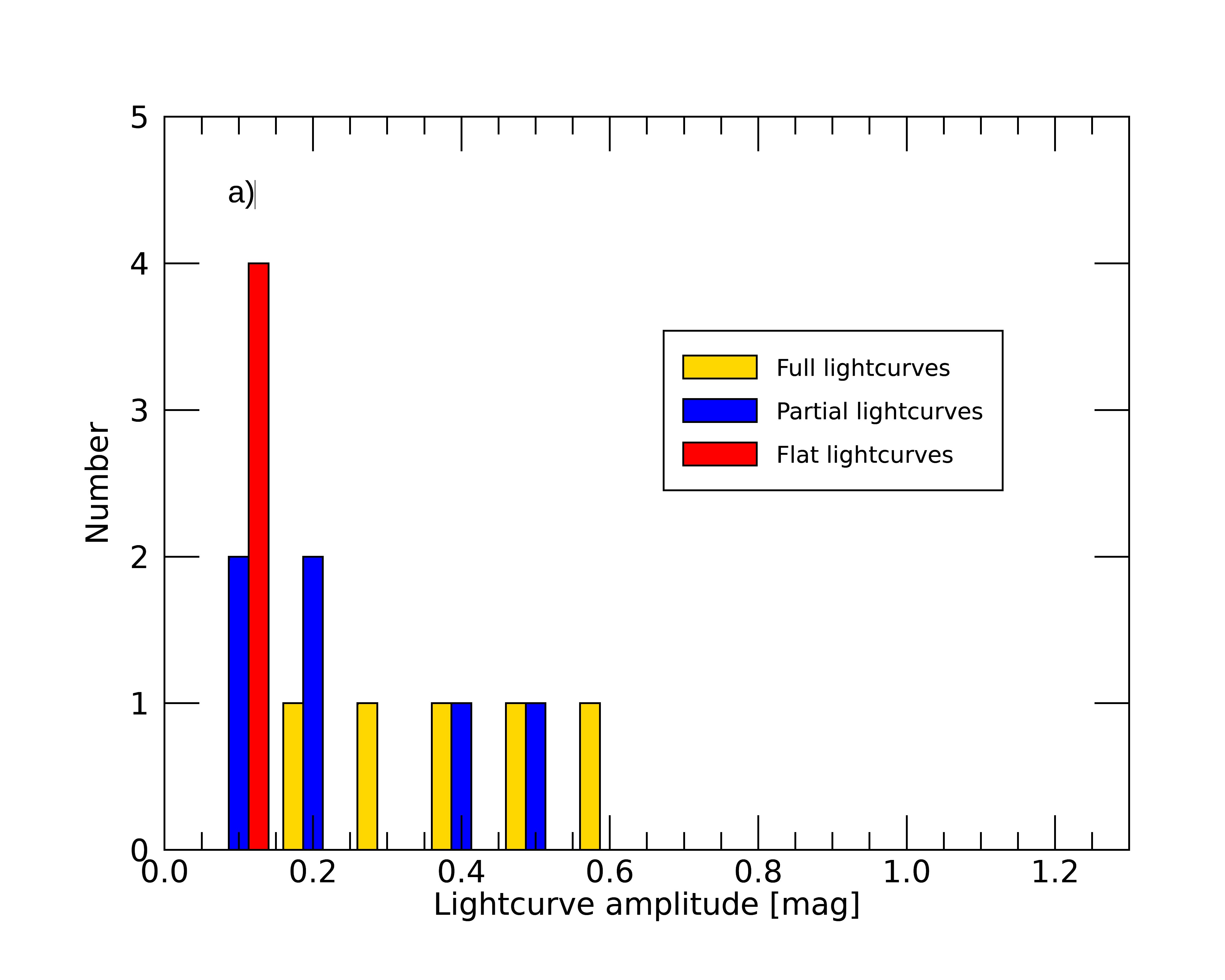}
 \includegraphics[width=9cm, angle=0]{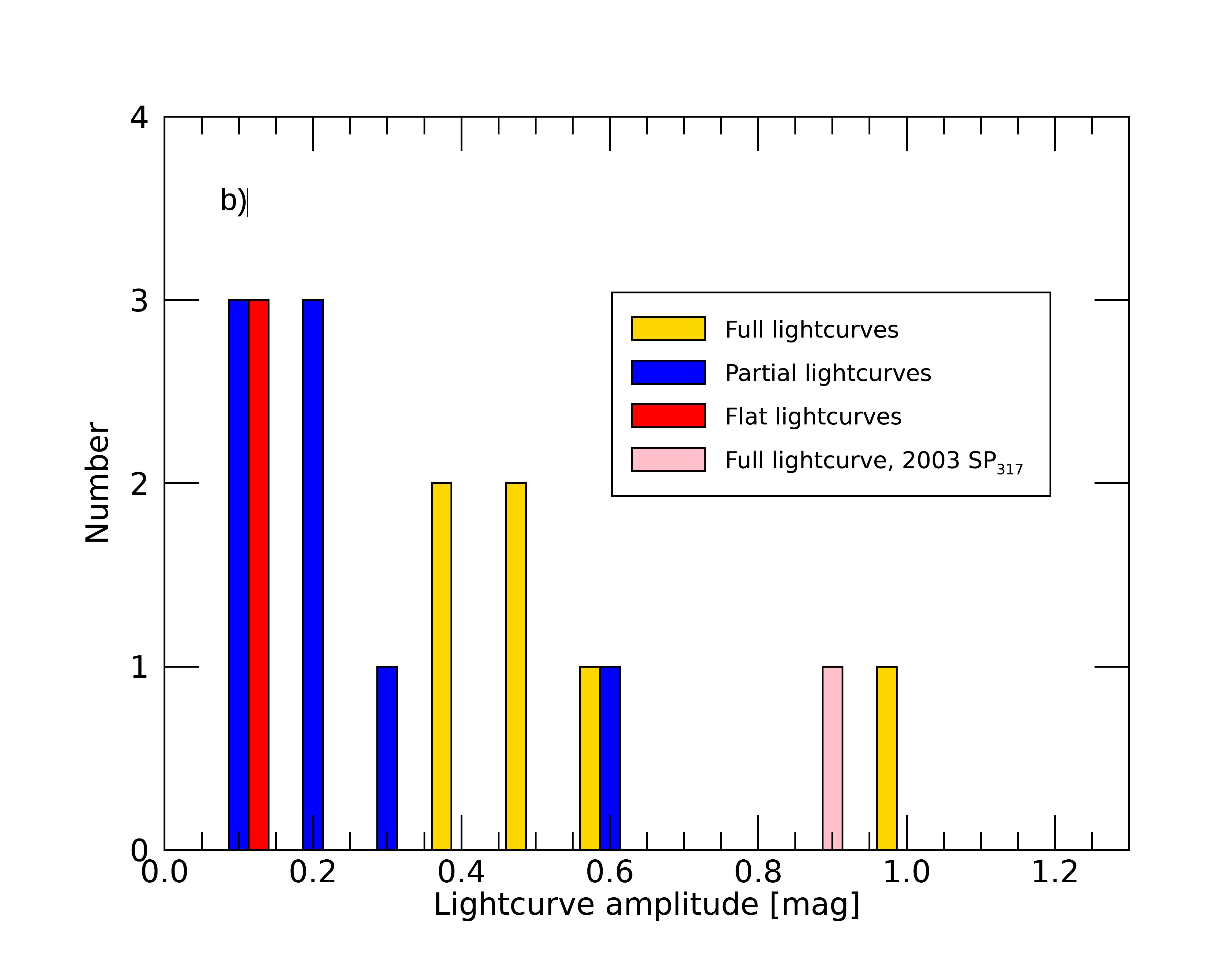}
 \includegraphics[width=9.0cm, angle=0]{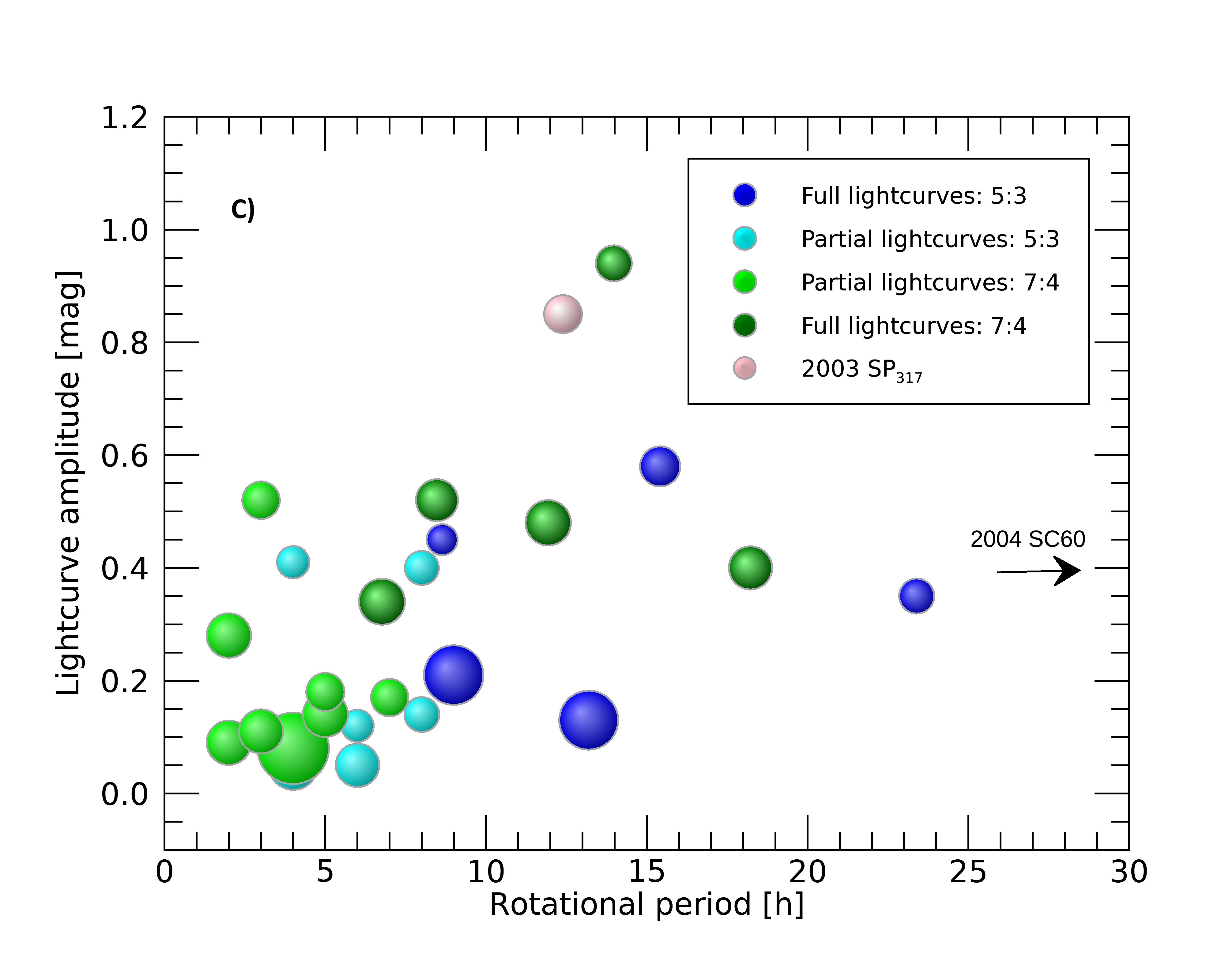}
 \caption{The amplitude distributions for the 5:3 and 7:4 resonances are plotted (Plots a) and b)). Due to the ambiguous dynamical class of 2003~SP$_{317}$, it is color-coded differently and added to the distribution of the 7:4 being the closest in semi-major axis. Only $\sim$23~\% have a flat lightcurve. Similarly, as for the Cold Classicals, there is a trend between amplitude and period (Plot c)) \citet{ThirouinSheppard2019a}). } 
 \label{fig:Histo}
 \end{figure*}

Following, we aim to compare the rotational frequency distributions of the objects studied in this paper to the most likely/likely close/contact binaries as well as to the dynamically Cold Classical population and ultimately to the other TNOs (Figure~ \ref{fig:Maxwellian}). As in \citet{Binzel1989}, we use a Maxwellian fit to match the rotational frequency distribution of these objects (Figure~\ref{fig:Maxwellian}). The fit gives a mean period$\footnote{If 2003~SP$_{317}$ is not considered, P$^{mean}_{Maxwellian}$=10.50$\pm$1.98~h}$ of P$^{mean}_{Maxwellian}$=10.67$\pm$1.93~h. This mean rotational period indicates that the resonant TNOs within the Classical belt rotate slowly in comparison to the other TNOs (P$^{mean}_{Maxwellian}$=8.74$\pm$0.66~h), and even potentially slower than the dynamically Cold Classicals (P$^{mean}_{Maxwellian}$=10.74$\pm$1.47~h) whose slow rotation was pointed out by \citet{ThirouinSheppard2019a}. However, the mean period uncertainties derived from the fits are large and thus we infer that both the resonant and Cold Classical TNOs have long rotations.

The bubble plot on Figure~\ref{fig:Histo} regroups the TNOs with partial and full lightcurves while the point size indicates the object's size. By looking at this plot, we can see a trend between period and amplitude with the slowest rotators having the highest variability. To confirm such a trend and find others, we run a correlation search using the program \texttt{ASURV} which allows datasets with upper and/or lower limits \citep{Spearman1904, Isobe1986}. (Anti-) Correlations are in Table~\ref{Correlations}. As suggested by the bubble plot, there is a strong correlation between amplitude and period with a Spearman value of 0.537 at a significance level of 99~\% for the combined 5:3, 7:4, and 2003~SP$_{317}$. By probing only the 5:3 and 7:4 resonances, this tendency is also present, but it is stronger in the 7:4 resonance. This trend is also present in the dynamically Cold Classicals \citep{ThirouinSheppard2019a}. The usual relation with amplitude and absolute magnitude (high amplitude at small size) which has been already identified in numerous studies is also found in our groupings (e.g., \citet{Sheppard2008, BenecchiSheppard2013, Thirouin2016, ThirouinSheppard2019a, Alexandersen2019}). In the 5:3 resonance, small bodies with higher variabilities are at low inclinations and slow rotators are at lower eccentricities. In the 7:4, slow rotators are at low inclinations and high eccentricities. The complete lightcurves favor high amplitudes at low inclinations. As we will discuss in Section~\ref{sec:colors}, low inclinations are dominated by Cold Classical TNOs in these resonances, and as demonstrated by \citet{ThirouinSheppard2019a}, the Cold Classicals are slow rotators with a higher amplitude compared to the other trans-neptunian populations. With low significance levels, it is possible that small objects rotate slowly in the 7:4 resonance as well as based on the complete lightcurve sample. Similitudes with the dynamically Cold Classicals and other resonant populations will be debated in Section~\ref{sec:comparison}.

 \startlongtable
\begin{deluxetable}{ccc}
\tablecaption{\label{Correlations} (Anti-) Correlation for the 5:3, 7:4 resonances, and 2003~SP$_{317}$. All trends, from very strong to non-existent are reported in this table. A trend is strong if the absolute value of the Spearman coefficient is larger than 0.3. If the absolute value is higher than 0.6, the trend is very strong, and if the absolute value is less than 0.3, there is no trend. Trends with ascending node, aphelion, and perihelion distance are likely due to observational biases. If the significance level is higher than 99~\%, the trend is very strong, but if the significance level is higher than 97.5~\% / 95~\%, the trend is strong / reasonably strong. See \citet{Spearman1904} for more details.}  
\tablewidth{0pt}
\tablehead{Values & Spearman & Significance  \\
                 & coefficient & level [\%]  }
\startdata
\multicolumn{3}{c}{\textit{All objects, \#$_{total}$=26}}\\                                                                 
Amplitude/Semimajor axis  & 0.110 & 42 \\ 
Amplitude/Absolute magnitude   & 0.322 & 89 \\ %
Amplitude/Inclination  & -0.307 & 87 \\ %
Amplitude/Eccentricity &0.077 & 30 \\ %
Amplitude/Ascending node  & -0.045 & 18 \\ 
Amplitude/Argument of perihelion  & -0.352 & 92 \\ 
Amplitude/Perihelion distance  & 0.011 & 4 \\ 
Amplitude/Aphelion distance  & 0.249 & 79 \\ 
Amplitude/Period & 0.537 & 99 \\ 
Period/Semimajor axis  & -0.173 & 61 \\ 
Period/Absolute magnitude  & 0.247 & 78 \\ %
Period/Inclination  & -0.167 & 60 \\ 
Period/Eccentricity  & 0.203 & 69 \\ %
Period/Ascending node  & 0.198 & 68 \\ 
Period/Argument of perihelion  & -0.138 &51 \\ 
Period/Perihelion distance  & -0.219 & 73 \\ 
Period/Aphelion distance  & 0.134 & 50 \\ 
\hline
\multicolumn{3}{c}{\textit{5:3 resonants, \#$_{total}$=11}}\\                                                                 
Amplitude/Semimajor axis  & 0.067 & 17 \\ 
Amplitude/Absolute magnitude   & 0.302 &66 \\ %
Amplitude/Inclination  & -0.687 & 97 \\ %
Amplitude/Eccentricity &0.168 & 40 \\ %
Amplitude/Ascending node  & 0.021 & 5 \\ 
Amplitude/Argument of perihelion  & -0.459 & 85 \\ 
Amplitude/Perihelion distance  & -0.177& 42 \\ 
Amplitude/Aphelion distance  & 0.168 & 40 \\ 
Amplitude/Period & 0.405 & 81 \\ 
Period/Semimajor axis  & -0.763 & 98 \\ 
Period/Absolute magnitude  & -0.202 & 48 \\ %
Period/Inclination  & 0.302 & 66 \\ 
Period/Eccentricity  & -0.241 & 55 \\ %
Period/Ascending node  & -0.329 & 70 \\ 
Period/Argument of perihelion  & -0.406 & 80 \\ 
Period/Perihelion distance  & 0.201 & 47 \\ 
Period/Aphelion distance  & -0.241 & 55 \\ 
\hline
\multicolumn{3}{c}{\textit{7:4 resonants, \#$_{total}$=14}}\\                                                                 
Amplitude/Semimajor axis  & -0.361 & 81 \\ 
Amplitude/Absolute magnitude   & 0.588 & 97 \\ %
Amplitude/Inclination  & -0.130 & 36 \\ %
Amplitude/Eccentricity &0.139 & 38 \\ %
Amplitude/Ascending node  & -0.213 & 56 \\ 
Amplitude/Argument of perihelion  & -0.293 & 71 \\ 
Amplitude/Perihelion distance  & -0.163 & 44 \\ 
Amplitude/Aphelion distance  & 0.176 & 47 \\ 
Amplitude/Period &0.595 & 97 \\ 
Period/Semimajor axis  & 0.105 & 29 \\ 
Period/Absolute magnitude  & 0.409 & 86 \\ %
Period/Inclination  & -0.443& 89 \\ 
Period/Eccentricity  & 0.476 & 91 \\ %
Period/Ascending node  & 0.450 &  90\\ 
Period/Argument of perihelion  & -0.036 &10 \\ 
Period/Perihelion distance  &-0.477& 91 \\ 
Period/Aphelion distance  & 0.475 & 91 \\ 
\hline
\multicolumn{3}{c}{\textit{Complete lightcurves, \#$_{total}$=12}}\\                                                                 
Amplitude/Semimajor axis  & 0.294 & 67 \\ 
Amplitude/Absolute magnitude   & 0.557 & 94 \\ %
Amplitude/Inclination  & -0.587 & 95 \\ %
Amplitude/Eccentricity & 0.301 & 68 \\ %
Amplitude/Ascending node  & -0.112 & 29 \\ 
Amplitude/Argument of perihelion  & -0.636 & 97 \\ 
Amplitude/Perihelion distance  & -0.112 & 29 \\ 
Amplitude/Aphelion distance  & 0.545 & 93 \\ 
Amplitude/Period & 0.056 & 15 \\ 
Period/Semimajor axis  & -0.126 & 32 \\ 
Period/Absolute magnitude  & 0.343 & 75 \\ %
Period/Inclination  & -0.203 & 50 \\ 
Period/Eccentricity  & 0.413 & 83 \\ %
Period/Ascending node  & 0.329 & 72 \\ 
Period/Argument of perihelion  & -0.154 & 39 \\ 
Period/Perihelion distance  & -0.385 & 80 \\ 
Period/Aphelion distance  & 0.280 & 65 \\ 
\hline
\hline
\enddata
\end{deluxetable}

 \subsection{Contact binaries}

In Section~\ref{sec:res}, we report that 2013~FR$_{28}$ is a most likely contact binary, while 2003~SP$_{317}$, 2004~SC$_{60}$, 2006~CJ$_{69}$ and 2013~BN$_{82}$ are likely contact binaries, and we also classify 1999~HT$_{11}$, 2004~VE$_{131}$, and 2014~OL$_{394}$ as potential contact binaries due to their variabilities. \citet{Rabinowitz2020} suggests that the primary of Manw\"e-Thorondor is a contact binary. Because the lightcurve of Manw\"e-Thorondor is below the 0.9~mag limit and for consistency along this paper,  we consider Manw\"e-Thorondor as a likely contact binary. In total, nine objects are showing signs of contact binarity. 

The range of variability for the contact binaries infers a diversity of the system's geometries and characteristics (Table~\ref{modeling}). All contact binary candidates have a moderate eccentricity with e$>$0.2 and a low inclination with i$<$10$^\circ$, except for 2006~CJ$_{69}$ whose inclination is 17.9$^\circ$ as well Manw\"e-Thorondor and 1999~HT$_{11}$ whose eccentricities are e$<$0.2. Out of all the contact binaries discovered with the lightcurve and occultations techniques, most of them have an inclination lower than 10$^\circ$ \citep{ThirouinSheppard2019b, Leiva2020, Buie2020, Leiva2023}. In this paper, contact binaries are small TNOs with absolute magnitude H$>$7.2~mag, aside from 2013~BN$_{82}$ whose absolute magnitude is 6.73~mag. Similarly, in the 3:2 and 2:1 resonances, contact binaries have a H$>$7~mag, but they are bigger in the Cold Classical dynamical group \citep{ThirouinSheppard2018, ThirouinSheppard2019a, ThirouinSheppard2022}. 

With a complete sample composed of our survey and the literature, the lower estimate of the fraction of nearly equal-sized contact binaries in several object groupings can be estimated. \citet{ThirouinSheppard2022} summarizes the approach and here we follow it to derive fractions based on Equation~1 (f$^{Eq.1}$) and on Equation~2 (f$^{Eq.2}$). As already mentioned in \citet{SheppardJewitt2004, ThirouinSheppard2022}, Equation~1 assumes objects with axes such as a$>$b, and b=c whereas Equation~2 considers triaxial objects with a$\geq$b=c. We list our fraction estimates in Table~\ref{CBfractions}. In the first step, only the amplitude, and not our interpretation in Section~\ref{sec:res} has been considered. As an example; in the 7:4 resonance, lightcurve information is known for seventeen TNOs and there are six objects with a $\Delta m$$\geq$0.4~mag which gives f$^{Eq.1}$$\sim$47~\% and f$^{Eq.2}$$\sim$55~\%. But, one of these five objects is 2015~FP$_{345}$ which we have classified as an elongated object. So, our second step is to consider the amplitude cut-off and our interpretation, and in such a case, the 7:4 resonance has five non-elongated objects with a $\Delta m$$\geq$0.4~mag which gives f$^{Eq.1}$$\sim$~39~\% and f$^{Eq.2}$$\sim$46~\%. 

Based on the different values reported in Table~\ref{CBfractions}, we conclude that the 5:3 resonance has a lower estimate of $\sim$10-50~\% of (nearly) equal-sized contact binary, while the 7:4 resonance has a percentage of $\sim$20-55~\%, and the full sample infers a fraction of $\sim$10-50~\%. In conclusion, both resonances have about the same fraction of contact binaries. Overall, the fractions overlap with the predicted one by \citet{Nesvorny2019} at 10 to 30~\%. 

Though there is overlap in the expected fraction of nearly equal-sized contact binaries in the 5:3, 7:4, and Cold Classical regions, in general, it appears the 5:3 and 7:4 resonances have more contact binaries than the Cold Classicals \citep{ThirouinSheppard2019a}. If further observations continue to show that the number of contact binaries in the Cold Classical belt is lower than in the resonances, it is possible that contact binaries are preferentially formed when objects are gravitationally perturbed while they escape the Cold Classical population or once they are trapped into the resonances \citep{ThirouinSheppard2019a, ThirouinSheppard2019b, Nesvorny2019}.

 \subsubsection{Contact binaries with/without moon}

Both Manw\"e-Thorondor and 2003~SP$_{317}$ have been imaged with the \textit{HST} to search for moon(s) around them. No moon was detected for 2003~SP$_{317}$, but the resolved satellite Thorondor was found around Manw\"e. Other contact binaries in our sample have never been searched for (widely) separated moon(s) (Table~\ref{Summary_photo}). The number of resolved wide binaries in both resonances is low with only two binaries out of nineteen observed TNOs ((385446) Manw\"e-Thorondor and (525816) 2005~SF$_{278}$) in the 7:4, and one ((469420) 2001~XP$_{254}$) out of thirteen in the 5:3 resonance (\textcolor{blue}{W. M. Grundy, private communication}). With $\sim$8 and $\sim$11~\% of resolved binaries in the 5:3 and 7:4 resonances respectively, contact binaries seem to be the favored kind of binaries.  

 \subsubsection{Rotation periods of contact binaries}

Rotational periods of most likely/likely contact binaries are from $\sim$6~h to $\sim$60~h, but most have periods between 10 and 20~h (\citet{SheppardJewitt2004, Lacerda2011, Lacerda2014, Thirouin2017, ThirouinSheppard2017, ThirouinSheppard2018, ThirouinSheppard2019a, Rabinowitz2020, ThirouinSheppard2022}, and this work). Using the seventeen likely/most likely contact binaries with a complete lightcurve, we fit a Maxwellian distribution with a P$^{mean}_{Maxwellian}$=11.21$\pm$1.76~h (Figure~\ref{fig:Maxwellian}). This mean period is similar to the one of the resolved binary population (10.11$\pm$1.19~h) derived by \citet{Thirouin2014} inferring that resolved and contact binaries are on average slower rotators than the more general trans-neptunian population. Using numerical simulations including Kozai effects, tidal friction and giant planets' perturbations on a synthetic population, \citet{Brunini2023} is able to reproduce the excess of contact binaries in the 3:2 (Plutino) resonance. One interesting outcome of the modeling is the rotational period of the 3:2 contact binaries (Figure 4 of \citet{Brunini2023}). Even if this distribution is only for the 3:2 TNOs, we note some similitudes with the period distribution of all contact binaries (i.e., in all sub-populations) with 76~\% of contact binaries with a periodicity less than 20~h, 18~\% with periods between 20 and 40~h, and 6~\% with periods longer than 40~h.  

 \startlongtable
\begin{deluxetable}{ccccc}
\tablecaption{\label{CBfractions} Contact binary fractions in several object groupings and with different magnitude cut-offs.   }
\tablewidth{0pt}
\tablehead{Sample &$\Delta m$$^{cut}$ &  \# & f$^{Eq.1}$ & f$^{Eq.2}$ \\
& [mag]& & [\%]& [\%]    }
\startdata
\multicolumn{5}{c}{\textit{All objects, \#$_{total}$=15}}\\                                                                 
5:3   &$\geq$0.9 & 0 & 0 & 0 \\%
    &$\geq$0.5 &  1 & $\sim$11 & $\sim$12 \\%
    &$\geq$0.4 &  4 & $\sim$36 & $\sim$42 \\%
    &$\geq$0.3 &  5 & $\sim$38 & $\sim$46 \\%
\hline 
\multicolumn{5}{c}{\textit{All objects, \#$_{total}$=17}}\\                                                                 
7:4   &$\geq$0.9 &  1 & $\sim$35& $\sim$20 \\ %
   &$\geq$0.5 &  3 & $\sim$30 & $\sim$31 \\%
    &$\geq$0.4$^{a}$ &  6 & $\sim$47 & $\sim$55 \\%
\hline
\multicolumn{5}{c}{\textit{All objects, \#$_{total}$=33}}\\                                                                 
5:3, 7:4, 2003~SP$_{317}$  &$\geq$0.9 &  1 & $\sim$18 & $\sim$10 \\%
    &$\geq$0.8 &  2 & $\sim$23 & $\sim$17 \\%
    &$\geq$0.5 &  4 & $\sim$26 & $\sim$27 \\%
    &$\geq$0.4 &  10 & $\sim$44 & $\sim$52 \\ %
    &$\geq$0.3 &  11 & $\sim$42 & $\sim$51\\%
\hline
\hline
\multicolumn{5}{c}{\textit{All objects minus elongated ones, \#$_{total}$=15}}\\                                                                 
5:3    &$\geq$0.4 &  2 & $\sim$18 & $\sim$21 \\ %
          &$\geq$0.3 &  3 & $\sim$23 & $\sim$28 \\%
\hline
\multicolumn{5}{c}{\textit{All objects minus elongated ones, \#$_{total}$=17}}\\                                                                 
7:4   &$\geq$0.9 &  1 & $\sim$35 & $\sim$20 \\ %
      &$\geq$0.5 &  2 & $\sim$20 & $\sim$21 \\%
      &$\geq$0.4 &  5 & $\sim$39 & $\sim$46 \\%
\hline
\multicolumn{5}{c}{\textit{All objects minus elongated ones, \#$_{total}$=33}}\\                                                                 
5:3, 7:4, 2003~SP$_{317}$    &$\geq$0.9 &  1 & $\sim$18 & $\sim$10 \\%
    &$\geq$0.8 &  2 & $\sim$23 & $\sim$17 \\%
    &$\geq$0.5 &  3 & $\sim$15 & $\sim$16 \\ %
    &$\geq$0.4 &  8 & $\sim$32 & $\sim$38 \\%
    &$\geq$0.3 &  9 & $\sim$31 & $\sim$38 \\ %
\hline
\hline
\enddata
\tablenotetext{a}{Only one 7:4 resonant object, 2001~KJ$_{76}$, has an amplitude between 0.3 and 0.4~mag, but as its lightcurve is uncertain, we do not consider the cut-off at 0.3~mag for the 7:4 resonance.  }
\end{deluxetable}

\subsection{Colors}
\label{sec:colors}

\begin{figure*}
 \includegraphics[width=9cm, angle=0]{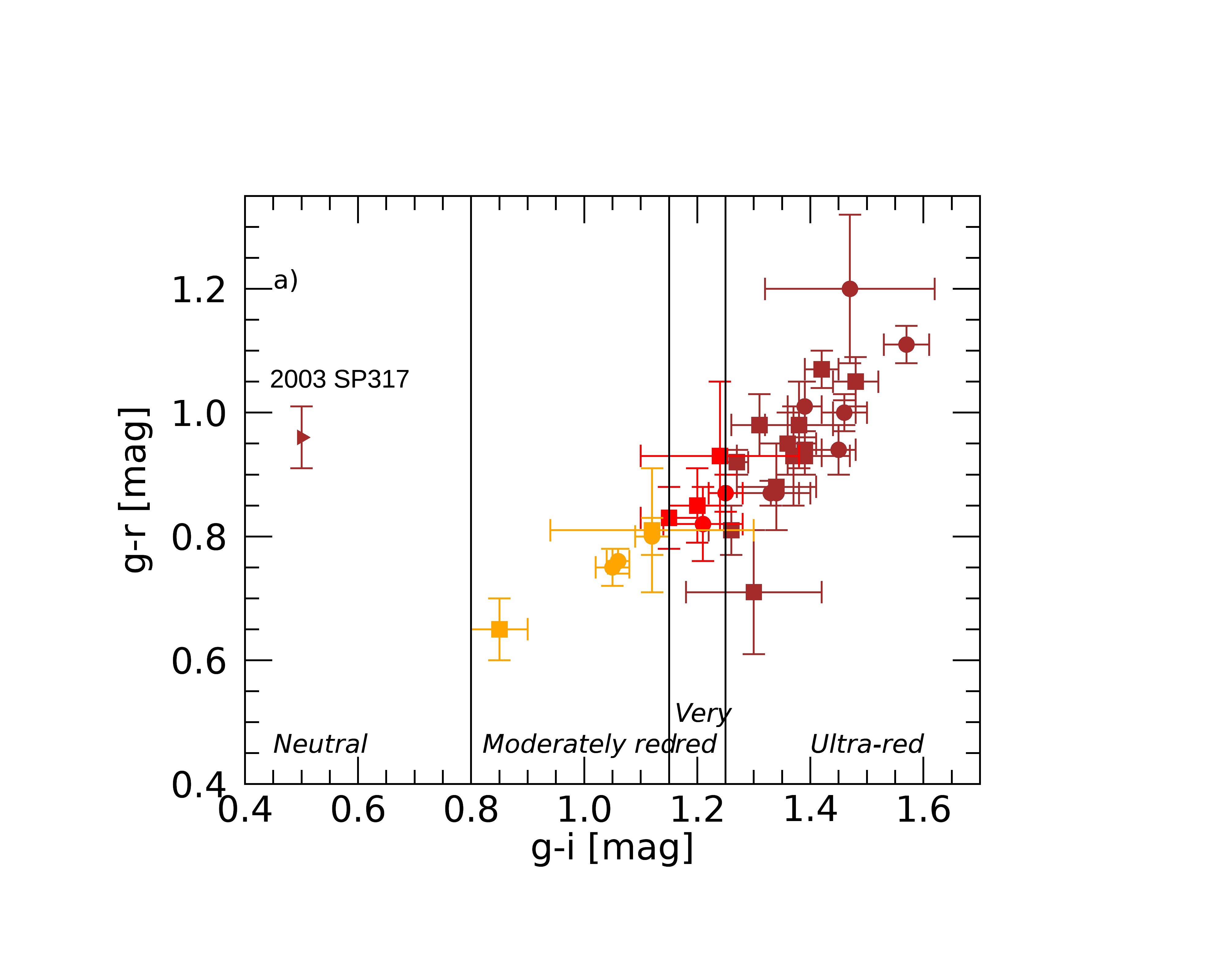}
 \includegraphics[width=9cm, angle=0]{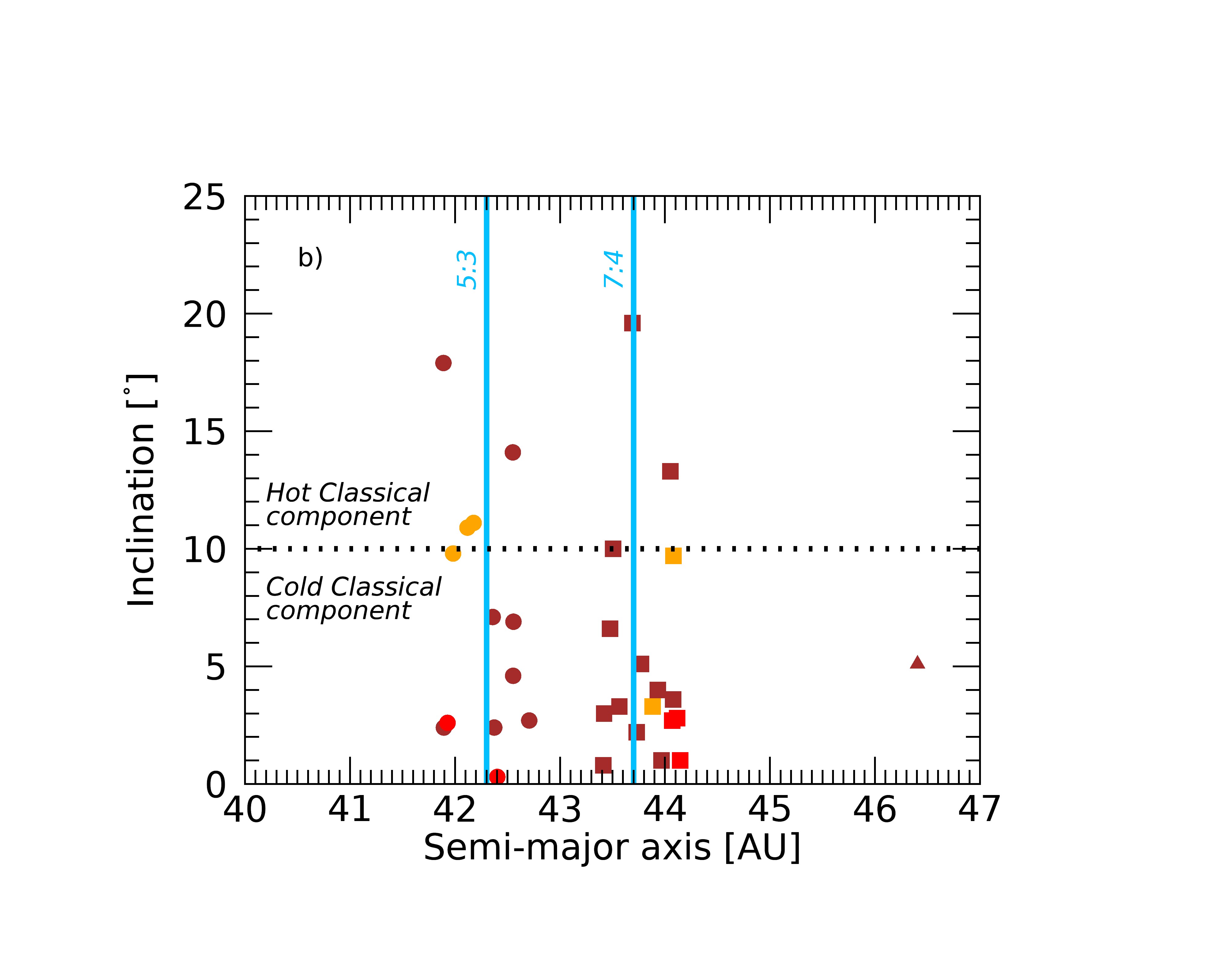}
 \includegraphics[width=9cm, angle=0]{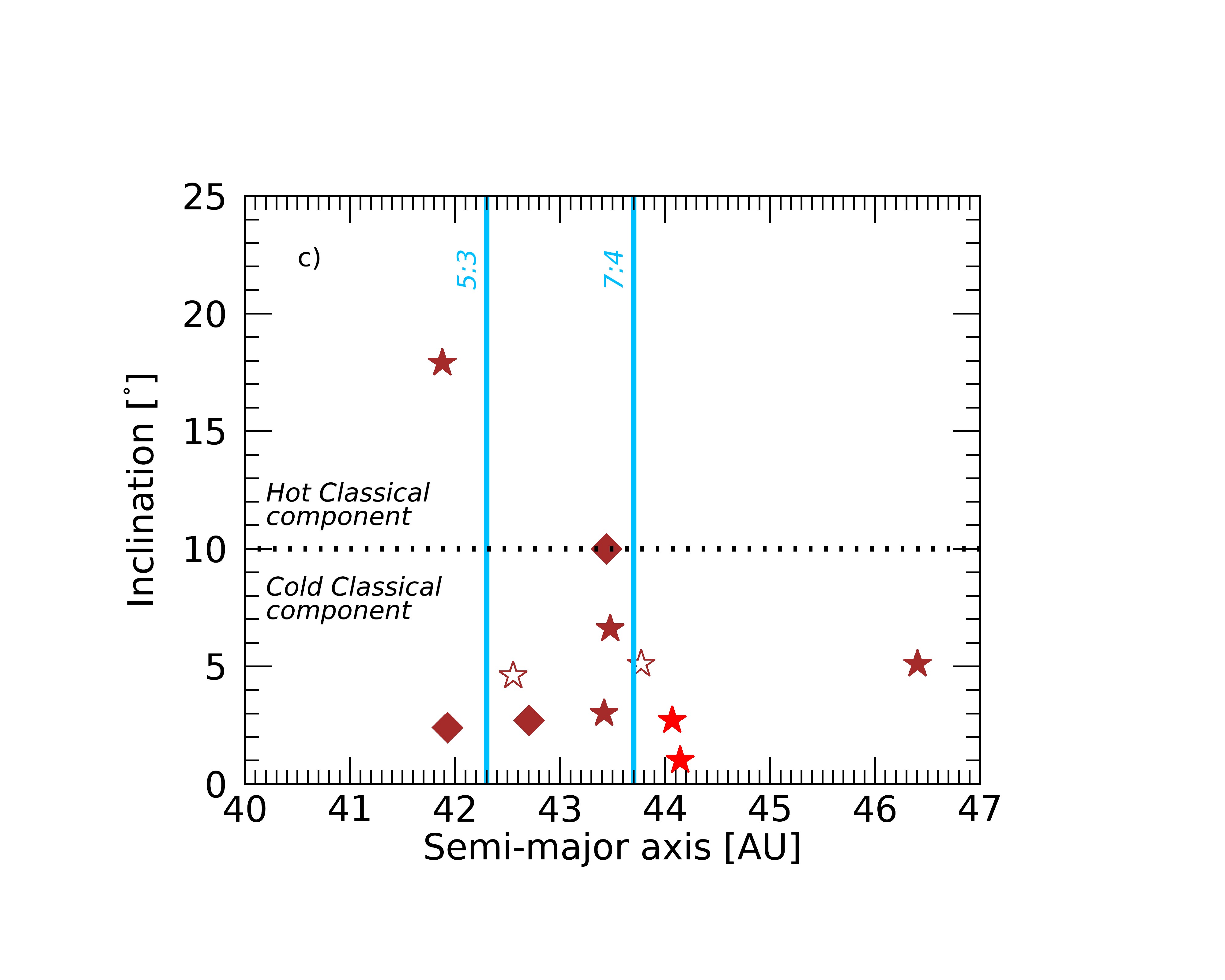} 
 \includegraphics[width=9cm, angle=0]{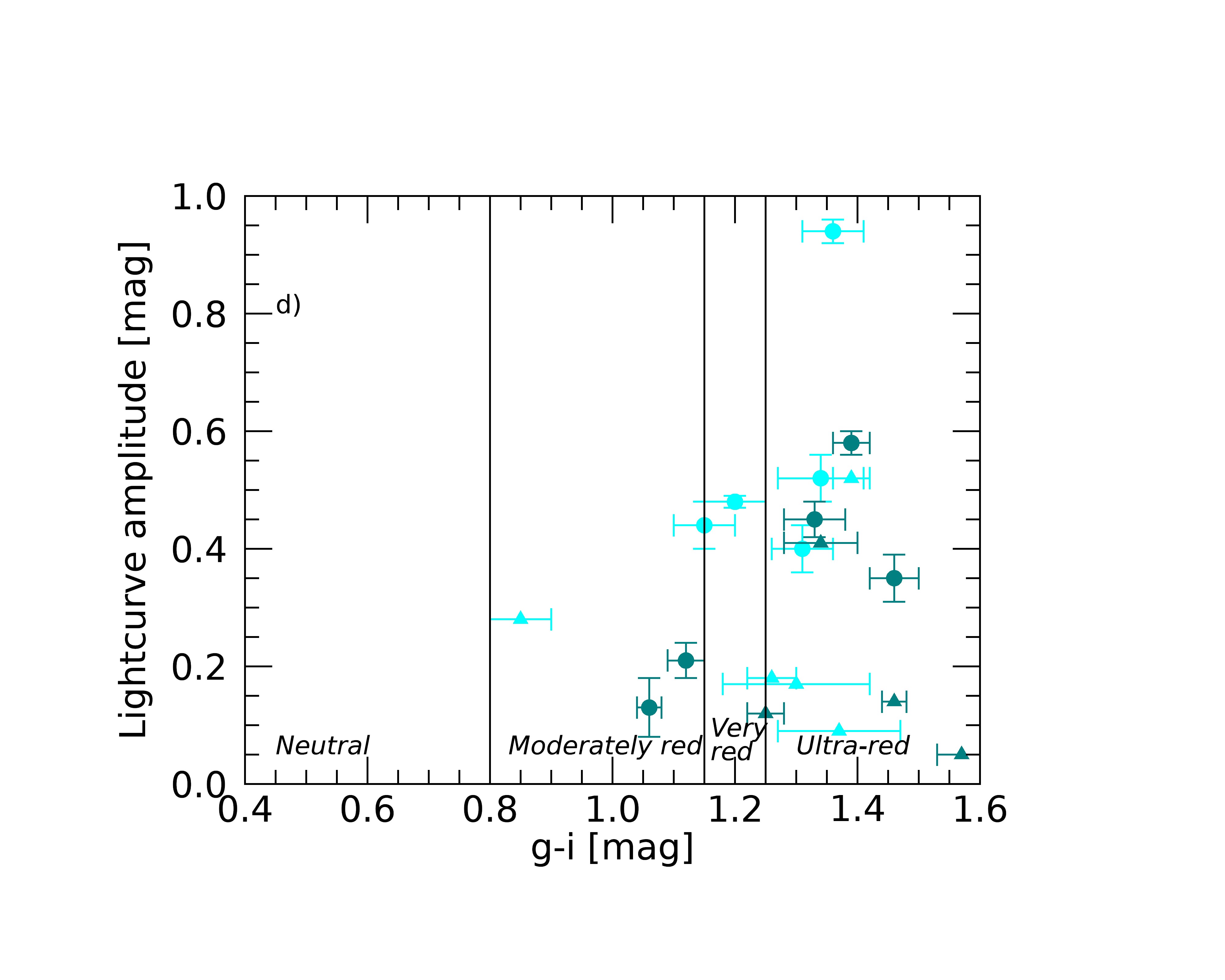} 
\caption{Surface colors of small bodies trapped in the 5:3 (circles) and in the 7:4 (squares) resonances are color-coded such as orange for moderately red, red for very red, and brown for ultra-red surfaces (plot a) and b)). The limit at an inclination of 10$^\circ$ separating the Cold from the Hot classical component is the dotted line \citep{Sheppard2012}. Contact binaries (filled stars for most likely and likely, open star for potential) and elongated objects (diamonds) are ultra-red objects, except Manw\"e-Thorondor (plot c)) In plot d), cyan and teal circles are for 7:4 and 5:3 resonants (respectively) with a full lightcurve whereas the triangles indicate a lower limit for the lightcurve amplitude. The i'-band magnitude of 2003~SP$_{317}$ (brown triangle in Plot a) and b)) is unknown therefore this object is not included in Plot d) and only the g'-r' is taken into account in Plot a).  } 
\label{fig:Colors}
\end{figure*}

\citet{Sheppard2012} conducted a comprehensive color survey of resonant TNOs, including the two (7:4 and 5:3) resonances targeted in this work. He concluded that the 5:3 and 7:4 resonances, especially at low inclinations, are dominated by ultra-red material which is common of the dynamically Cold Classical TNOs while a handful of non-ultra-red objects were identified at higher inclinations (i$>$10$^\circ$). \citet{Sheppard2012} suggested that these resonances have a low-inclination Cold Classical component dominated by Cold Classical TNOs which are now trapped in these resonances whereas the objects at higher inclination are likely from the dynamically Hot Classical population. In Figure~\ref{fig:Colors}, we plot all the published$\footnote{We updated the datasets from \citet{Sheppard2012} as some observed objects are not classified as resonant TNOs anymore.}$ colors of 5:3 and 7:4 resonants as well as colors from this work \citep{Sheppard2012, Peixinho2015, Pike2017, ThirouinSheppard2019b}. 

 \startlongtable
\begin{deluxetable}{cccc}
\tablecaption{\label{tabcolors} Colors of likely/most likely contact binaries and elongated objects. }
\tablewidth{0pt}
\tablehead{ Small &  g'-r' & g'-i'  & Ref. \\
     body & [mag]& [mag] &   }
\startdata
\multicolumn{4}{c}{\textit{Likely and most likely contact binaries}}\\                                                                 
1999~HT$_{11}$   & 0.94$\pm$0.04&1.39$\pm$0.03 & \citet{Sheppard2012}\\
2003~QW$_{111}$   & 0.85$\pm$0.06&1.20$\pm$0.05 & \citet{Sheppard2012}\\
2003~SP$_{317}$    & 0.96$\pm$0.05 & - & \citet{Pike2023} \\
2004~SC$_{60}$    &0.83$\pm$0.05& 1.15$\pm$0.05&This work\\
2006~CJ$_{69}$    & 1.00$\pm$0.03& 1.46$\pm$0.04& \citet{Sheppard2012}\\
2013~BN$_{82}$    &0.98$\pm$0.05 & 1.31$\pm$0.05 & This work\\
2013~FR$_{28}$    &0.95$\pm$0.05 &1.36$\pm$0.05&
 \citet{ThirouinSheppard2019b}\\
2014~OL$_{394}$    &0.87$\pm$0.06& 1.34$\pm$0.06& This work\\
\multicolumn{4}{c}{\textit{Elongated objects}}\\                                                                 
2001~QF$_{331}$    &0.87$\pm$0.02&1.33$\pm$0.05 & \citet{Pike2017}\\
2003~YW$_{179}$    & 1.01$\pm$0.04 &1.39$\pm$0.03 & \citet{Sheppard2012}\\
2015~FP$_{345}$    &0.88$\pm$0.07&1.34$\pm$0.07& This work\\
\hline
\hline
\enddata
\end{deluxetable}

Table~\ref{tabcolors} summarizes the g'r'i' colors\footnote{Colors of 2014~DK$_{143}$ are not included in Table~\ref{tabcolors} because this object has a moderate lightcurve amplitude.} of the contact binaries and elongated objects (with $\Delta m$$>$0.3~mag) in the 5:3 and 7:4 resonances as well as 2003~SP$_{317}$. With the exception of 2003~QW$_{111}$ (a.k.a., Manw\"e-Thorondor) and 2004~SC$_{60}$ whose surfaces are still very red, all the contact binaries and elongated objects have an ultra-red surface. This color suggests that their origins are in the dynamically Cold Classical population which is reinforced by the fact that nearly all these objects have inclinations i$\leq$10$^\circ$. The only anomaly is the ultra-red likely contact binary 2006~CJ$_{69}$ with i=17.9$^\circ$ (Figure~\ref{fig:Colors}). But, as demonstrated by \citet{Volk2011, Lykawka2005}, resonant TNOs can be excited to higher inclinations than their original ones. Therefore, the dynamically Cold Classical population may create a Cold Classical component in the resonances but some objects trapped in the resonances may also be dynamically excited to higher inclinations from initially lower inclinations. Also, as illustrated in Figure~\ref{fig:Colors}, objects with larger lightcurve amplitude tend to have high g'-i' colors demonstrating that TNOs with an origin in the Cold Classical belt are more deformed/elongated \citep{ThirouinSheppard2019a}.

\subsection{Resonant versus Cold Classical trans-neptunian objects}
\label{sec:comparison}

Following, we compare the rotational properties of the TNOs trapped in several resonances and the Cold Classical. The lightcurve survey published in \citet{ThirouinSheppard2018, ThirouinSheppard2019a, ThirouinSheppard2022} updated with newly published lightcurves (e.g., \citet{Alexandersen2019, Kecskemethy2023}) are considered. 

\citet{Kecskemethy2023} report the photometry of sixty-six TNOs in several sub-populations using \textit{Kepler 2} data, and by taking advantage of the long and continuous observing blocks, they retrieve long periodicities which are difficult from the ground because they require an extremely large amount of telescope time. Lightcurves from \citet{Kecskemethy2023} are smoothed lightcurves. Assuming that they followed the methodology in \citet{Pal2015}, we speculate that they have weighted the means of the photometry within a box with a certain number of points. \citet{Pal2015} used a box of sixty-eight points to derive a weighted mean magnitude and construct the entire smoothed (or binned) lightcurve. Because there is no detail in \citet{Kecskemethy2023}, we do not know if the same number of points per box was used nor if several numbers of points per box were tested to see if it significantly affects the smoothed lightcurve. In some occurrences, \citet{Kecskemethy2023} confirm rotational periods and amplitudes derived with ground-based lightcurves but sometimes the results are quite different. As an illustration, 2014~JQ$_{80}$ is a likely contact binary with a period of 12.16~h and a variability of 0.76~mag based on a ground-based lightcurve obtained by \citet{ThirouinSheppard2018}. \citet{Kecskemethy2023} retrieve a consistent period but an amplitude of only 0.39~mag. Data from \citet{ThirouinSheppard2018} are from May-June 2017 and September-October 2017 for \citet{Kecskemethy2023}, therefore there is no reason to expect a drastic change of the system's geometry (and so amplitude change) over about 5 months. \citet{Kecskemethy2023} argue that the inconsistent amplitudes have to be expected because the object is faint and because of the achievable \textit{Kepler 2} accuracy. However, the mean \textit{Kepler 2} magnitude of 2014~JQ$_{80}$ is $\sim$22.2~mag and thus it is not one of the faintest TNOs in their sample.  

In \citet{Kecskemethy2023}, the lightcurves of two Cold Classicals, 2003~YS$_{179}$ and (420356) 2012~BX$_{85}$ Praamzius$\footnote{The lightcurve of 2003~YS$_{179}$ is considered tentative while the lightcurve of Praamzius is secured \citep{Kecskemethy2023}.}$, catch our attention because of their extreme variabilities of 1.278$\pm$0.197~mag and 1.433$\pm$0.181~mag, respectively. If these estimates are correct, these two small bodies would have the largest variability ever recorded in the trans-neptunian belt. Unfortunately, the authors do not discuss, nor interpret or even flag the amplitudes of these objects. Such large variabilities can only be caused by contact binaries but the smoothed lightcurves do not have U-/V-shapes. Also, the individual \textit{Kepler 2} data are scattered over nearly 6~mag for Praamzius and about 3~mag for 2003~YS$_{179}$. With mean \textit{Kepler 2} magnitudes around 22.5~mag and 22~mag for 2003~YS$_{179}$ and Praamzius, respectively, they are in the same range as 2014~JQ$_{80}$. As the authors expressed their concerns regarding the 2014~JQ$_{80}$ results, we can speculate that the same concerns can be expressed for 2003~YS$_{179}$ and Praamzius. Following, results from \citet{Kecskemethy2023} are not included in our analysis as their paper presents some debatable results. 

           \begin{figure}
  \includegraphics[width=9cm, angle=0]{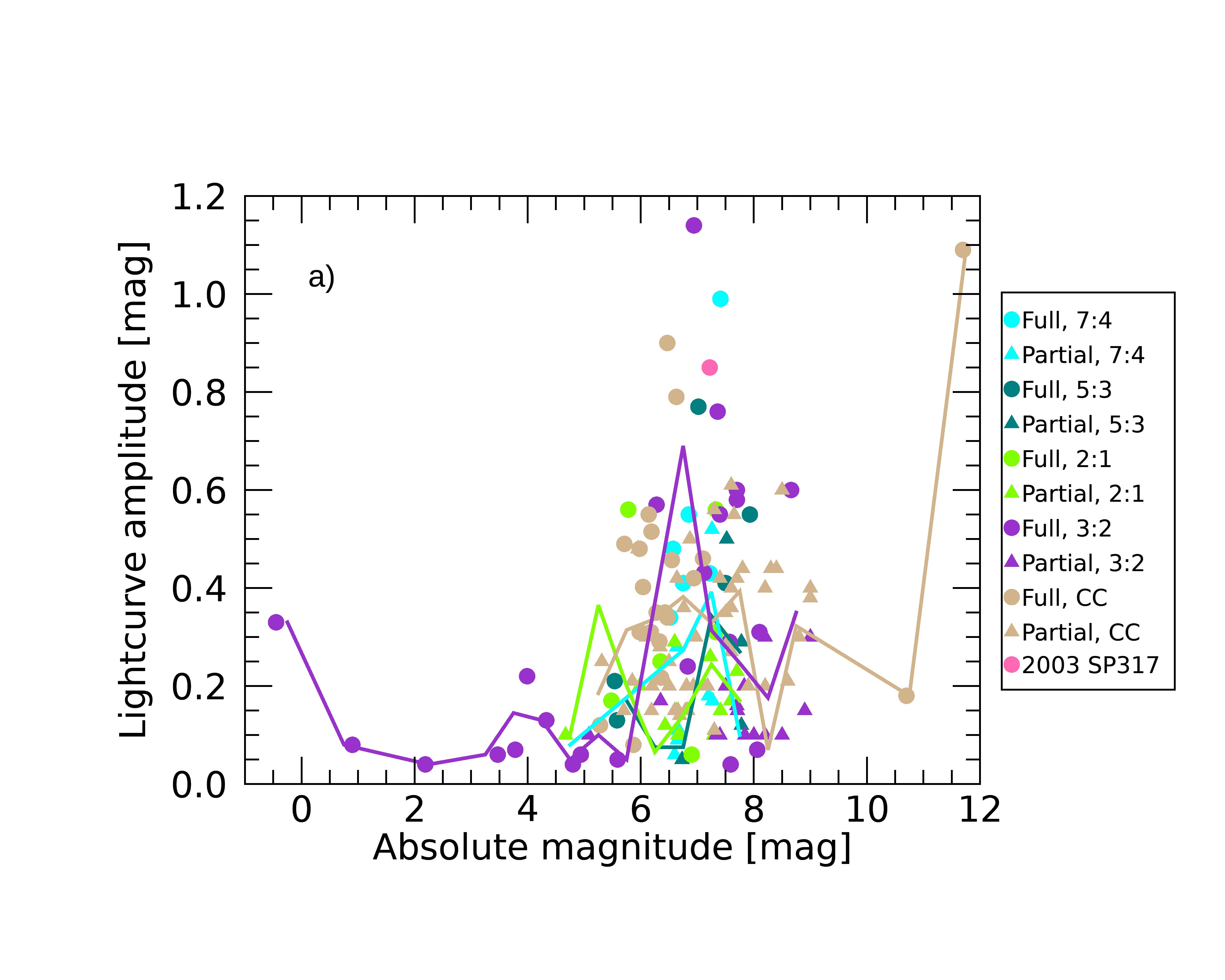}
          \includegraphics[width=9cm, angle=0]{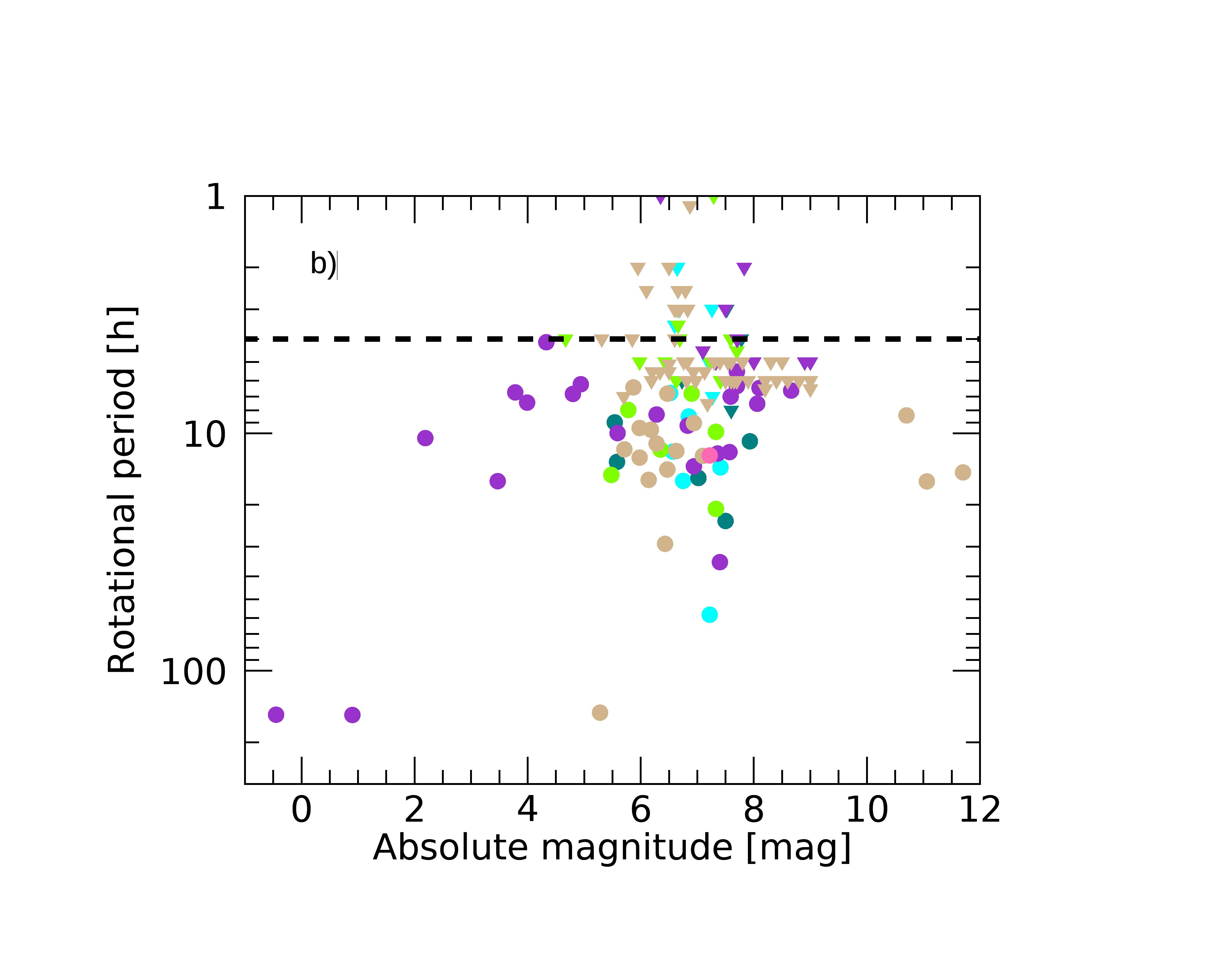}
          \includegraphics[width=9cm, angle=0]{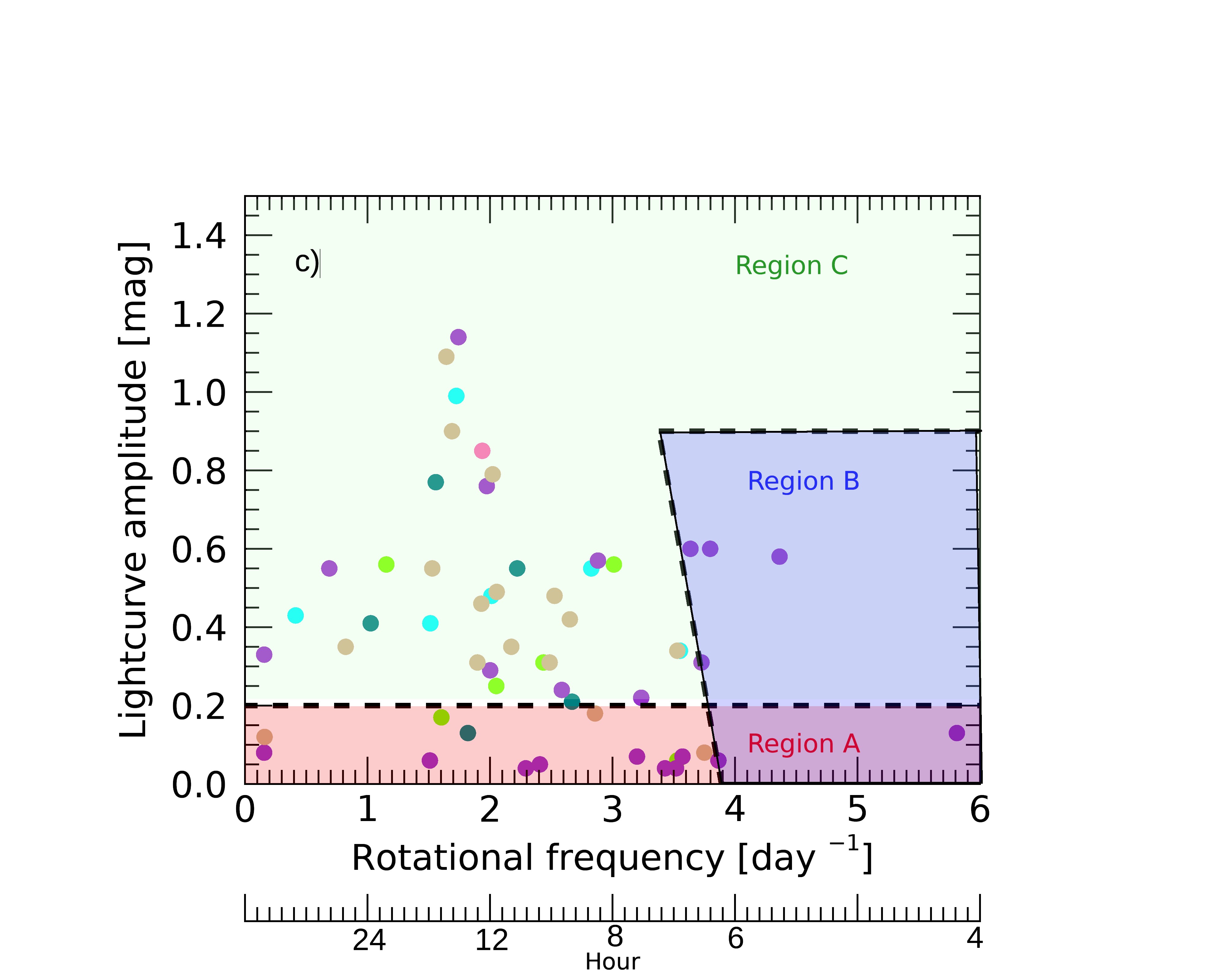}
\caption{We compare the rotational properties of the 2:1, 3:2, 7:4, 5:3 and Cold Classical (CC) TNOs (Plots a), b) and c)). Legend is the same for all plots. Running means show that small objects are more deformed in all sub-populations (Plot a)). Regions A, B, and C are defined in \citet{SheppardJewitt2004} and described in Section~\ref{sec:interpretation}. Most TNOs are in the Region C and only a handful of them have an amplitude larger than 0.9~mag (Plot c)). For clarity, error bars are not plotted.  }
\label{fig:Comparison}
\end{figure}

In Figure~\ref{fig:Comparison}, all 3:2, 5:3, 7:4, and 2:1 resonant TNOs as well as Cold Classicals (CCs) with some rotational information are plotted. Circles highlight objects with a complete lightcurve while the triangles indicate limits for the period and amplitude. Most of the observed TNOs have an absolute magnitude from 5 to 9~mag. Running means (continuous lines in Figure~\ref{fig:Comparison}) indicate that the amplitude increases at higher absolute magnitudes for all groups with a significant increase starting at H$\sim$6-7~mag based on the full and partial lightcurves (flat ones are not included). We run a 2D Kolmogorov–Smirnov (KS) as implemented in \citet{Press1992} test to check if a specific sub-population is significantly different from another using only the full lightcurves. The test returns two parameters, D which is the deviation between the cumulative distribution of the two samples, and a probability (Pr) between 0 and 1. If the two samples are not different from each other the probability will be 1. We use the Cold Classical TNOs measurements reported in \citet{ThirouinSheppard2019a} as the main reference population, as the Cold Classical TNOs are believed to be the most primordial bodies in the Solar System. Amplitudes of the 3:2 and Cold Classical TNOs are significantly different with a D=0.41 and Pr=0.06, similarly, D=0.43 and Pr=0.29 for the 2:1 compared to the Cold Classicals. Therefore, we can consider that the 3:2, and 2:1 are not from the Cold Classical. But, a D=0.18/0.36 and Pr=1/0.51 for the 5:3/7:4 suggest that both resonances can be linked to the Cold Classical population. By comparing the resonances, we note that the 3:2 and 2:1 are similar with a D=0.29 and Pr=0.75. However, the 3:2 and 2:1 are different from the 7:4 with D=0.68/0.67 and Pr=0.01/0.08, respectively. By comparing the 5:3 with the 3:2 and 2:1 resonant TNOs, there is D=0.41/0.27 and Pr=0.40/0.97.

A trans-neptunian spin barrier at $\sim$4~h has been inferred by \citet{Thirouin2010} and since then, no TNO rotating faster than this limit has been found. In this work, the fastest rotator is the 3:2 resonant (455502) 2003~UZ$_{413}$ with a period of 4.13$\pm$0.05~h while the slowest is the Cold Classical (385437) 2003~GH$_{55}$ rotating in 210.526$\pm$3.693~h \citep{Perna2009, Kecskemethy2023}. This time, we use the rotational periods to run some 2D-KS tests between the samples. Based on periods, the 7:4 and Cold Classicals are likely from the same population as D=0.20 and Pr=0.99. Surprisingly, the 2:1 and Cold Classical have a D=0.22 and Pr=0.97 suggesting that they are similar but the 5:3 and Cold Classical are less similar with D=0.25 and Pr=0.94. The 2D-KS test confirms that the 5:3 and 7:4 are likely from the same population (D=0.33, Pr=0.85) whereas the 5:3 and 3:2 are from different ones (D=0.55, Pr=0.11). 

Based on the description in Section~\ref{sec:interpretation}, the three regions to distinguish the lightcurve cause(s) are plotted in Figure~\ref{fig:Comparison}. Only resonant and Cold Classicals with complete lightcurves are considered. Most of the small bodies are in the Region C and only a handful of them are in the Region B and A. Due to the expected low lightcurve amplitude in Region A, we are probably facing an observational bias as low variability objects require a lot of telescope time and tend to not be reported in the literature. Region B is also scarce in objects showing that these TNOs tend to rotate slowly. Only six objects have a $\Delta m$$\geq$0.9~mag: 2001~QG$_{298}$ \citep{SheppardJewitt2004}, 2013~FR$_{28}$ (this work), 2003~YS$_{179}$ and 2012~BX$_{85}$ \citep{Kecskemethy2023}, the satellite of 2003~QY$_{90}$ \citep{Kern2006QY90}, and 2003~BF$_{91}$ \citep{TrillingBernstein2006}. In this Section, we expressed some concerns about 2003~YS$_{179}$ and 2012~BX$_{85}$. \citet{Kern2006QY90} published the lightcurves of 2003~QY$_{90}$ (primary, secondary, and combined lightcurves) using six data points. The secondary's lightcurve is too sparse to detect a V- or U-shape and the amplitude is highly uncertain at 0.90$\pm$0.36~mag. The Cold Classical 2003~BF$_{91}$ is a very small object (H=11.7~mag) detected with the \textit{HST}. \citet{TrillingBernstein2006} found a lightcurve amplitude of 1.09$\pm$0.25~mag without U-/V-shapes but their dataset is very noisy and they do not discuss if this object can be a contact binary. \citet{ThirouinSheppard2019a} already discussed this issue regarding this dataset and we follow their conclusions. In summary, only 2001~QG$_{298}$ and 2013~FR$_{28}$ are most likely contact binaries \citep{Lacerda2011, SheppardJewitt2004}.      

\section{Conclusion} 
\label{sec:conclusion}

Our lightcurve study of the trans-neptunian objects trapped in the 5:3 and 7:4 mean motion resonances with Neptune produces the following conclusions: 

\begin{itemize}

\item We report several elongated objects with asymmetric lightcurves that we can interpret as albedo spot(s) on the object's surface: 2001~QF$_{331}$, 2014~DK$_{143}$, and 2015~FP$_{345}$.

\item With a full lightcurve amplitude of 0.94$\pm$0.02~mag, 2013~FR$_{28}$ in the 7:4 resonance is the small body in the trans-neptunian belt with the second-largest lightcurve amplitude detected with ground-based observations after (139775) 2001~QG$_{298}$ whose variability was 1.14$\pm$0.04~mag in 2002-2003 \citep{SheppardJewitt2004}. Modeling suggests that 2013~FR$_{28}$ has a mass ratio of approximately 1 for a density of around 1~g/cm$^{3}$, and that the primary-secondary separation is less than 200~km. 

\item Aside from the most likely contact binary 2013~FR$_{28}$, we also find four likely contact binary -- 2003~SP$_{317}$, 2004~SC$_{60}$, 2006~CJ$_{69}$, and 2013~BN$_{82}$ -- and three (highly) elongated TNOs -- 2001~QF$_{331}$,  2003~YW$_{179}$, and 2015~FP$_{345}$ --. All elongated and contact binaries are at low inclinations, i$\leq$10$^\circ$, except for 2006~CJ$_{69}$ and have an ultra-red surface, except for Manw\"e-Thorondor and 2004~SC$_{60}$. We flag 1999~HT$_{11}$, 2004~VE$_{131}$, and 2014~OL$_{394}$ as potential contact binaries necessitating more data for an in-depth study.
 
\item Using the literature and our survey, we infer that there are 10-50~\% of (nearly-) equal-sized contact binaries in the 5:3 and 20-55~\% in the 7:4 resonances. The fraction of contact binaries is compatible with the one predicted by \citet{Nesvorny2019}, and it appears to be higher than the Cold Classical fraction \citep{ThirouinSheppard2019a}.  

\item Contact binary rotational periods span a large range of values from about 6~h to nearly 60~h, but most of them have a periodicity of 10 to 15~h. Using the lightcurves of all known contact binaries across the trans-neptunian belt, we report that their rotational frequency distribution follows a Maxwellian distribution. The mean rotational period of the contact binaries based on a Maxwellian distribution fit is 11.21$\pm$1.76~h which is slow compared to the rest of the trans-neptunian belt. One results from \citet{Brunini2023} is the rotational period distribution of the 3:2 resonant contact binaries, and despite the fact that this distribution is only for one sub-population, the complete contact binary population seems to follow the \citet{Brunini2023} distribution.  

\item Overall, the rotational properties of the 5:3 and 7:4 resonants are similar to the ones of the dynamically Cold Classicals because 1) by fitting a Maxwellian fit to the rotational frequency distribution, the mean period of our sample and the literature is 10.67$\pm$1.93~h, which infers that these resonants are slow rotators as the dynamically Cold Classicals are \citep{ThirouinSheppard2019a}; 2) the high average lightcurve amplitude at 0.47~mag suggests that the resonants are also far from spheroidal objects; and 3) there is a very strong correlation between lightcurve amplitude and rotational period which is also in the Cold Classical belt, but not reported in other sub-populations.
 
\item 2D Kolmogorov–Smirnov tests suggest that the 5:3 and 7:4 are drawn from the same parent population which is likely the Cold Classical one. Based on the lightcurve amplitude, the 5:3 is related to the Cold Classical, but the 7:4 is closely related to the Cold Classical based on rotational periods. The amplitude and period of the 3:2 TNOs are inconsistent with an origin as Cold Classicals. The amplitude of the 2:1 is similar to the 3:2 and inconsistent with Cold Classicals, but periods of the 2:1 are similar to the Cold Classical. 

\item Based on surface colors \citep{Sheppard2012}, binary fractions \citep{Noll2020}, and rotational properties (this work), we strengthen the case that the Cold Classicals and several resonant TNOs are linked. By studying different sub-populations, we can provide a global picture of the trans-neptunian belt, as well as find links between sub-populations, but also probe Neptune's migration and capture mechanisms \citep{MurrayClay2011}. 

\item By discovering and characterizing close/contact binaries across the trans-neptunian belt, we aim to first of all increase the number of them, but also probe their characteristics in several sub-populations as well as provide inputs for formation and evolution models which is still an open question for these systems \citep{Brunini2023}. 

\end{itemize}


\acknowledgments

This paper is dedicated to Josiane and Edmond Bouchand. This paper includes data gathered with the 6.5~m \textit{Magellan-Baade} Telescope located at Las Campanas Observatory, Chile. This research is based on data obtained at the \textit{Lowell Discovery Telescope} (LDT). Lowell Observatory is a private, non-profit institution dedicated to astrophysical research and public appreciation of astronomy and operates the LDT in partnership with Boston University, the University of Maryland, the University of Toledo, Northern Arizona University, and Yale University. Partial support of the LDT was provided by Discovery Communications. LMI was built by Lowell Observatory using funds from the National Science Foundation (AST-1005313). We are grateful to the Magellan and LDT staffs. Authors thank William M. Grundy for sharing the list of 7:4 and 5:3 resonant TNOs observed with the \textit{Hubble Space Telescope}. AT would like to thank her observing assistants, Thierry, Annick, and Zorro, for keeping her fed, highly- (borderline over-) caffeinated, and awake/entertained while observing remotely from France with a 9~h time difference. Authors thank two reviewers for their careful reading of this paper.  \\
The authors acknowledge support from the National Science Foundation with the grant 1734484 awarded to the ``Comprehensive Study of the Most Pristine Objects Known in the Outer Solar System'' and the grant 2109207 awarded to the ``Resonant Contact Binaries in the Trans-Neptunian Belt''.  

%

\vspace{5mm}
\facilities{Lowell Discovery Telescope (LDT), Magellan-Baade Telescope}





\bibliography{biblio}{}

\begin{thebibliography}{}
\expandafter\ifx\csname natexlab\endcsname\relax\def\natexlab#1{#1}\fi
\providecommand{\url}[1]{\href{#1}{#1}}
\providecommand{\dodoi}[1]{doi:~\href{http://doi.org/#1}{\nolinkurl{#1}}}
\providecommand{\doeprint}[1]{\href{http://ascl.net/#1}{\nolinkurl{http://ascl.net/#1}}}
\providecommand{\doarXiv}[1]{\href{https://arxiv.org/abs/#1}{\nolinkurl{https://arxiv.org/abs/#1}}}

\bibitem[{{Alexandersen} {et~al.}(2019){Alexandersen}, {Benecchi}, {Chen},
  {Eduardo}, {Thirouin}, {Schwamb}, {Lehner}, {Wang}, {Bannister}, {Gladman},
  {Gwyn}, {Kavelaars}, {Petit}, \& {Volk}}]{Alexandersen2019}
{Alexandersen}, M., {Benecchi}, S.~D., {Chen}, Y.-T., {et~al.} 2019, \apjs,
  244, 19, \dodoi{10.3847/1538-4365/ab2fe4}

\bibitem[{{Bannister} {et~al.}(2018){Bannister}, {Gladman}, {Kavelaars},
  {Petit}, {Volk}, {Chen}, {Alexandersen}, {Gwyn}, {Schwamb}, {Ashton},
  {Benecchi}, {Cabral}, {Dawson}, {Delsanti}, {Fraser}, {Granvik},
  {Greenstreet}, {Guilbert-Lepoutre}, {Ip}, {Jakubik}, {Jones}, {Kaib},
  {Lacerda}, {Van Laerhoven}, {Lawler}, {Lehner}, {Lin}, {Lykawka}, {Marsset},
  {Murray-Clay}, {Pike}, {Rousselot}, {Shankman}, {Thirouin}, {Vernazza}, \&
  {Wang}}]{Bannister2018}
{Bannister}, M.~T., {Gladman}, B.~J., {Kavelaars}, J.~J., {et~al.} 2018, \apjs,
  236, 18, \dodoi{10.3847/1538-4365/aab77a}

\bibitem[{{Benecchi} \& {Sheppard}(2013)}]{BenecchiSheppard2013}
{Benecchi}, S.~D., \& {Sheppard}, S.~S. 2013, \aj, 145, 124,
  \dodoi{10.1088/0004-6256/145/5/124}

\bibitem[{{Binzel} {et~al.}(1989){Binzel}, {Farinella}, {Zappala}, \&
  {Cellino}}]{Binzel1989}
{Binzel}, R.~P., {Farinella}, P., {Zappala}, V., \& {Cellino}, A. 1989, in
  Asteroids II, ed. R.~P. {Binzel}, T.~{Gehrels}, \& M.~S. {Matthews}, 416--441

\bibitem[{{Brunini}(2023)}]{Brunini2023}
{Brunini}, A. 2023, \mnras, 524, L45, \dodoi{10.1093/mnrasl/slad076}

\bibitem[{{Buie} {et~al.}(2020){Buie}, {Porter}, {Tamblyn}, {Terrell},
  {Parker}, {Baratoux}, {Kaire}, {Leiva}, {Verbiscer}, {Zangari}, {Colas},
  {Diop}, {Samaniego}, {Wasserman}, {Benecchi}, {Caspi}, {Gwyn}, {Kavelaars},
  {Ocampo Ur{\'\i}a}, {Rabassa}, {Skrutskie}, {Soto}, {Tanga}, {Young},
  {Stern}, {Andersen}, {Arango P{\'e}rez}, {Arredondo}, {Artola}, {B{\^a}},
  {Ballet}, {Blank}, {Bop}, {Bosh}, {Camino L{\'o}pez}, {Carter},
  {Castro-Chac{\'o}n}, {Caycedo Desprez}, {Caycedo Guerra}, {Conard},
  {Dauvergne}, {Dean}, {Dean}, {Desmars}, {Dieng}, {Bousso Dieng}, {Diouf},
  {Dorego}, {Dunham}, {Dunham}, {Durantini Luca}, {Edwards}, {Erasmus}, {Faye},
  {Faye}, {Ferrario}, {Ferrell}, {Finley}, {Fraser}, {Friedli}, {Galvez Serna},
  {Garcia-Migani}, {Genade}, {Getrost}, {Gil-Hutton}, {Gimeno}, {Golub},
  {Gonz{\'a}lez Murillo}, {Grusin}, {Gurovich}, {Hanna}, {Henn}, {Hinton},
  {Hughes}, {Josephs}, {Joya}, {Kammer}, {Keeney}, {Keller}, {Kramer},
  {Levine}, {Lisse}, {Lovell}, {Mackie}, {Makarchuk}, {Manzano}, {Mbaye},
  {Mbaye}, {Melia}, {Moreno}, {Moss}, {Ndaiye}, {Ndiaye}, {Nelson}, {Olkin},
  {Olsen}, {Ospina Moreno}, {Pasachoff}, {Pereyra}, {Person}, {Pinz{\'o}n},
  {Pulver}, {Quintero}, {Regester}, {Resnick}, {Reyes-Ruiz}, {Rolfsmeier},
  {Ruhland}, {Salmon}, {Santos-Sanz}, {Santucho}, {Sep{\'u}lveda Ni{\~n}o},
  {Sickafoose}, {Silva}, {Singer}, {Skipper}, {Slivan}, {Smith}, {Spagnotto},
  {Stephens}, {Strabala}, {Tamayo}, {Throop}, {Torres Ca{\~n}as}, {Toure},
  {Traore}, {Tsang}, {Turner}, {Vanegas}, {Venable}, {Wilson}, {Zuluaga}, \&
  {Zuluaga}}]{Buie2020}
{Buie}, M.~W., {Porter}, S.~B., {Tamblyn}, P., {et~al.} 2020, \aj, 159, 130,
  \dodoi{10.3847/1538-3881/ab6ced}

\bibitem[{{Chandrasekhar}(1987)}]{Chandrasekhar1987}
{Chandrasekhar}, S. 1987, {Ellipsoidal figures of equilibrium}

\bibitem[{{Degewij} {et~al.}(1979){Degewij}, {Tedesco}, \&
  {Zellner}}]{Degewij1979}
{Degewij}, J., {Tedesco}, E.~F., \& {Zellner}, B. 1979, \icarus, 40, 364,
  \dodoi{10.1016/0019-1035(79)90029-0}

\bibitem[{{Dias-Oliveira} {et~al.}(2017){Dias-Oliveira}, {Sicardy}, {Ortiz},
  {Braga-Ribas}, {Leiva}, {Vieira-Martins}, {Benedetti-Rossi}, {Camargo},
  {Assafin}, {Gomes-J{\'u}nior}, {Baug}, {Chandrasekhar}, {Desmars}, {Duffard},
  {Santos-Sanz}, {Ergang}, {Ganesh}, {Ikari}, {Irawati}, {Jain}, {Liying},
  {Richichi}, {Shengbang}, {Behrend}, {Benkhaldoun}, {Brosch}, {Daassou},
  {Frappa}, {Gal-Yam}, {Garcia-Lozano}, {Gillon}, {Jehin}, {Kaspi}, {Klotz},
  {Lecacheux}, {Mahasena}, {Manfroid}, {Manulis}, {Maury}, {Mohan}, {Morales},
  {Ofek}, {Rinner}, {Sharma}, {Sposetti}, {Tanga}, {Thirouin}, {Vachier},
  {Widemann}, {Asai}, {Hayato}, {Hiroyuki}, {Owada}, {Yamamura}, {Hayamizu},
  {Bradshaw}, {Kerr}, {Tomioka}, {Andersson}, {Dangl}, {Haymes}, {Naves}, \&
  {Wortmann}}]{DiasOliveira2017}
{Dias-Oliveira}, A., {Sicardy}, B., {Ortiz}, J.~L., {et~al.} 2017, \aj, 154,
  22, \dodoi{10.3847/1538-3881/aa74e9}

\bibitem[{{Elliot} {et~al.}(2005){Elliot}, {Kern}, {Clancy}, {Gulbis},
  {Millis}, {Buie}, {Wasserman}, {Chiang}, {Jordan}, {Trilling}, \&
  {Meech}}]{Elliot2005}
{Elliot}, J.~L., {Kern}, S.~D., {Clancy}, K.~B., {et~al.} 2005, \aj, 129, 1117,
  \dodoi{10.1086/427395}

\bibitem[{{Gladman} {et~al.}(2008){Gladman}, {Marsden}, \&
  {Vanlaerhoven}}]{Gladman2008}
{Gladman}, B., {Marsden}, B.~G., \& {Vanlaerhoven}, C. 2008, in The Solar
  System Beyond Neptune, ed. M.~A. {Barucci}, H.~{Boehnhardt}, D.~P.
  {Cruikshank}, A.~{Morbidelli}, \& R.~{Dotson}, 43--57

\bibitem[{{Gladman} {et~al.}(2012){Gladman}, {Lawler}, {Petit}, {Kavelaars},
  {Jones}, {Parker}, {Van Laerhoven}, {Nicholson}, {Rousselot}, {Bieryla}, \&
  {Ashby}}]{Gladman2012}
{Gladman}, B., {Lawler}, S.~M., {Petit}, J.~M., {et~al.} 2012, \aj, 144, 23,
  \dodoi{10.1088/0004-6256/144/1/23}

\bibitem[{{Grundy} {et~al.}(2014){Grundy}, {Benecchi}, {Porter}, \&
  {Noll}}]{Grundy2014}
{Grundy}, W.~M., {Benecchi}, S.~D., {Porter}, S.~B., \& {Noll}, K.~S. 2014,
  \icarus, 237, 1, \dodoi{10.1016/j.icarus.2014.04.021}

\bibitem[{{Isobe} {et~al.}(1986){Isobe}, {Feigelson}, \& {Nelson}}]{Isobe1986}
{Isobe}, T., {Feigelson}, E.~D., \& {Nelson}, P.~I. 1986, \apj, 306, 490,
  \dodoi{10.1086/164359}

\bibitem[{{Jeans}(1919)}]{Jeans1919}
{Jeans}, J.~H. 1919, {Problems of cosmogony and stellar dynamics}

\bibitem[{{Jewitt} \& {Sheppard}(2002)}]{JewittSheppard2002}
{Jewitt}, D.~C., \& {Sheppard}, S.~S. 2002, \aj, 123, 2110,
  \dodoi{10.1086/339557}

\bibitem[{{Kecskem{\'e}thy} {et~al.}(2023){Kecskem{\'e}thy}, {Kiss},
  {Szak{\'a}ts}, {P{\'a}l}, {Szab{\'o}}, {Moln{\'a}r}, {S{\'a}rneczky},
  {Vink{\'o}}, {Szab{\'o}}, {Marton}, {Farkas-Tak{\'a}cs}, {Kalup}, \&
  {Kiss}}]{Kecskemethy2023}
{Kecskem{\'e}thy}, V., {Kiss}, C., {Szak{\'a}ts}, R., {et~al.} 2023, \apjs,
  264, 18, \dodoi{10.3847/1538-4365/ac9c67}

\bibitem[{{Kern}(2006)}]{Kern2006}
{Kern}, S.~D. 2006, PhD thesis, Massachusetts Institute of Technology

\bibitem[{{Kern} \& {Elliot}(2006)}]{Kern2006QY90}
{Kern}, S.~D., \& {Elliot}, J.~L. 2006, \icarus, 183, 179,
  \dodoi{10.1016/j.icarus.2006.01.002}

\bibitem[{{Lacerda}(2011)}]{Lacerda2011}
{Lacerda}, P. 2011, The Astronomical Journal, vol. 142,, pp. 90.,
  \dodoi{10.1088/00046256/142/3/90}

\bibitem[{{Lacerda} {et~al.}(2008){Lacerda}, {Jewitt}, \&
  {Peixinho}}]{Lacerda2008}
{Lacerda}, P., {Jewitt}, D., \& {Peixinho}, N. 2008, The Astronomical Journal,
  vol. 135,, pp. 1749, \dodoi{10.1088/00046256/135/5/1749}

\bibitem[{{Lacerda} \& {Jewitt}(2007)}]{LacerdaJewitt2007}
{Lacerda}, P., \& {Jewitt}, D.~C. 2007, The Astronomical Journal, 133, 1393,
  \dodoi{10.1086/511772}

\bibitem[{{Lacerda} {et~al.}(2014{\natexlab{a}}){Lacerda}, {McNeill}, \&
  {Peixinho}}]{Lacerda2014}
{Lacerda}, P., {McNeill}, A., \& {Peixinho}, N. 2014{\natexlab{a}}, \mnras,
  437, 3824, \dodoi{10.1093/mnras/stt2180}

\bibitem[{{Lacerda} {et~al.}(2014{\natexlab{b}}){Lacerda}, {Fornasier},
  {Lellouch}, {Kiss}, {Vilenius}, {Santos-Sanz}, {Rengel}, {M{\"u}ller},
  {Stansberry}, {Duffard}, {Delsanti}, \&
  {Guilbert-Lepoutre}}]{Lacerda2014Albedo}
{Lacerda}, P., {Fornasier}, S., {Lellouch}, E., {et~al.} 2014{\natexlab{b}},
  \apjl, 793, L2, \dodoi{10.1088/2041-8205/793/1/L2}

\bibitem[{{Lawler} {et~al.}(2019){Lawler}, {Pike}, {Kaib}, {Alexandersen},
  {Bannister}, {Chen}, {Gladman}, {Gwyn}, {Kavelaars}, {Petit}, \&
  {Volk}}]{Lawler2019}
{Lawler}, S.~M., {Pike}, R.~E., {Kaib}, N., {et~al.} 2019, \aj, 157, 253,
  \dodoi{10.3847/1538-3881/ab1c4c}

\bibitem[{{Leiva} {et~al.}(2023){Leiva}, {Ortiz}, {Gomez-Limon}, {Perez},
  {Kretlow}, \& {Braga-Ribas}}]{Leiva2023}
{Leiva}, R., {Ortiz}, J., {Gomez-Limon}, J., {et~al.} 2023, in Asteroids,
  Comets, Meteors Conference 2023 (LPI Contrib. No. 2851)

\bibitem[{{Leiva} {et~al.}(2020){Leiva}, {Buie}, {Keller}, {Wasserman},
  {Kavelaars}, {Bridges}, {Haley}, {Strauss}, {Wilde}, {Weryk}, {Kervella},
  {Baker}, {Bock}, {Conway}, {Cota}, {Estes}, {Garc{\'\i}a}, {Kehrli},
  {McCandless}, {McCandless}, {Self}, {Settlemire}, {Swanson}, {Thompson}, \&
  {Wise}}]{Leiva2020}
{Leiva}, R., {Buie}, M.~W., {Keller}, J.~M., {et~al.} 2020, Planetary Science
  Journal, 1, 48, \dodoi{10.3847/PSJ/abb23d}

\bibitem[{{Leone} {et~al.}(1984){Leone}, {Paolicchi}, {Farinella}, \&
  {Zappala}}]{Leone1984}
{Leone}, G., {Paolicchi}, P., {Farinella}, P., \& {Zappala}, V. 1984, \aap,
  140, 265

\bibitem[{{Levison} {et~al.}(2008){Levison}, {Morbidelli}, {Van Laerhoven},
  {Gomes}, \& {Tsiganis}}]{Levison2008}
{Levison}, H.~F., {Morbidelli}, A., {Van Laerhoven}, C., {Gomes}, R., \&
  {Tsiganis}, K. 2008, \icarus, 196, 258, \dodoi{10.1016/j.icarus.2007.11.035}

\bibitem[{{Lomb}(1976)}]{Lomb1976}
{Lomb}, N.~R. 1976, \apss, 39, 447, \dodoi{10.1007/BF00648343}

\bibitem[{{Lykawka} \& {Mukai}(2005)}]{Lykawka2005}
{Lykawka}, P.~S., \& {Mukai}, T. 2005, \planss, 53, 1175,
  \dodoi{10.1016/j.pss.2004.12.015}

\bibitem[{{Malhotra}(1995)}]{Malhotra1995}
{Malhotra}, R. 1995, \aj, 110, 420, \dodoi{10.1086/117532}

\bibitem[{{Murray-Clay} \& {Schlichting}(2011)}]{MurrayClay2011}
{Murray-Clay}, R.~A., \& {Schlichting}, H.~E. 2011, \apj, 730, 132,
  \dodoi{10.1088/0004-637X/730/2/132}

\bibitem[{{Nesvorn{\'y}}(2021)}]{Nesvorny2021}
{Nesvorn{\'y}}, D. 2021, \apjl, 908, L47, \dodoi{10.3847/2041-8213/abe38f}

\bibitem[{{Nesvorn{\'y}} \& {Vokrouhlick{\'y}}(2019)}]{Nesvorny2019}
{Nesvorn{\'y}}, D., \& {Vokrouhlick{\'y}}, D. 2019, \icarus, 331, 49,
  \dodoi{10.1016/j.icarus.2019.04.030}

\bibitem[{{Nesvorn{\'y}} {et~al.}(2022){Nesvorn{\'y}}, {Vokrouhlick{\'y}}, \&
  {Fraser}}]{Nesvorny2022}
{Nesvorn{\'y}}, D., {Vokrouhlick{\'y}}, D., \& {Fraser}, W.~C. 2022, \aj, 163,
  137, \dodoi{10.3847/1538-3881/ac4bc9}

\bibitem[{{Noll} {et~al.}(2020){Noll}, {Grundy}, {Nesvorn{\'y}}, \&
  {Thirouin}}]{Noll2020}
{Noll}, K., {Grundy}, W.~M., {Nesvorn{\'y}}, D., \& {Thirouin}, A. 2020, in The
  Trans-Neptunian Solar System, ed. D.~{Prialnik}, M.~A. {Barucci}, \&
  L.~{Young}, 201--224, \dodoi{10.1016/B978-0-12-816490-7.00009-6}

\bibitem[{{Noll} {et~al.}(2006){Noll}, {Grundy}, {Stephens}, \&
  {Levison}}]{Noll2006}
{Noll}, K.~S., {Grundy}, W.~M., {Stephens}, D.~C., \& {Levison}, H.~F. 2006,
  \iaucirc, 8745, 1

\bibitem[{{Ortiz} {et~al.}(2006){Ortiz}, {Guti{\'e}rrez}, {Santos-Sanz},
  {Casanova}, \& {Sota}}]{Ortiz2006}
{Ortiz}, J.~L., {Guti{\'e}rrez}, P.~J., {Santos-Sanz}, P., {Casanova}, V., \&
  {Sota}, A. 2006, \aap, 447, 1131, \dodoi{10.1051/0004-6361:20053572}

\bibitem[{{P{\'a}l} {et~al.}(2015){P{\'a}l}, {Szab{\'o}}, {Szab{\'o}}, {Kiss},
  {Moln{\'a}r}, {S{\'a}rneczky}, \& {Kiss}}]{Pal2015}
{P{\'a}l}, A., {Szab{\'o}}, R., {Szab{\'o}}, G.~M., {et~al.} 2015, \apjl, 804,
  L45, \dodoi{10.1088/2041-8205/804/2/L45}

\bibitem[{{Peixinho} {et~al.}(2015){Peixinho}, {Delsanti}, \&
  {Doressoundiram}}]{Peixinho2015}
{Peixinho}, N., {Delsanti}, A., \& {Doressoundiram}, A. 2015, \aap, 577, A35,
  \dodoi{10.1051/0004-6361/201425436}

\bibitem[{{Perna} {et~al.}(2009){Perna}, {Dotto}, {Barucci}, {Rossi},
  {Fornasier}, \& {de Bergh}}]{Perna2009}
{Perna}, D., {Dotto}, E., {Barucci}, M.~A., {et~al.} 2009, \aap, 508, 451,
  \dodoi{10.1051/0004-6361/200911970}

\bibitem[{{Pike} {et~al.}(2017){Pike}, {Fraser}, {Schwamb}, {Kavelaars},
  {Marsset}, {Bannister}, {Lehner}, {Wang}, {Alexandersen}, {Chen}, {Gladman},
  {Gwyn}, {Petit}, \& {Volk}}]{Pike2017}
{Pike}, R.~E., {Fraser}, W.~C., {Schwamb}, M.~E., {et~al.} 2017, \aj, 154, 101,
  \dodoi{10.3847/1538-3881/aa83b1}

\bibitem[{{Pike} {et~al.}(2023){Pike}, {Fraser}, {Volk}, {Kavelaars},
  {Marsset}, {Peixinho}, {Schwamb}, {Bannister}, {Peltier}, {Buchanan},
  {Benecchi}, \& {Tan}}]{Pike2023}
{Pike}, R.~E., {Fraser}, W.~C., {Volk}, K., {et~al.} 2023, Planetary Science
  Journal, 4, 200, \dodoi{10.3847/PSJ/ace2c2}

\bibitem[{{Pirani} {et~al.}(2021){Pirani}, {Johansen}, \&
  {Mustill}}]{Pirani2021}
{Pirani}, S., {Johansen}, A., \& {Mustill}, A.~J. 2021, \aap, 650, A161,
  \dodoi{10.1051/0004-6361/202037465}

\bibitem[{{Press} {et~al.}(1992){Press}, {Teukolsky}, {Vetterling}, \&
  {Flannery}}]{Press1992}
{Press}, W.~H., {Teukolsky}, S.~A., {Vetterling}, W.~T., \& {Flannery}, B.~P.
  1992, {Numerical recipes in FORTRAN. The art of scientific computing}
  (Cambridge: University Press, 1992, 2nd ed.)

\bibitem[{{Rabinowitz} {et~al.}(2020){Rabinowitz}, {Benecchi}, {Grundy},
  {Verbiscer}, \& {Thirouin}}]{Rabinowitz2020}
{Rabinowitz}, D.~L., {Benecchi}, S.~D., {Grundy}, W.~M., {Verbiscer}, A.~J., \&
  {Thirouin}, A. 2020, \aj, 159, 27, \dodoi{10.3847/1538-3881/ab59d4}

\bibitem[{{Sheppard}(2004)}]{Sheppard2004}
{Sheppard}, S.~S. 2004, PhD thesis, University of Hawaii, Manoa

\bibitem[{{Sheppard}(2007)}]{Sheppard2007}
---. 2007, \aj, 134, 787, \dodoi{10.1086/519072}

\bibitem[{{Sheppard}(2012)}]{Sheppard2012}
---. 2012, \aj, 144, 169, \dodoi{10.1088/0004-6256/144/6/169}

\bibitem[{{Sheppard} \& {Jewitt}(2004)}]{SheppardJewitt2004}
{Sheppard}, S.~S., \& {Jewitt}, D. 2004, \aj, 127, 3023, \dodoi{10.1086/383558}

\bibitem[{{Sheppard} {et~al.}(2008){Sheppard}, {Lacerda}, \&
  {Ortiz}}]{Sheppard2008}
{Sheppard}, S.~S., {Lacerda}, P., \& {Ortiz}, J.~L. 2008, in The Solar System
  Beyond Neptune, ed. M.~A. {Barucci}, H.~{Boehnhardt}, D.~P. {Cruikshank},
  A.~{Morbidelli}, \& R.~{Dotson}, 129--142

\bibitem[{{Spearman}(1904)}]{Spearman1904}
{Spearman}, C. 1904, The American Journal of Psychology, vol. 15,, pp. 72

\bibitem[{{Thirouin}(2013)}]{Thirouin2013}
{Thirouin}, A. 2013, PhD thesis, University of Granada, Spain

\bibitem[{{Thirouin} {et~al.}(2014){Thirouin}, {Noll}, {Ortiz}, \&
  {Morales}}]{Thirouin2014}
{Thirouin}, A., {Noll}, K.~S., {Ortiz}, J.~L., \& {Morales}, N. 2014, \aap,
  569, A3, \dodoi{10.1051/0004-6361/201423567}

\bibitem[{{Thirouin} {et~al.}(2010){Thirouin}, {Ortiz}, {Duffard},
  {Santos-Sanz}, {Aceituno}, \& {Morales}}]{Thirouin2010}
{Thirouin}, A., {Ortiz}, J.~L., {Duffard}, R., {et~al.} 2010, \aap, 522, A93,
  \dodoi{10.1051/0004-6361/200912340}

\bibitem[{{Thirouin} \& {Sheppard}(2017)}]{ThirouinSheppard2017}
{Thirouin}, A., \& {Sheppard}, S.~S. 2017, \aj, 154, 241,
  \dodoi{10.3847/1538-3881/aa96fb}

\bibitem[{{Thirouin} \& {Sheppard}(2018)}]{ThirouinSheppard2018}
---. 2018, \aj, 155, 248, \dodoi{10.3847/1538-3881/aac0ff}

\bibitem[{{Thirouin} \& {Sheppard}(2019{\natexlab{a}})}]{ThirouinSheppard2019a}
---. 2019{\natexlab{a}}, \aj, 157, 228, \dodoi{10.3847/1538-3881/ab18a9}

\bibitem[{{Thirouin} \& {Sheppard}(2019{\natexlab{b}})}]{ThirouinSheppard2019b}
---. 2019{\natexlab{b}}, \aj, 158, 53, \dodoi{10.3847/1538-3881/ab27bc}

\bibitem[{{Thirouin} \& {Sheppard}(2022)}]{ThirouinSheppard2022}
---. 2022, Planetary Science Journal, 3, 178, \dodoi{10.3847/PSJ/ac7ab8}

\bibitem[{{Thirouin} {et~al.}(2017){Thirouin}, {Sheppard}, \&
  {Noll}}]{Thirouin2017}
{Thirouin}, A., {Sheppard}, S.~S., \& {Noll}, K.~S. 2017, \apj, 844, 135,
  \dodoi{10.3847/1538-4357/aa7ed3}

\bibitem[{{Thirouin} {et~al.}(2016){Thirouin}, {Sheppard}, {Noll}, {Moskovitz},
  {Ortiz}, \& {Doressoundiram}}]{Thirouin2016}
{Thirouin}, A., {Sheppard}, S.~S., {Noll}, K.~S., {et~al.} 2016, \aj, 151, 148,
  \dodoi{10.3847/0004-6256/151/6/148}

\bibitem[{{Trilling} \& {Bernstein}(2006)}]{TrillingBernstein2006}
{Trilling}, D.~E., \& {Bernstein}, G.~M. 2006, \aj, 131, 1149,
  \dodoi{10.1086/499228}

\bibitem[{{Volk} \& {Malhotra}(2011)}]{Volk2011}
{Volk}, K., \& {Malhotra}, R. 2011, \apj, 736, 11,
  \dodoi{10.1088/0004-637X/736/1/11}

\bibitem[{{Volk} \& {Malhotra}(2019)}]{Volk2019}
---. 2019, \aj, 158, 64, \dodoi{10.3847/1538-3881/ab2639}

\bibitem[{{Weidenschilling}(1980)}]{Weidenschilling1980}
{Weidenschilling}, S.~J. 1980, \icarus, 44, 807,
  \dodoi{10.1016/0019-1035(80)90147-5}

\end{thebibliography}
\bibliographystyle{aasjournal}



\appendix
 
\section{Appendix information}
\label{sec:appA}

\begin{figure*}
 \includegraphics[width=9.5cm, angle=0]{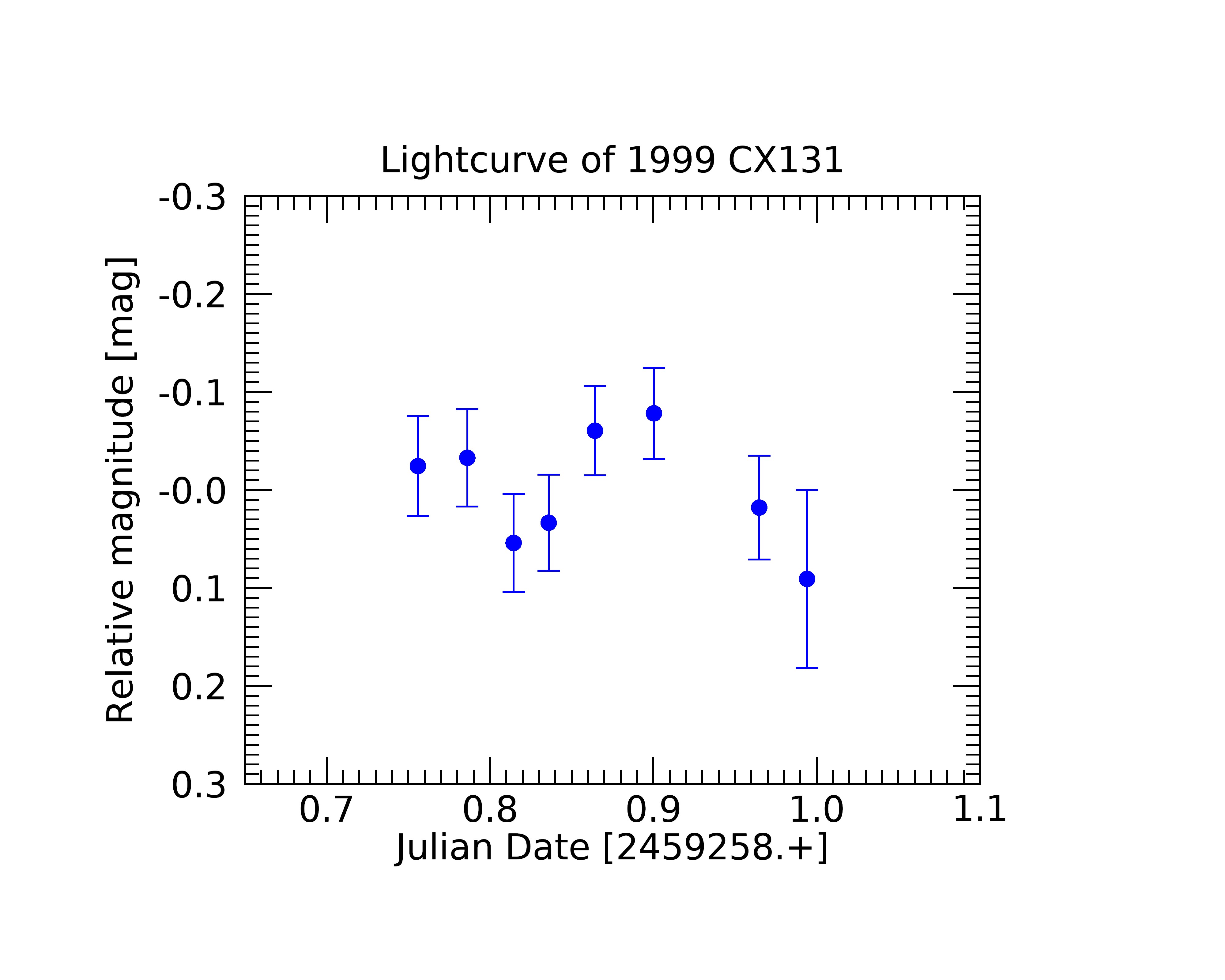}
 \includegraphics[width=9.5cm, angle=0]{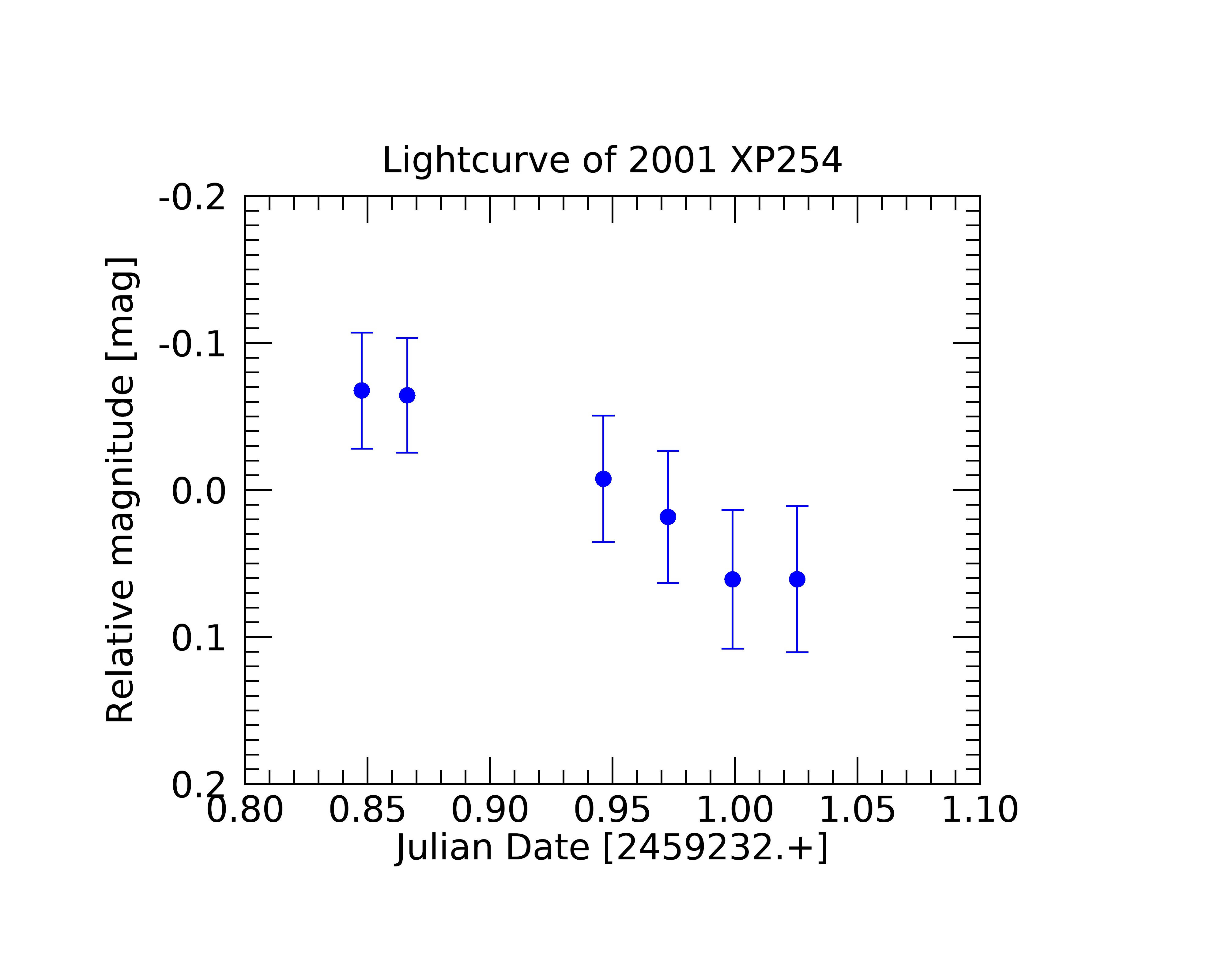}
 \includegraphics[width=9.5cm, angle=0]{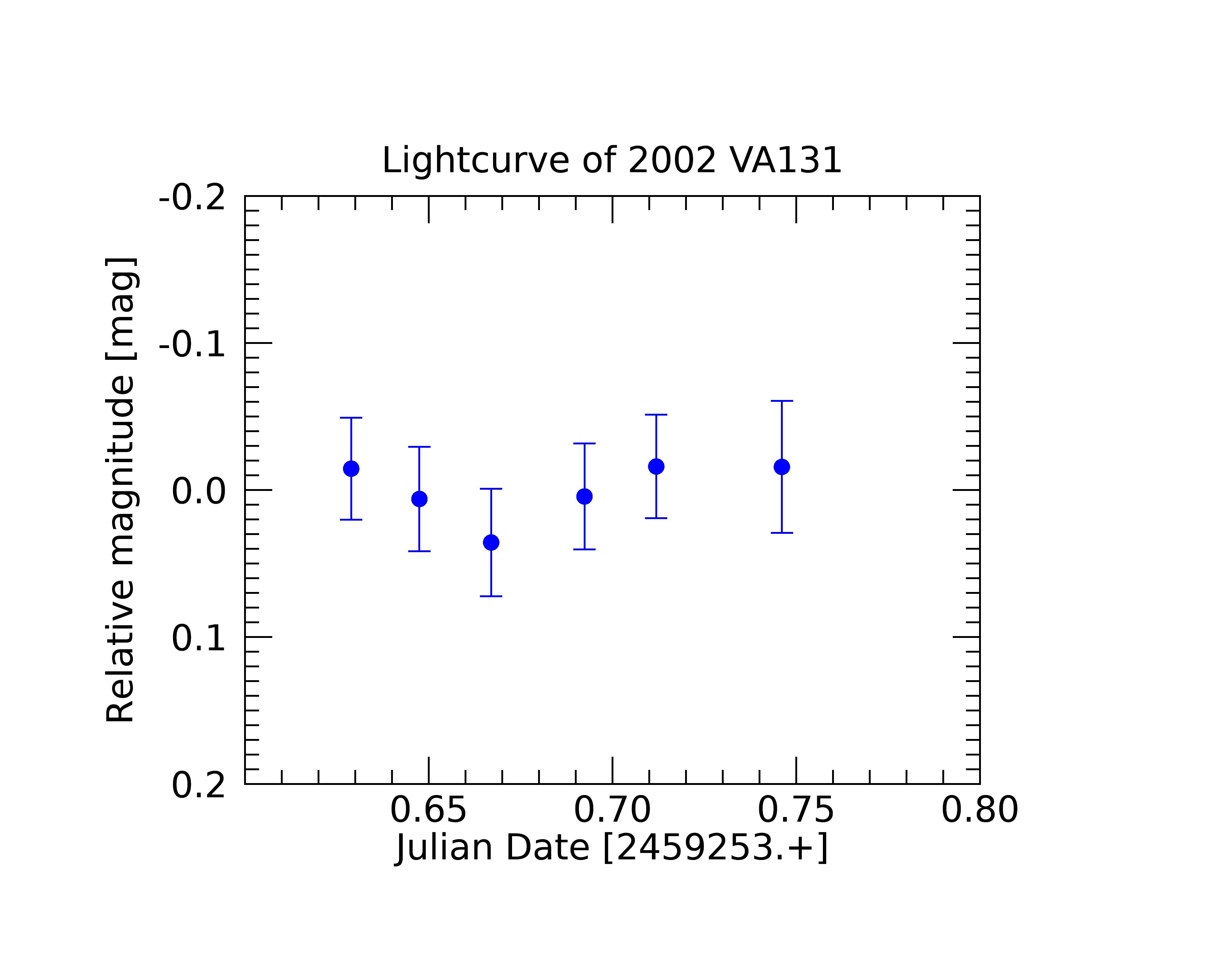}
 \includegraphics[width=9.5cm, angle=0]{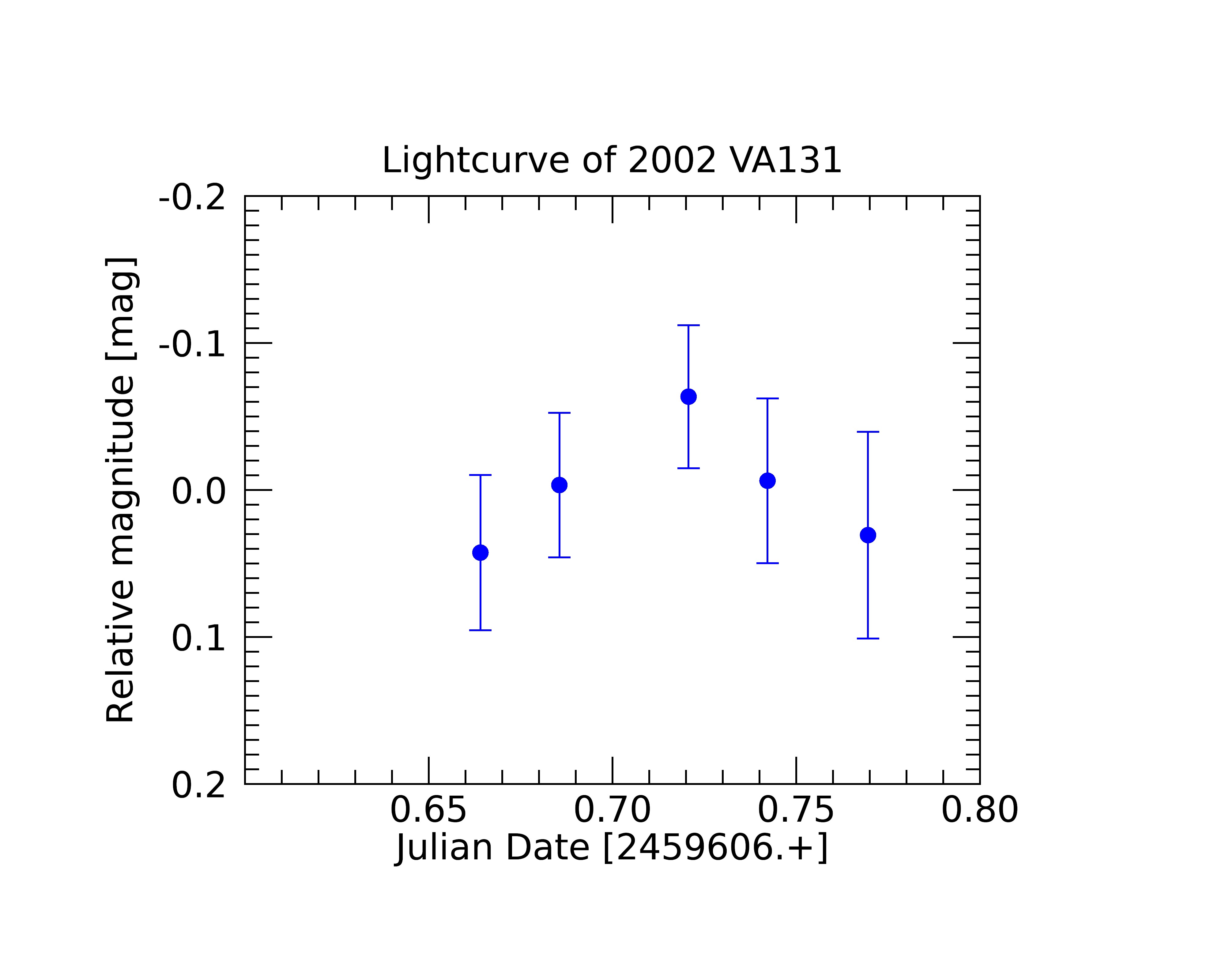}
 \includegraphics[width=9.5cm, angle=0]{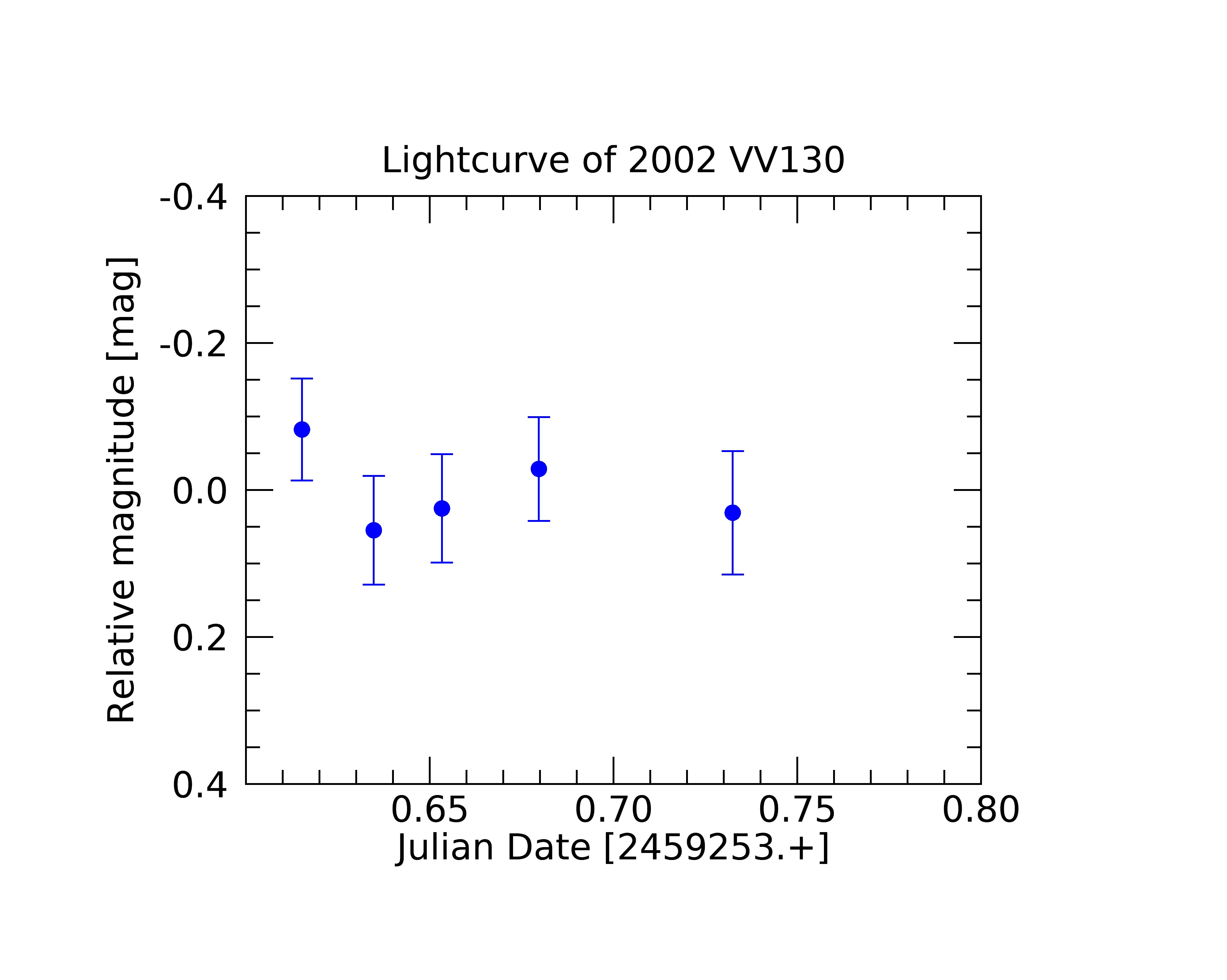}
 \includegraphics[width=9.5cm, angle=0]{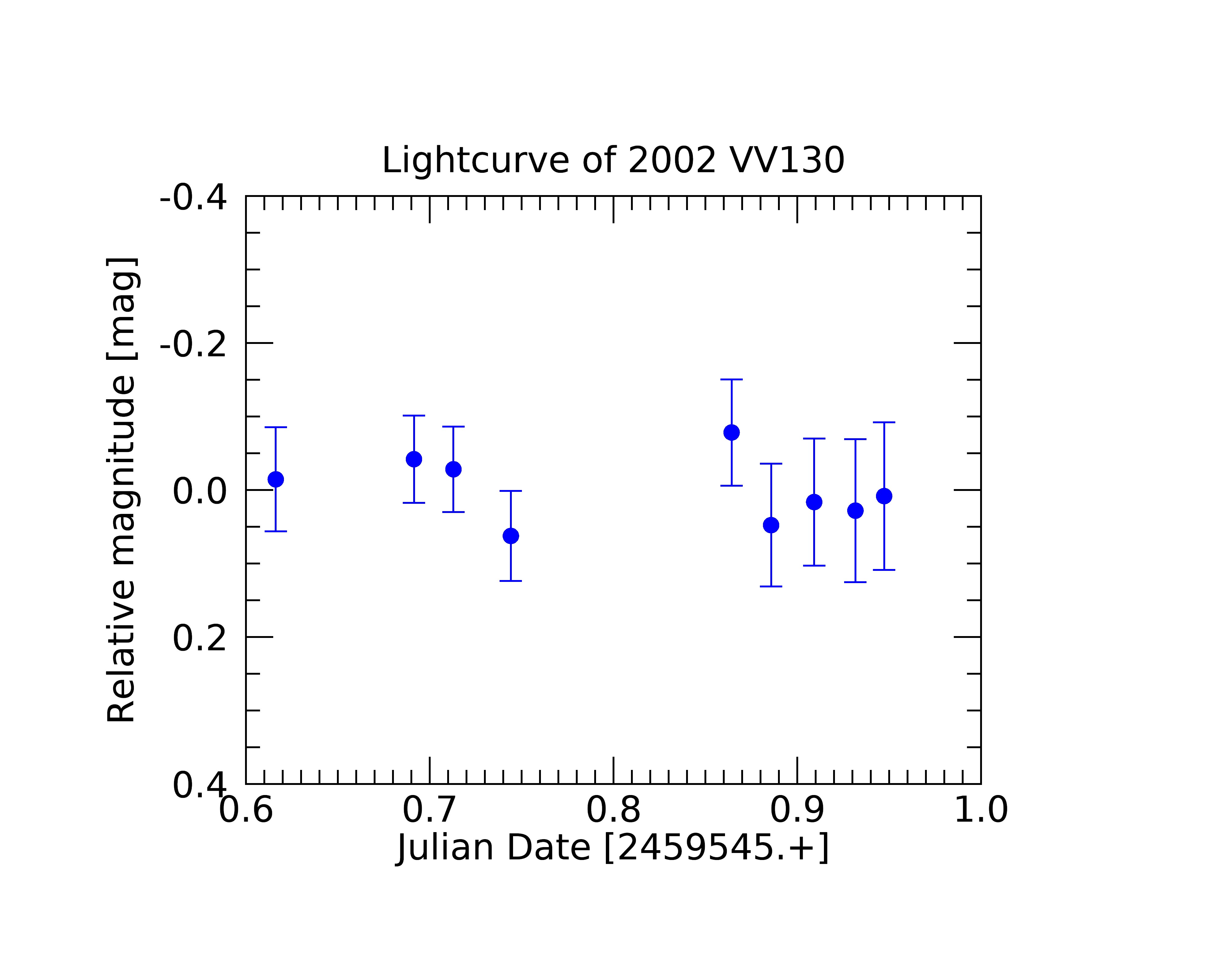}
\caption{\textit{Objects in the 5:3 mean motion resonance with Neptune }   }
\label{fig:LC53_1}
\end{figure*}

\begin{figure*}
 \includegraphics[width=9.5cm, angle=0]{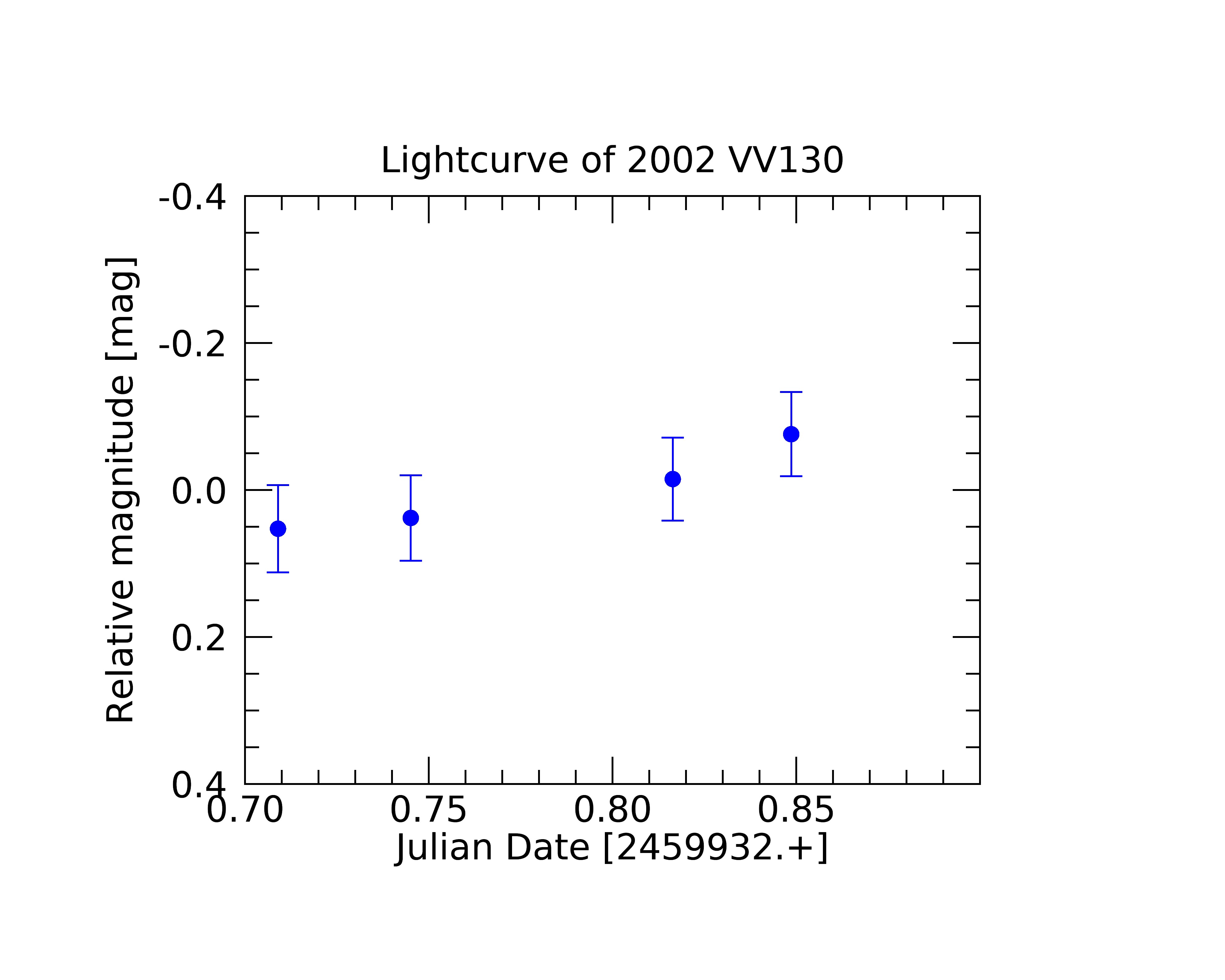}
 \includegraphics[width=9.5cm, angle=0]{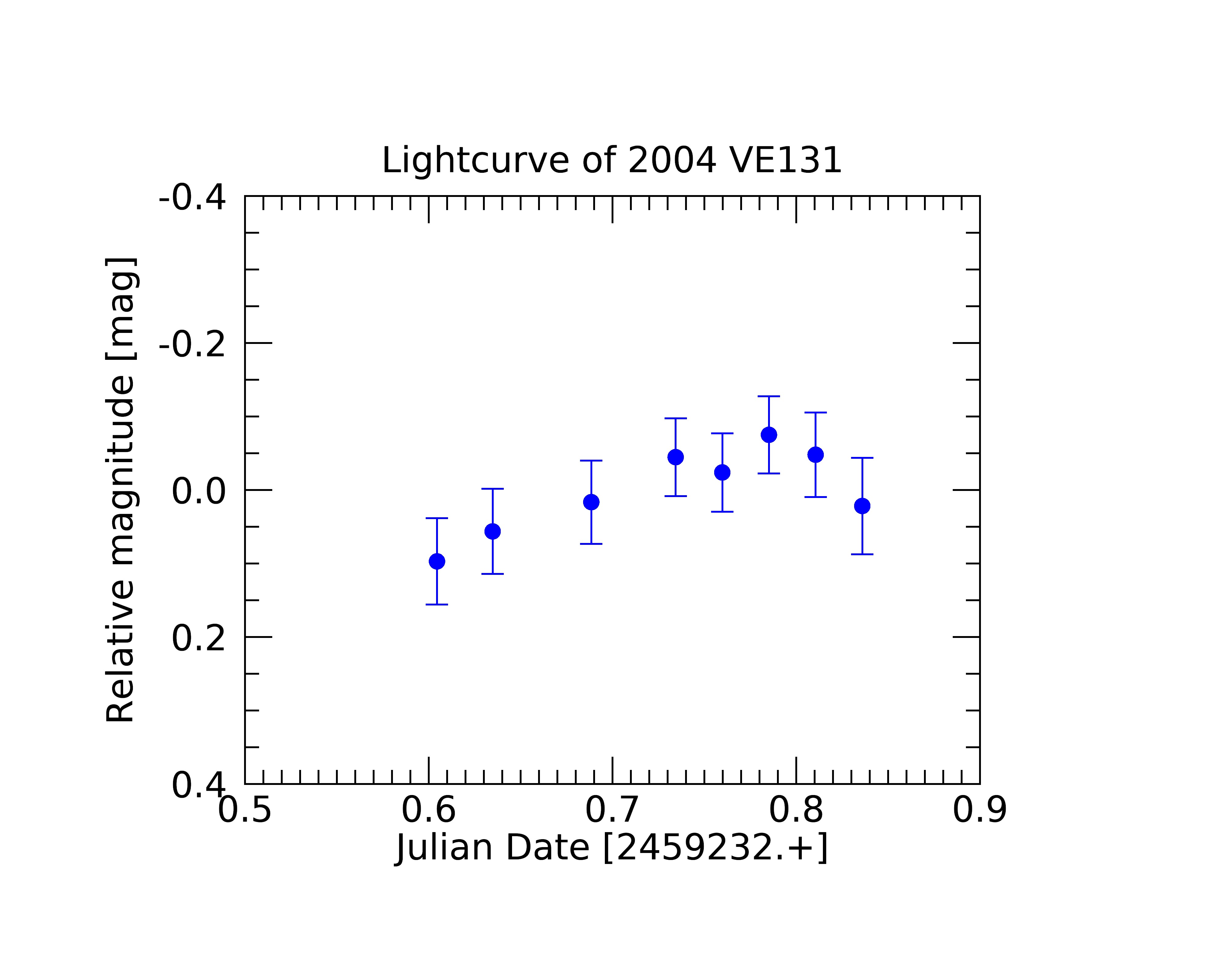}
 \includegraphics[width=9.5cm, angle=0]{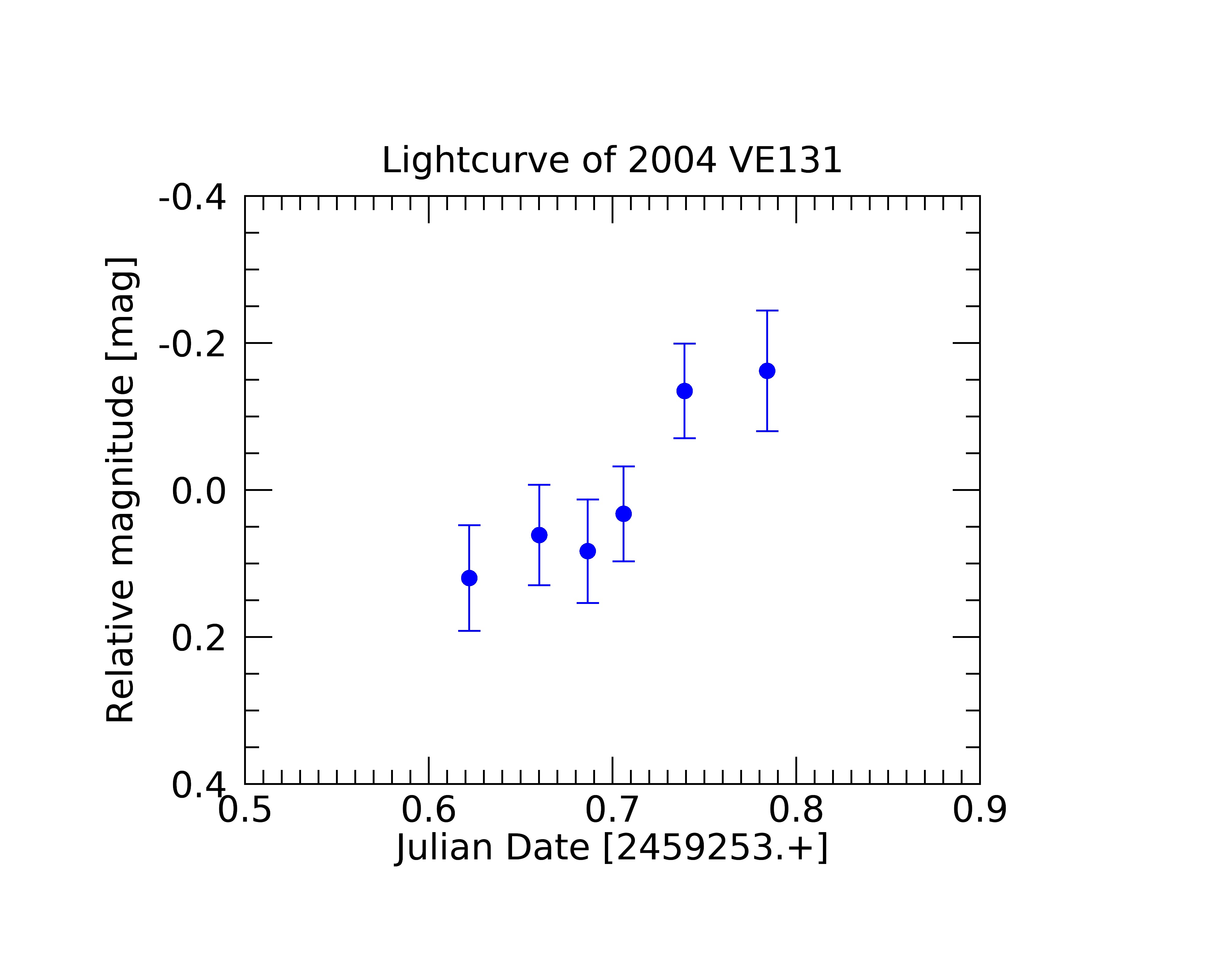}
 \includegraphics[width=9.5cm, angle=0]{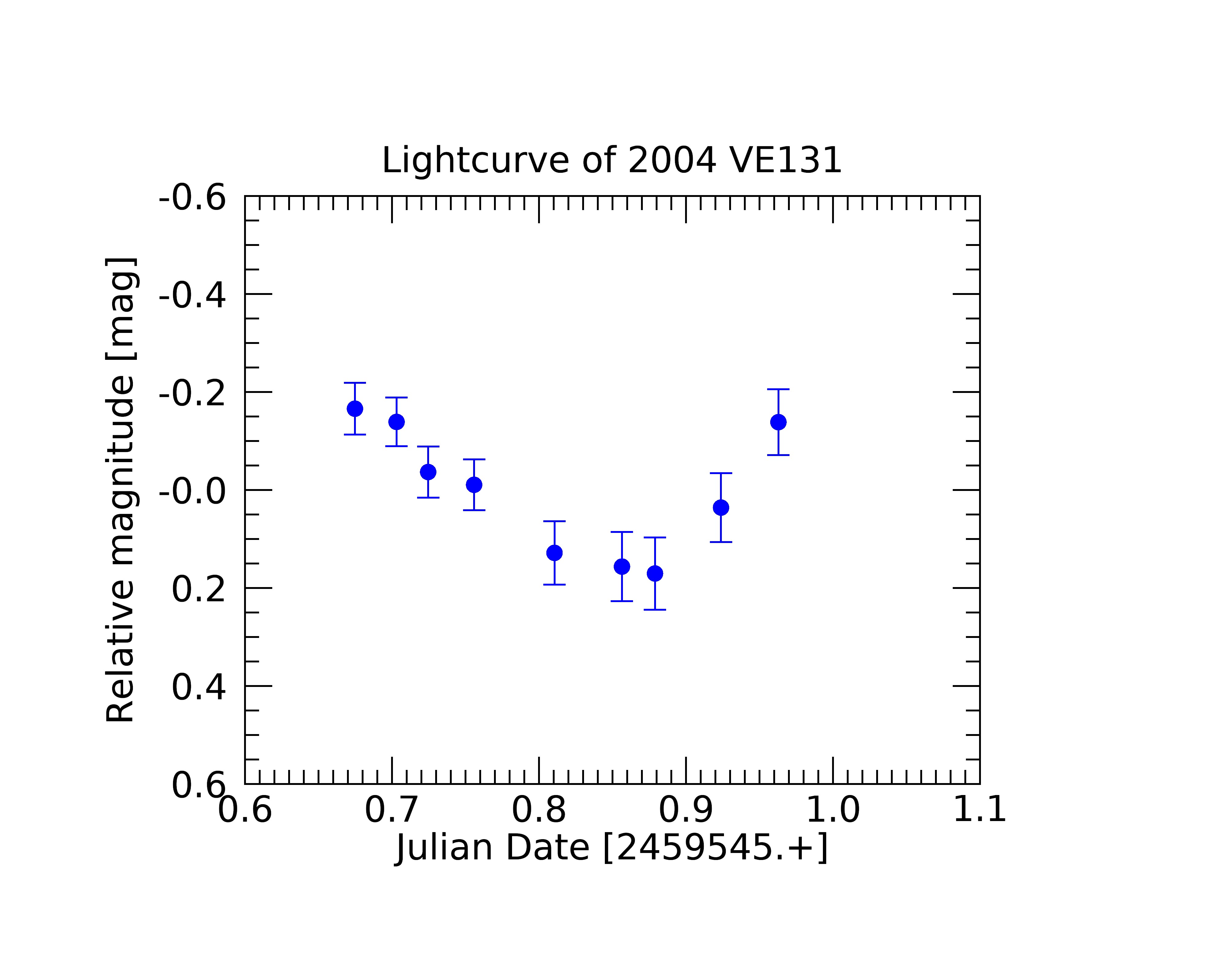}
 \includegraphics[width=9.5cm, angle=0]{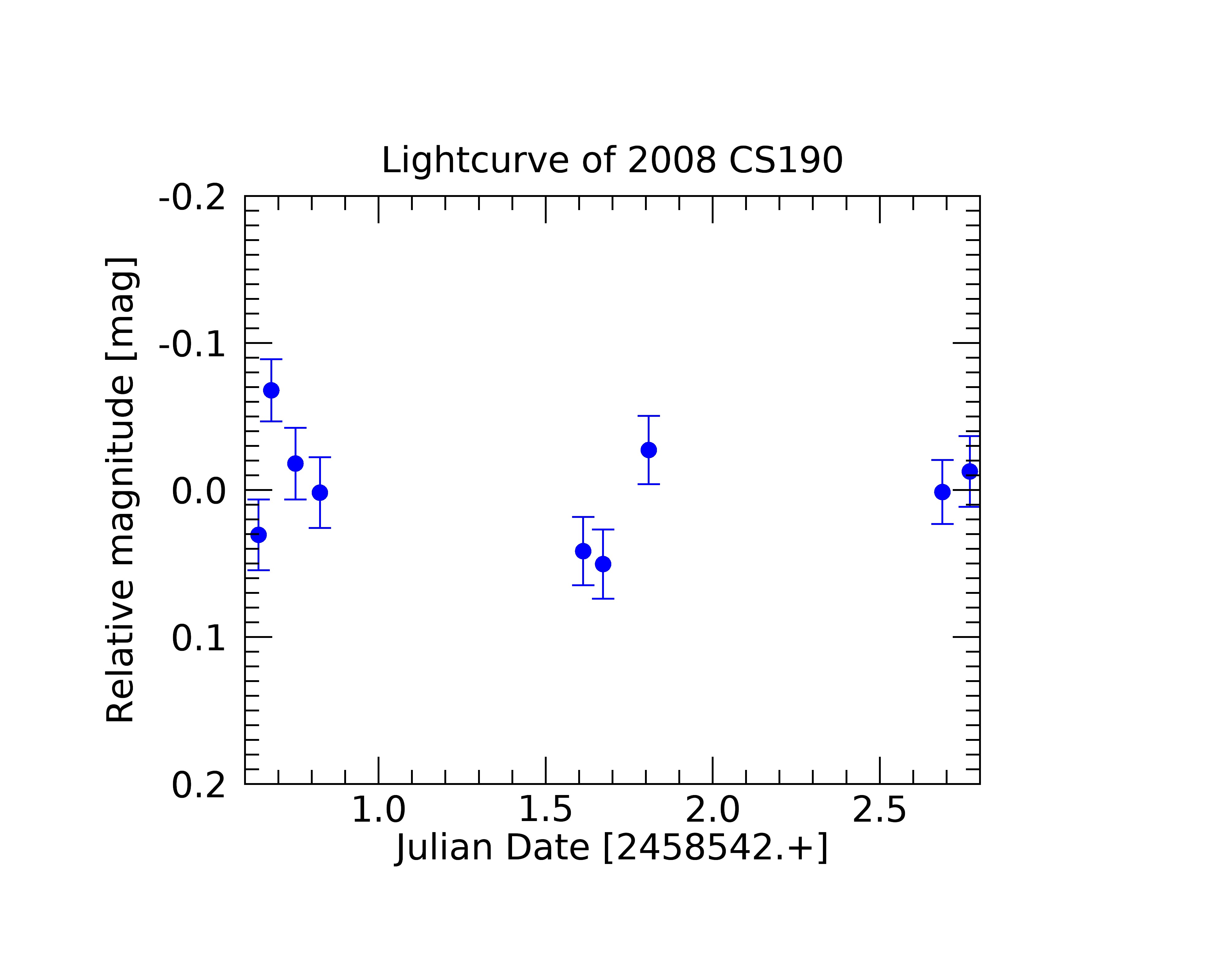}
 \includegraphics[width=9.5cm, angle=0]{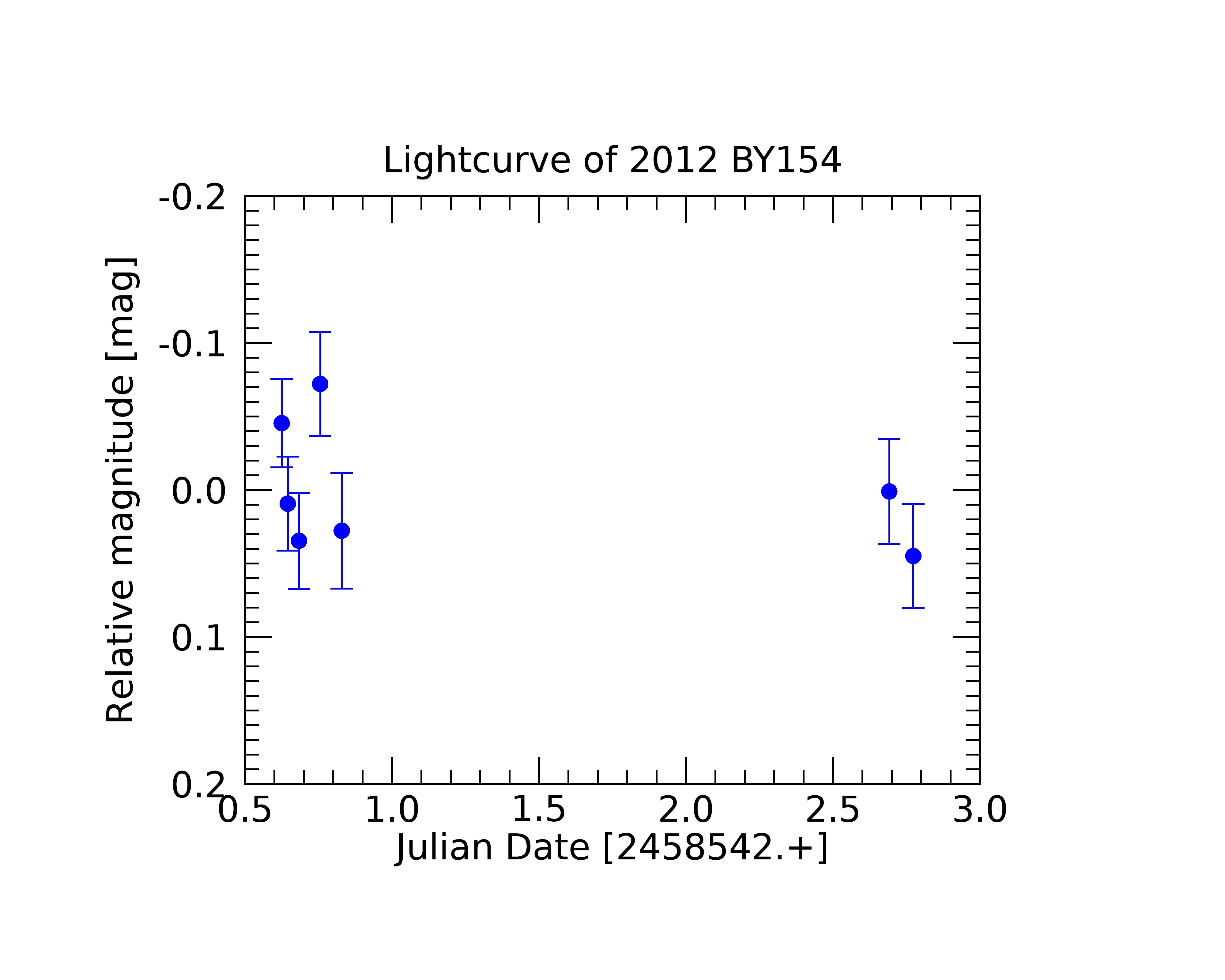}
\caption{\textit{Objects in the 5:3 mean motion resonance with Neptune }   }
\label{fig:LC53_2}
\end{figure*}

\begin{figure*}
 \includegraphics[width=9.5cm, angle=0]{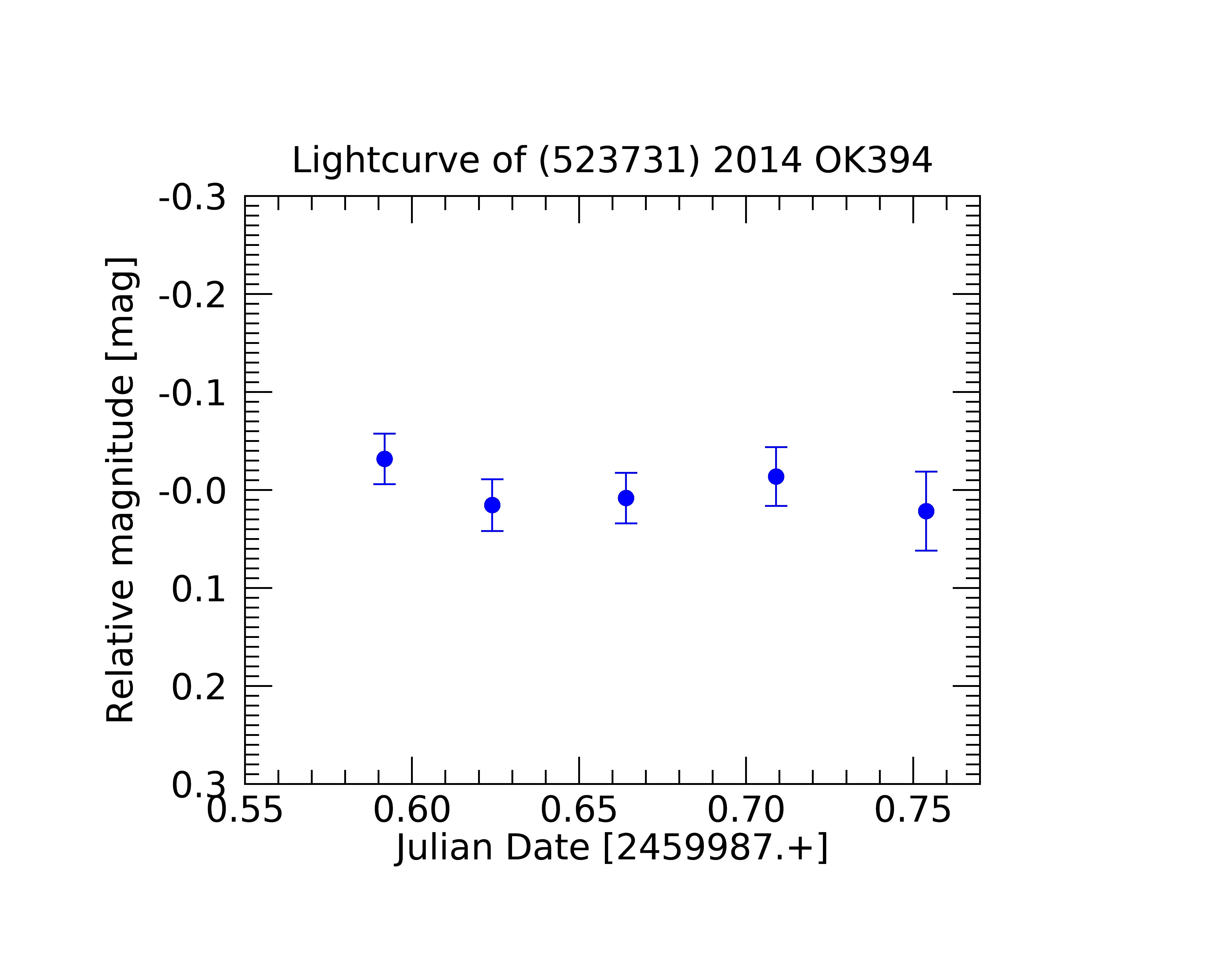}
 \includegraphics[width=9.5cm, angle=0]{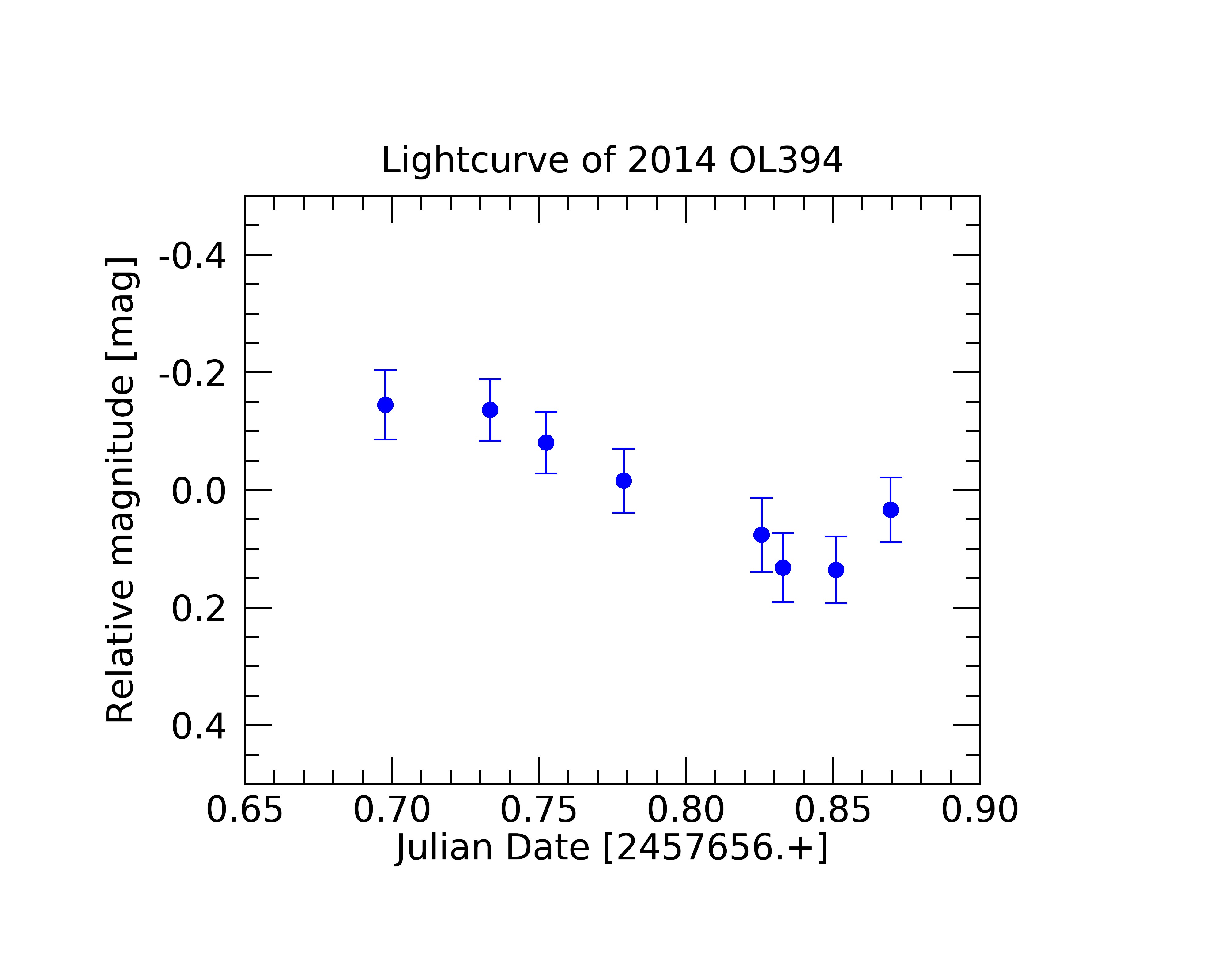}
 \includegraphics[width=9.5cm, angle=0]{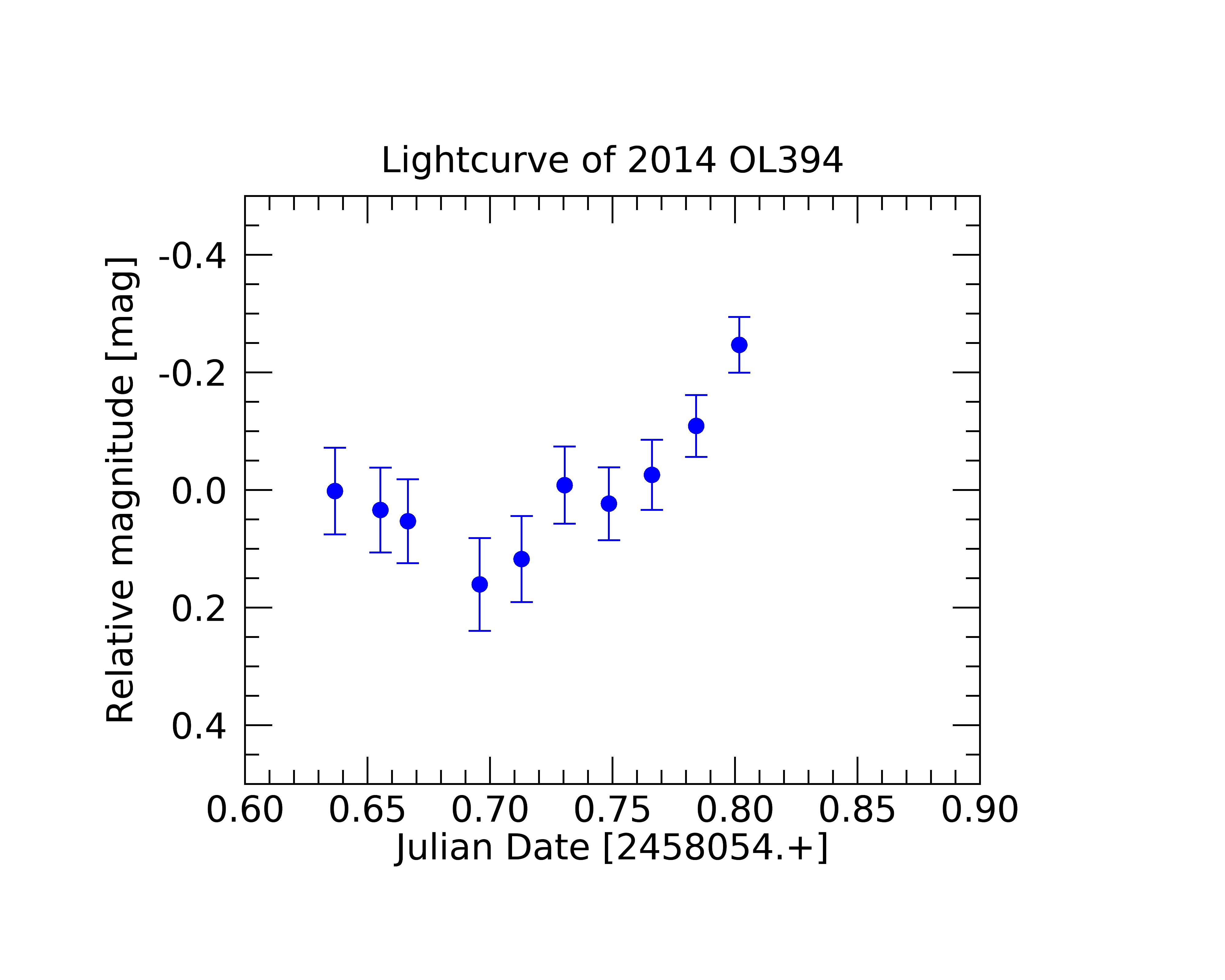}
\caption{\textit{Objects in the 5:3 mean motion resonance with Neptune }   }
\label{fig:LC53_3}
\end{figure*}

\begin{figure*}
 \includegraphics[width=9.5cm, angle=0]{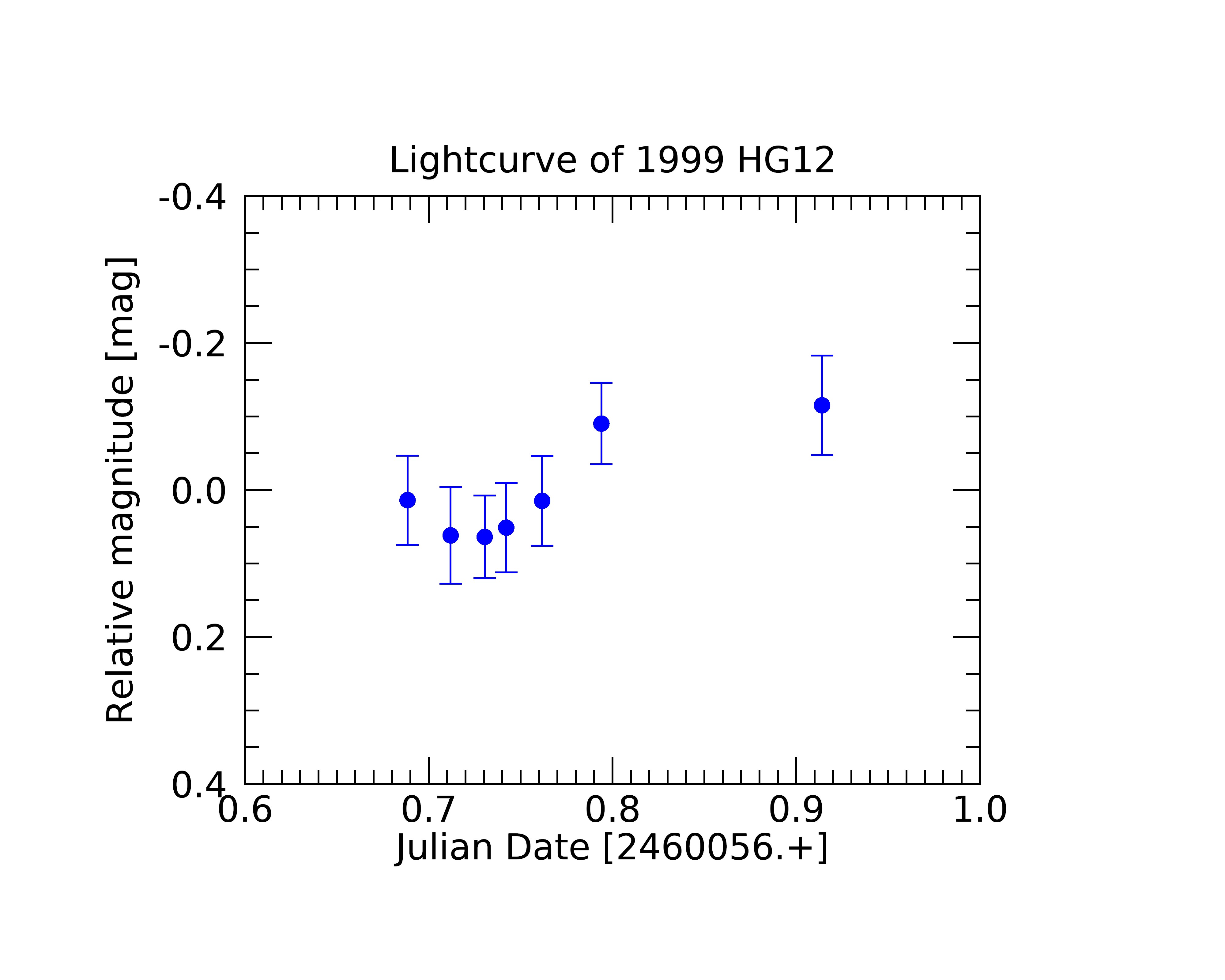}
 \includegraphics[width=9.5cm, angle=0]{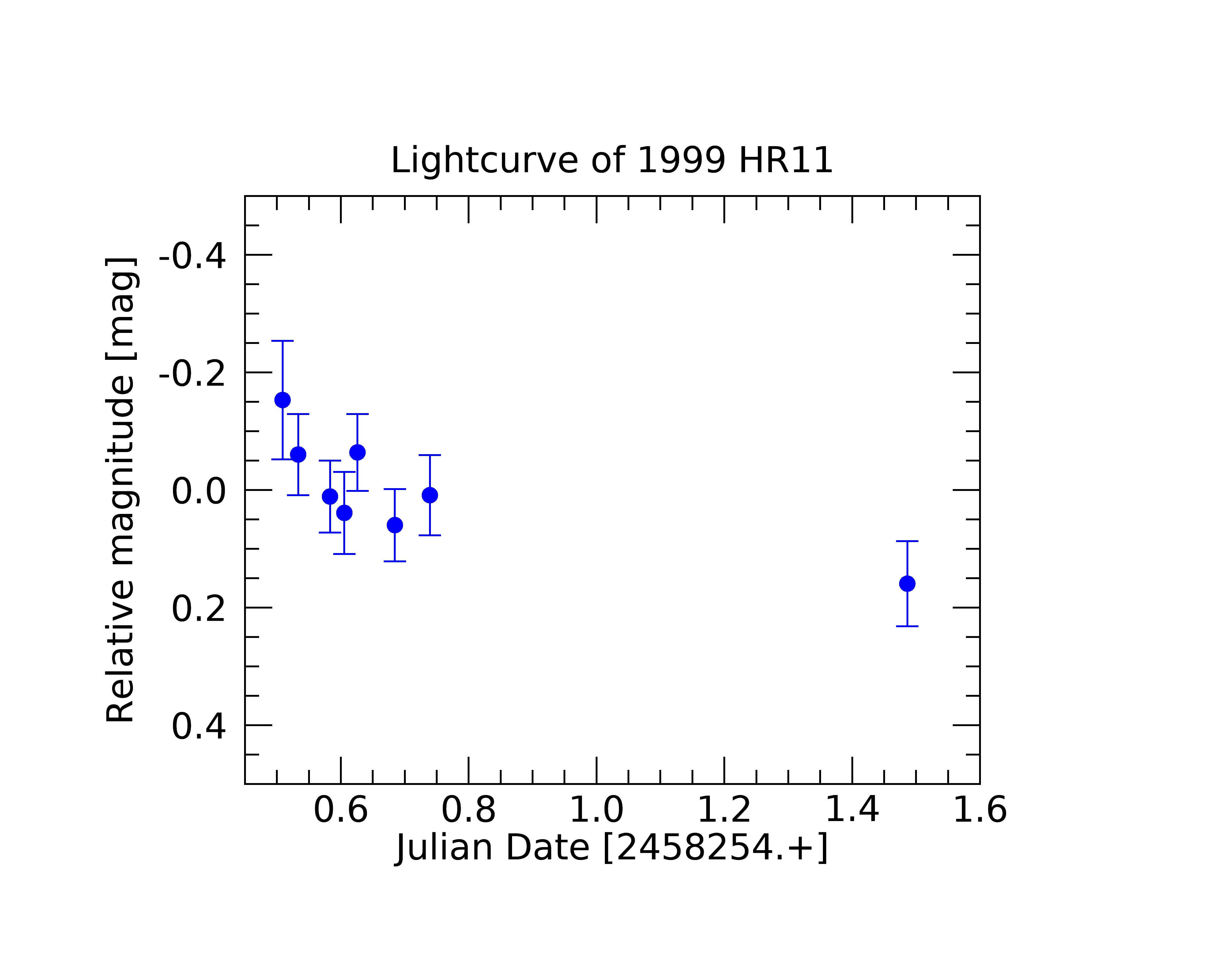}
  \includegraphics[width=9.5cm, angle=0]{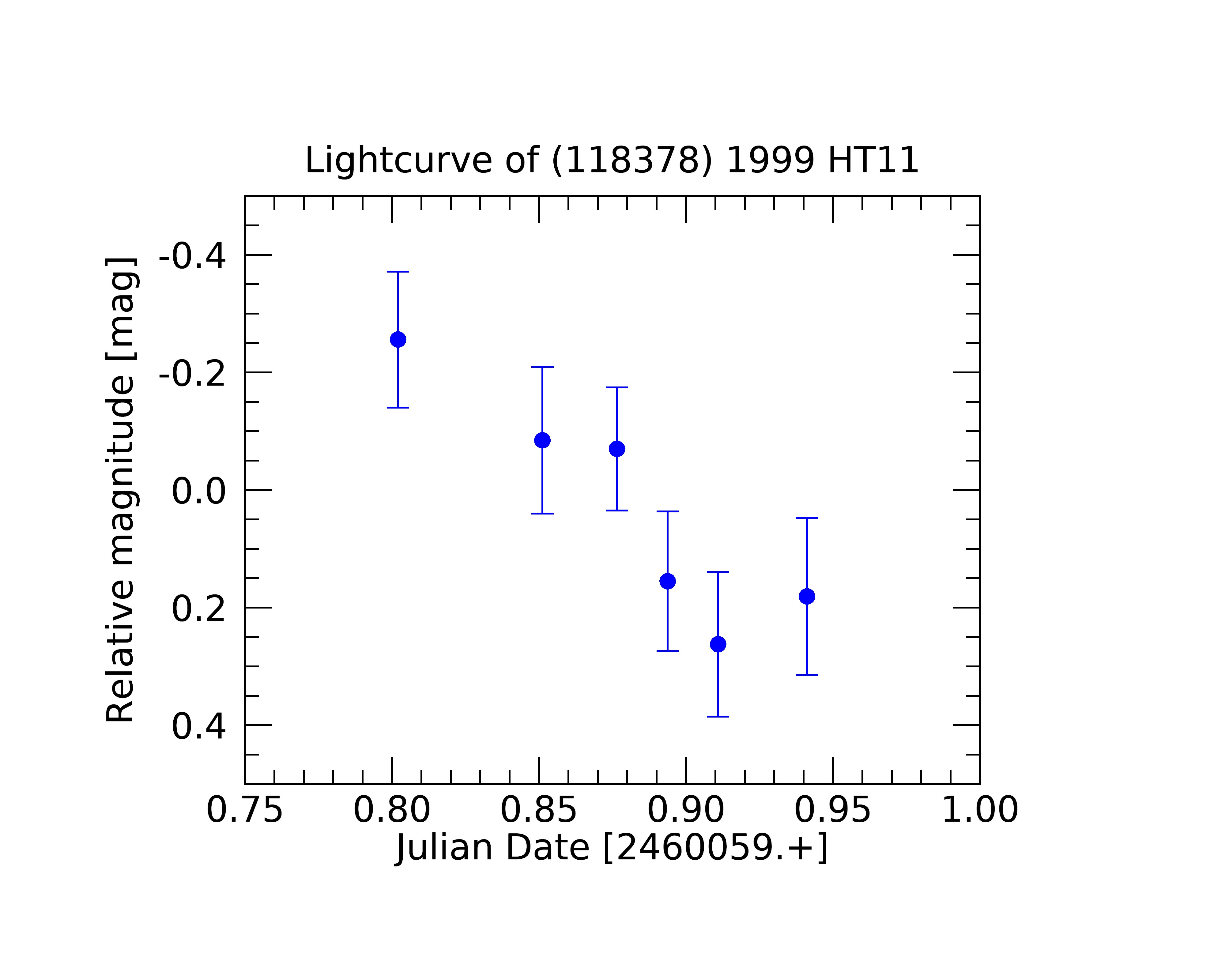}
  \includegraphics[width=9.5cm, angle=0]{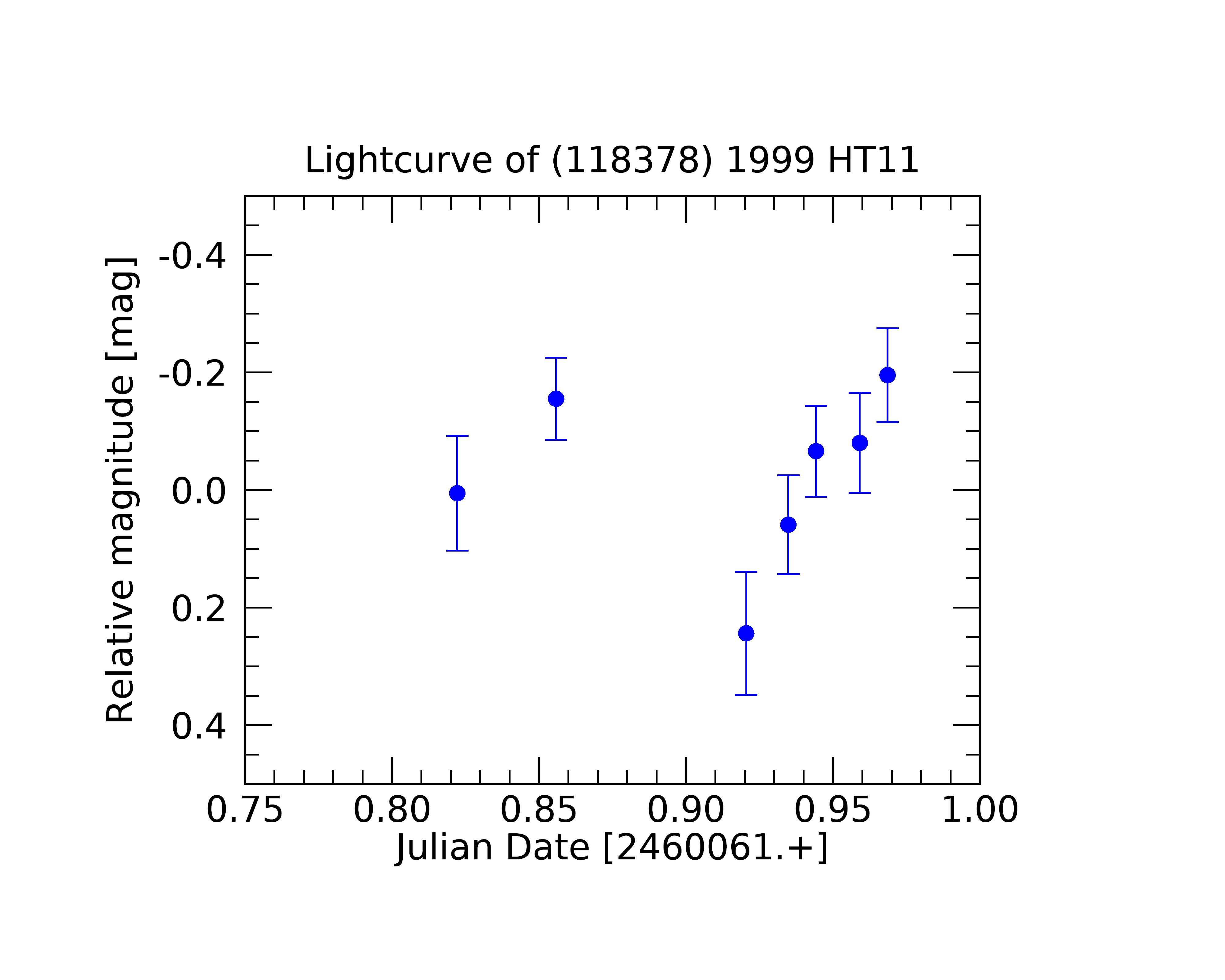} 
  \includegraphics[width=9.5cm, angle=0]{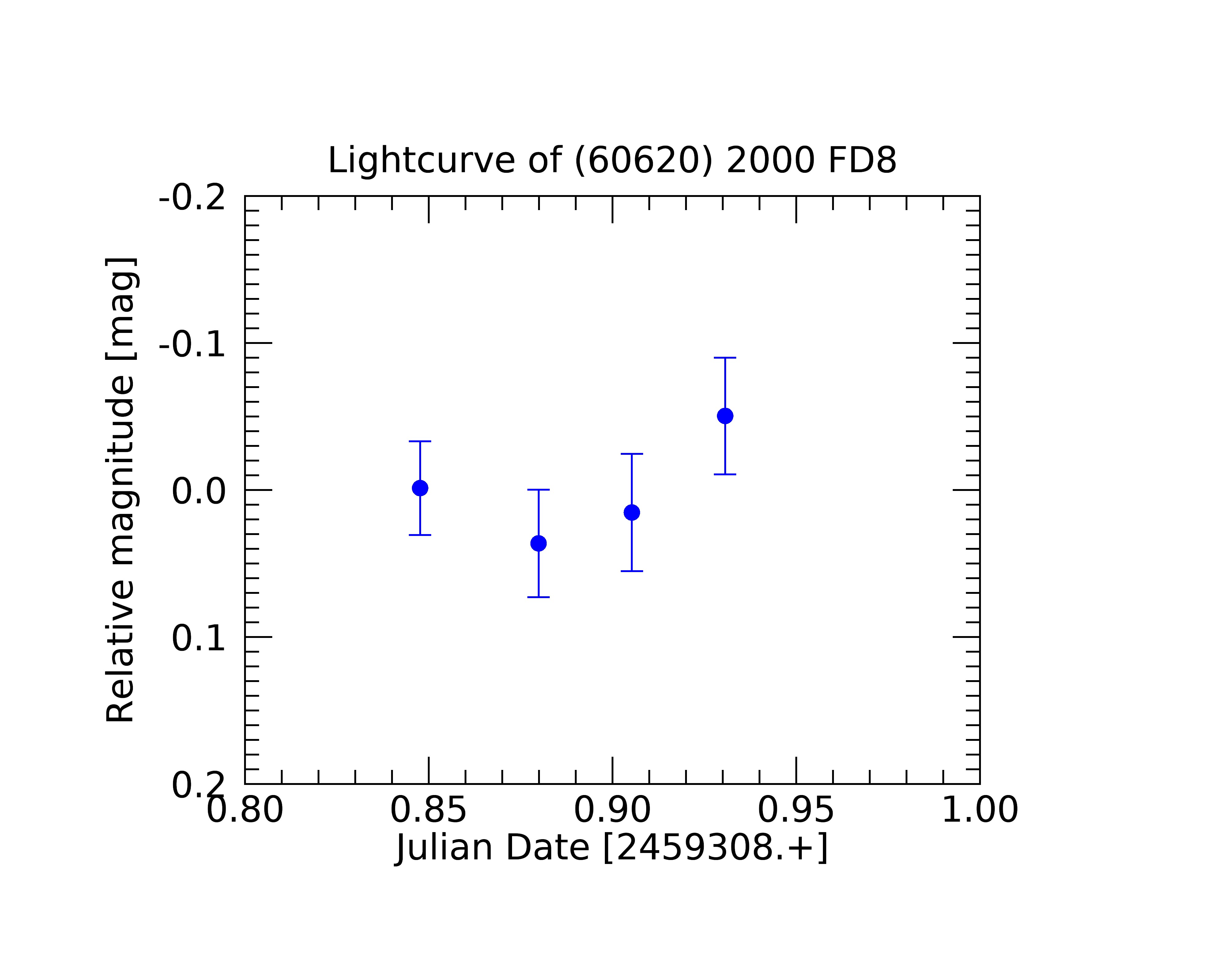}
  \includegraphics[width=9.5cm, angle=0]{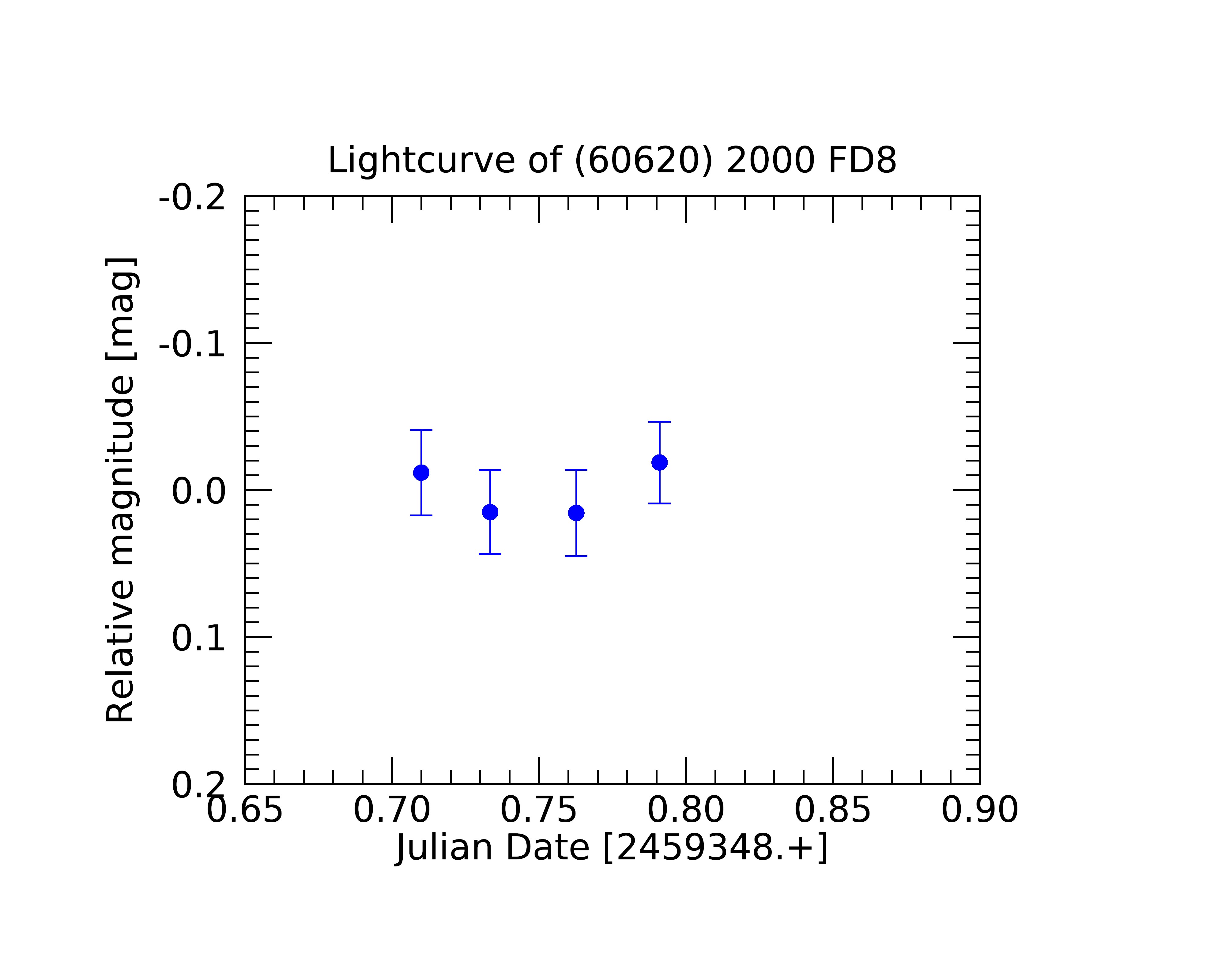} 
\caption{\textit{Objects in the 7:4 mean motion resonance with Neptune }   }
\label{fig:LC74_1}
\end{figure*}

\begin{figure*}
 \includegraphics[width=9.5cm, angle=0]{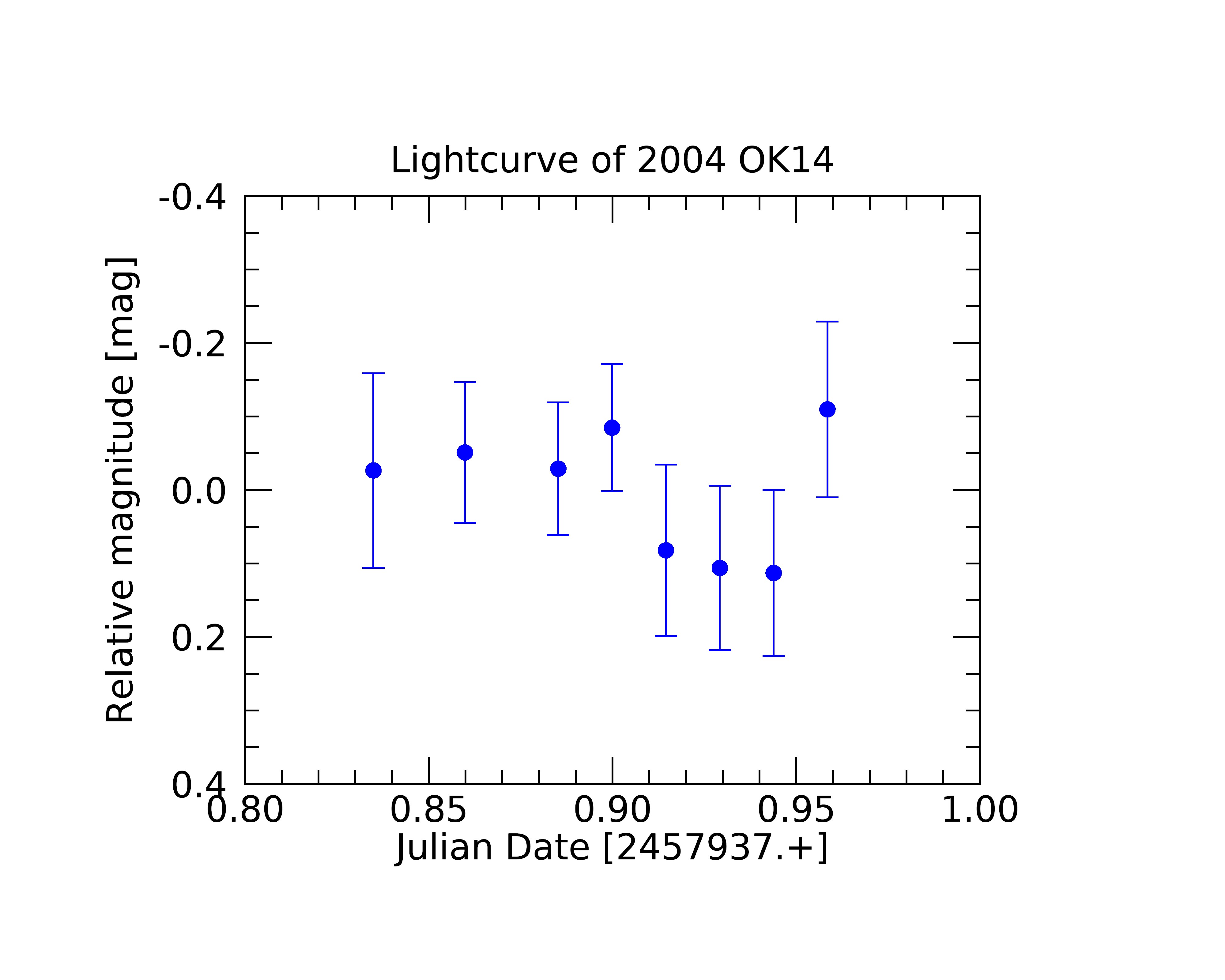}
  \includegraphics[width=9.5cm, angle=0]{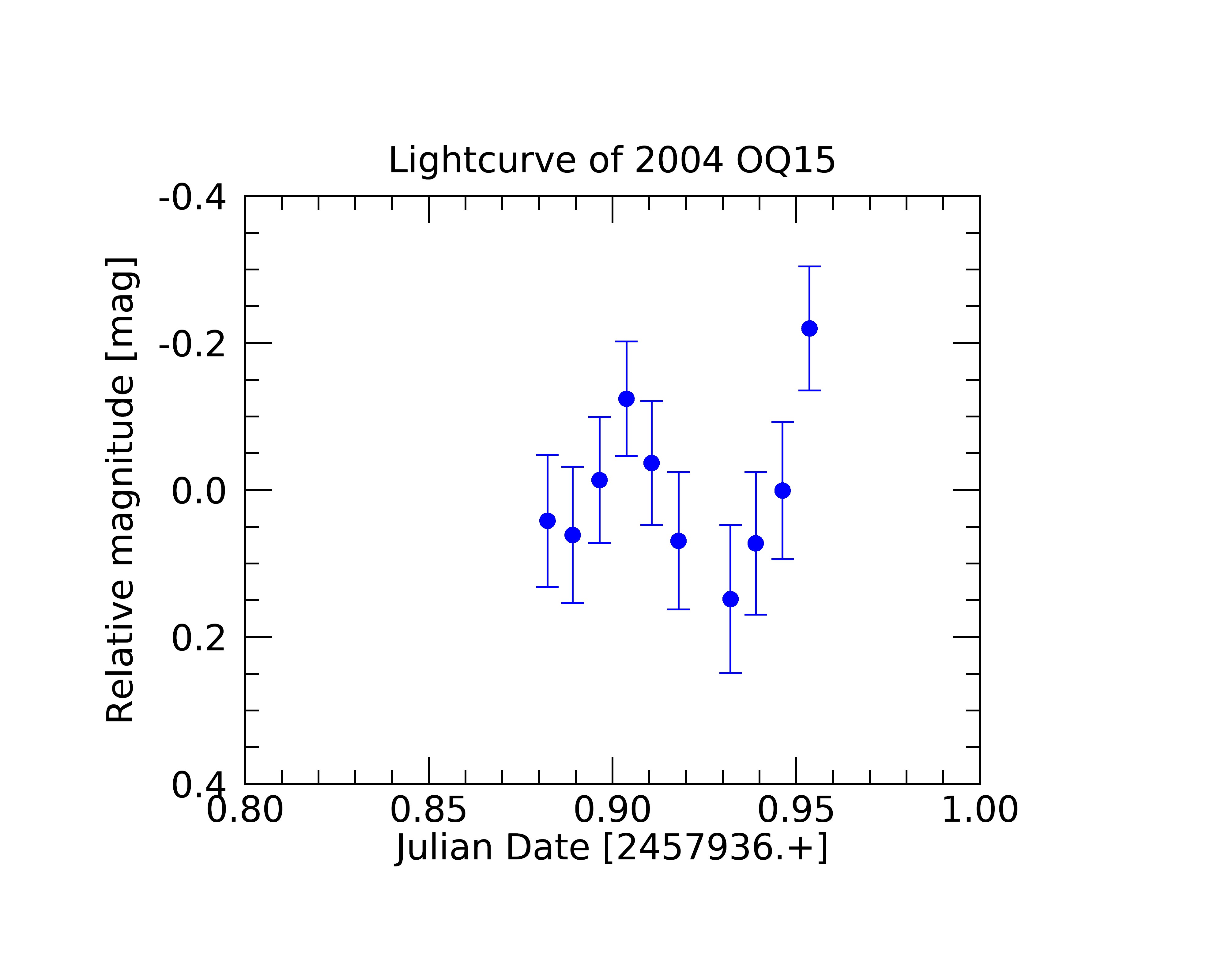}
  \includegraphics[width=9.5cm, angle=0]{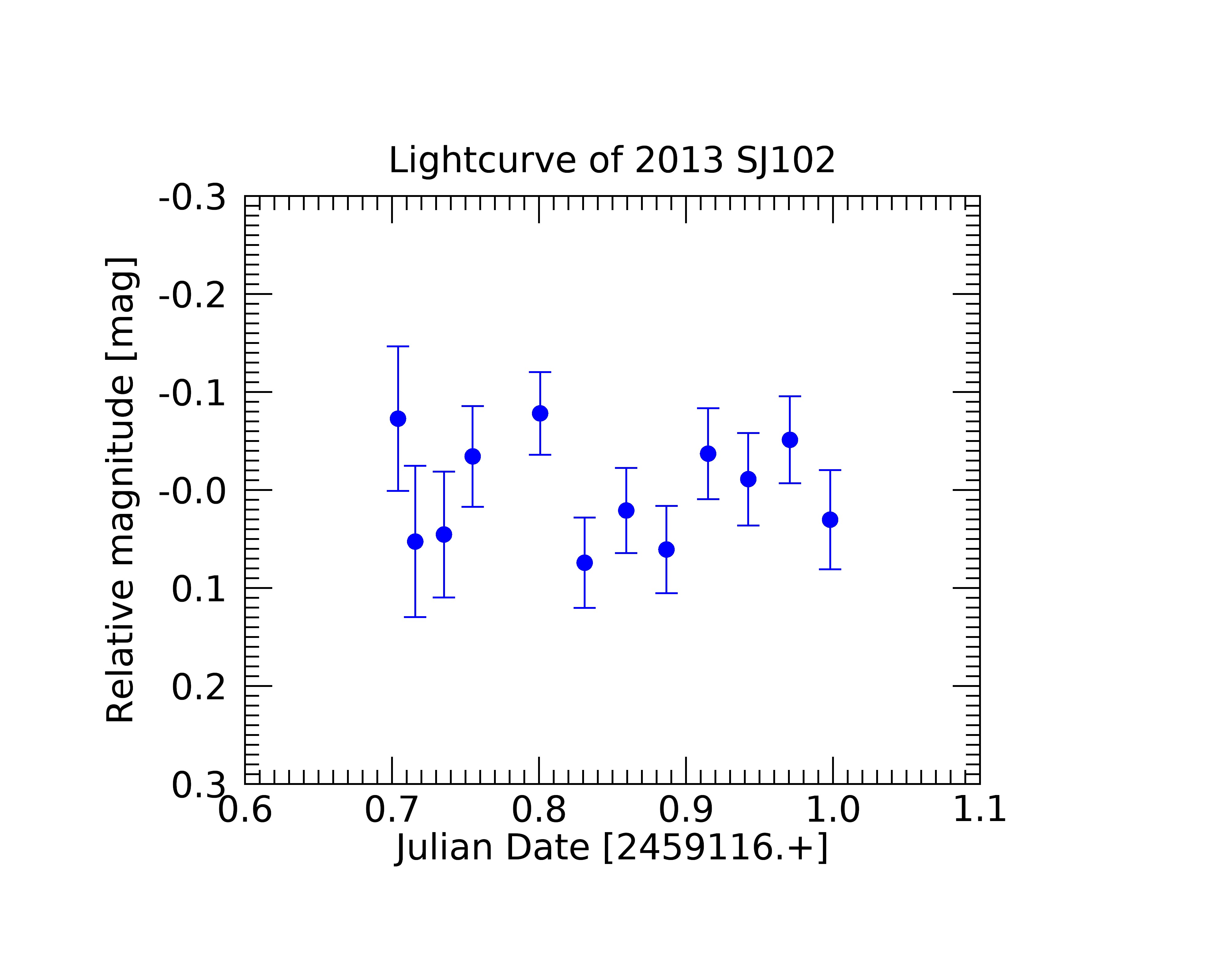}
  \includegraphics[width=9.5cm, angle=0]{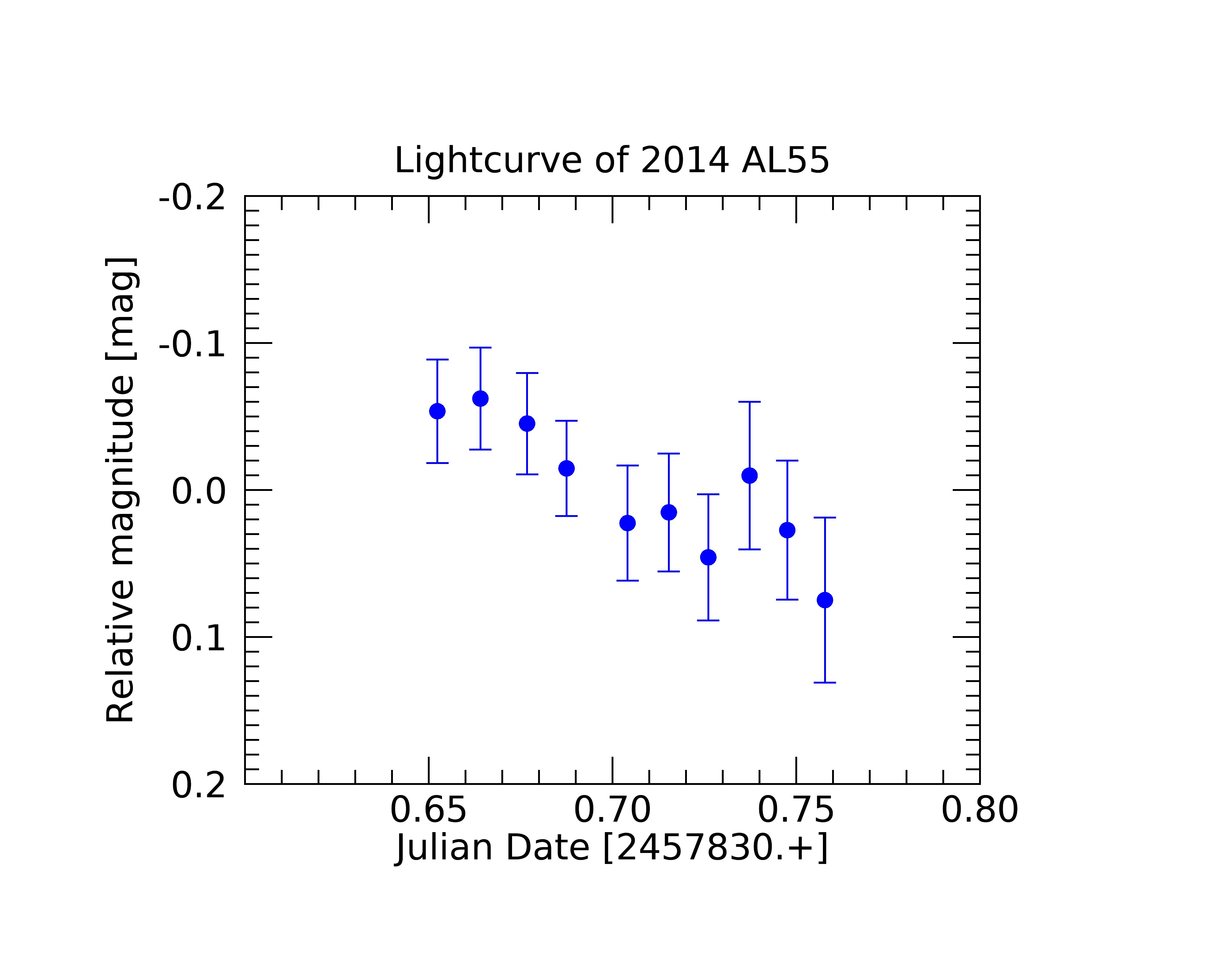}
  \includegraphics[width=9.5cm, angle=0]{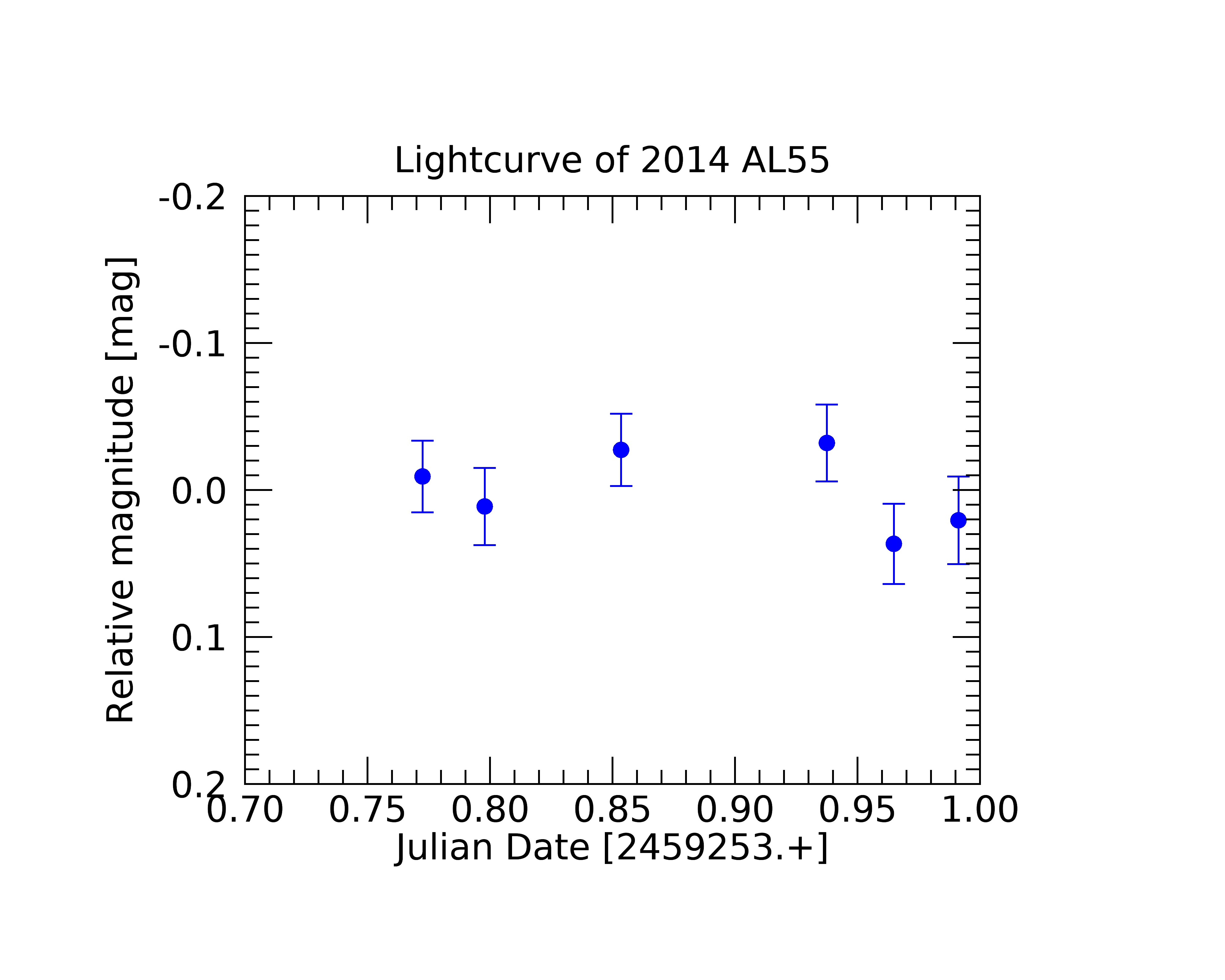} 
  \includegraphics[width=9.5cm, angle=0]{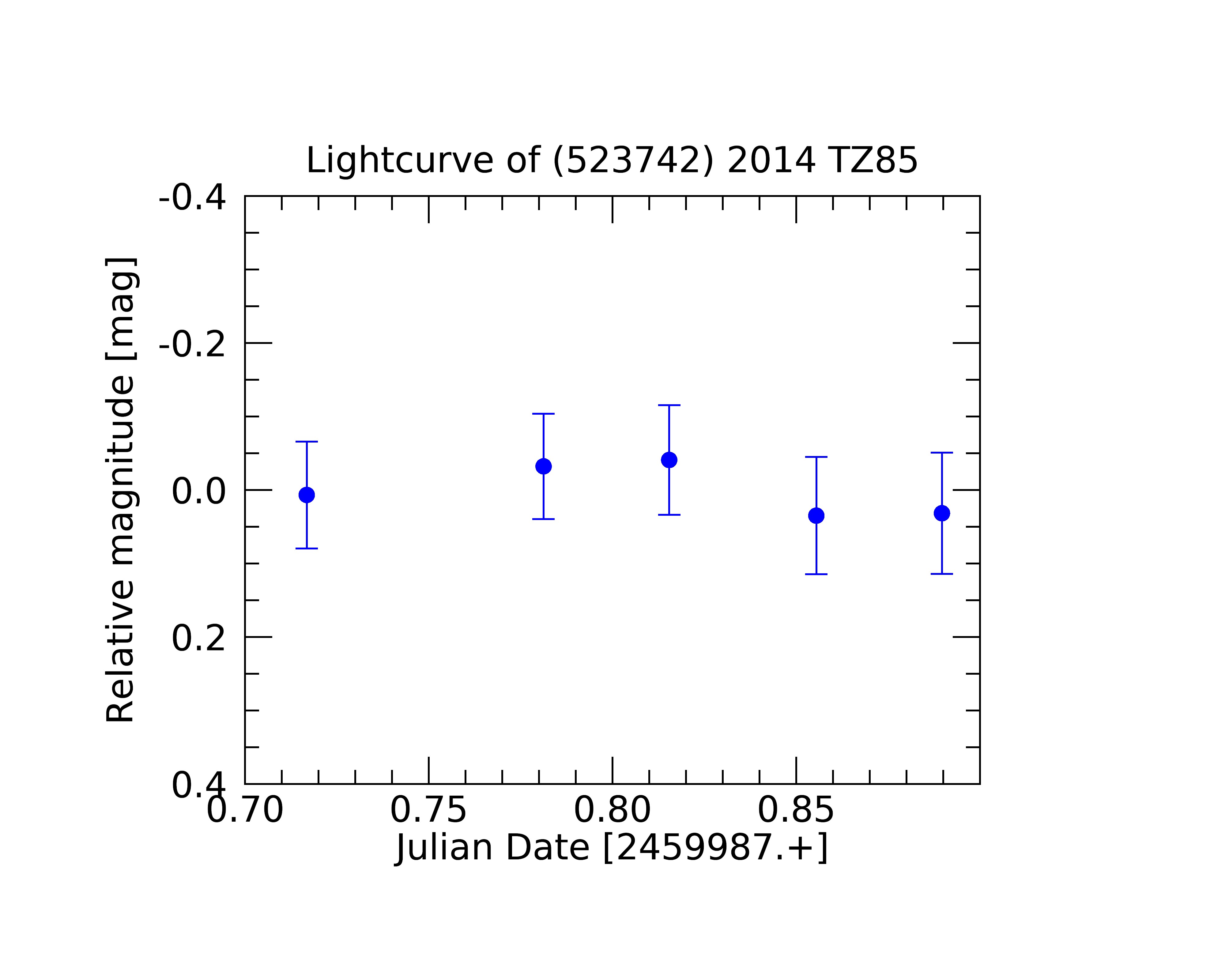} 
\caption{\textit{Objects in the 7:4 mean motion resonance with Neptune }   }
\label{fig:LC74_2}
\end{figure*}

\begin{figure*}
  \includegraphics[width=9.5cm, angle=0]{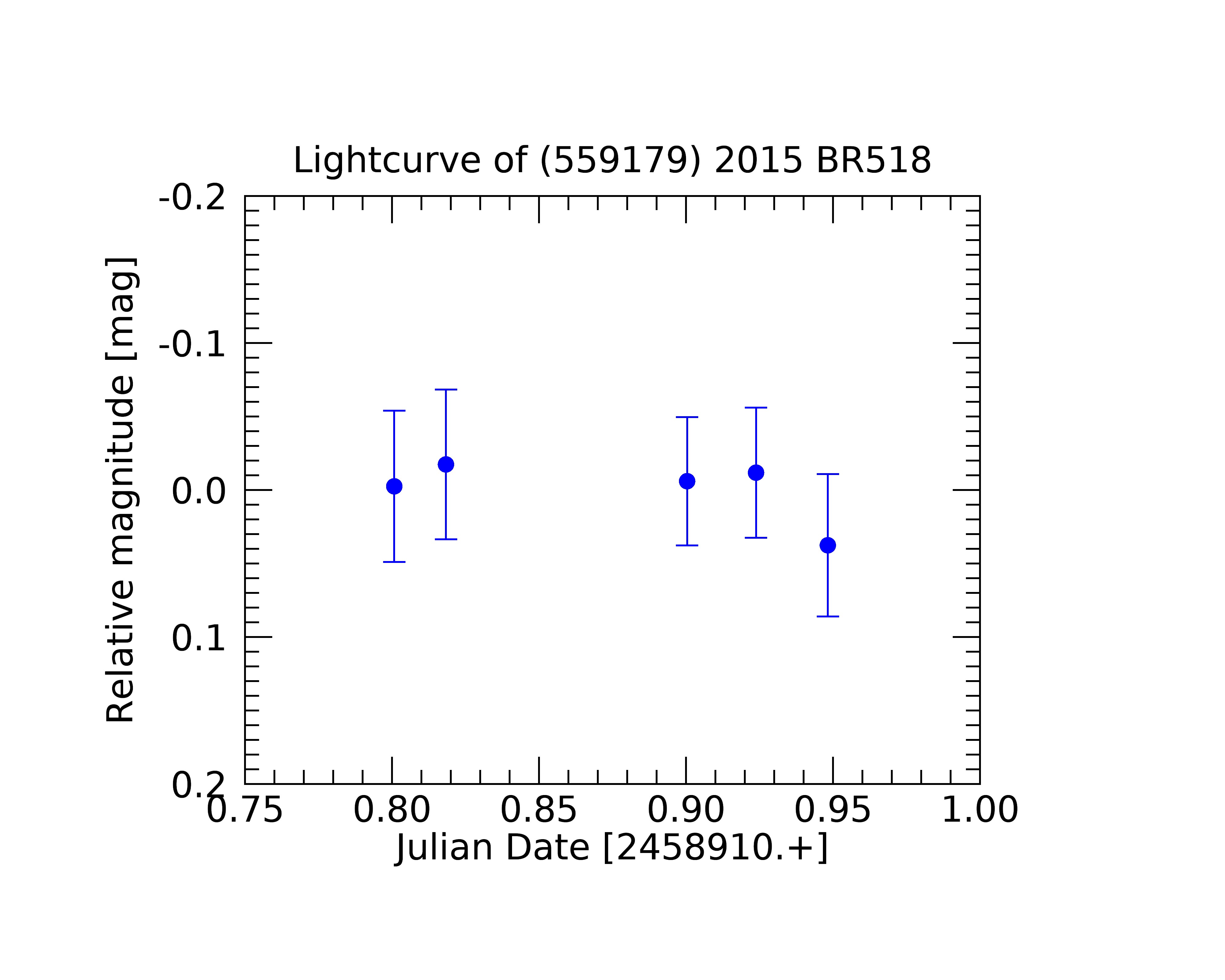} 
  \includegraphics[width=9.5cm, angle=0]{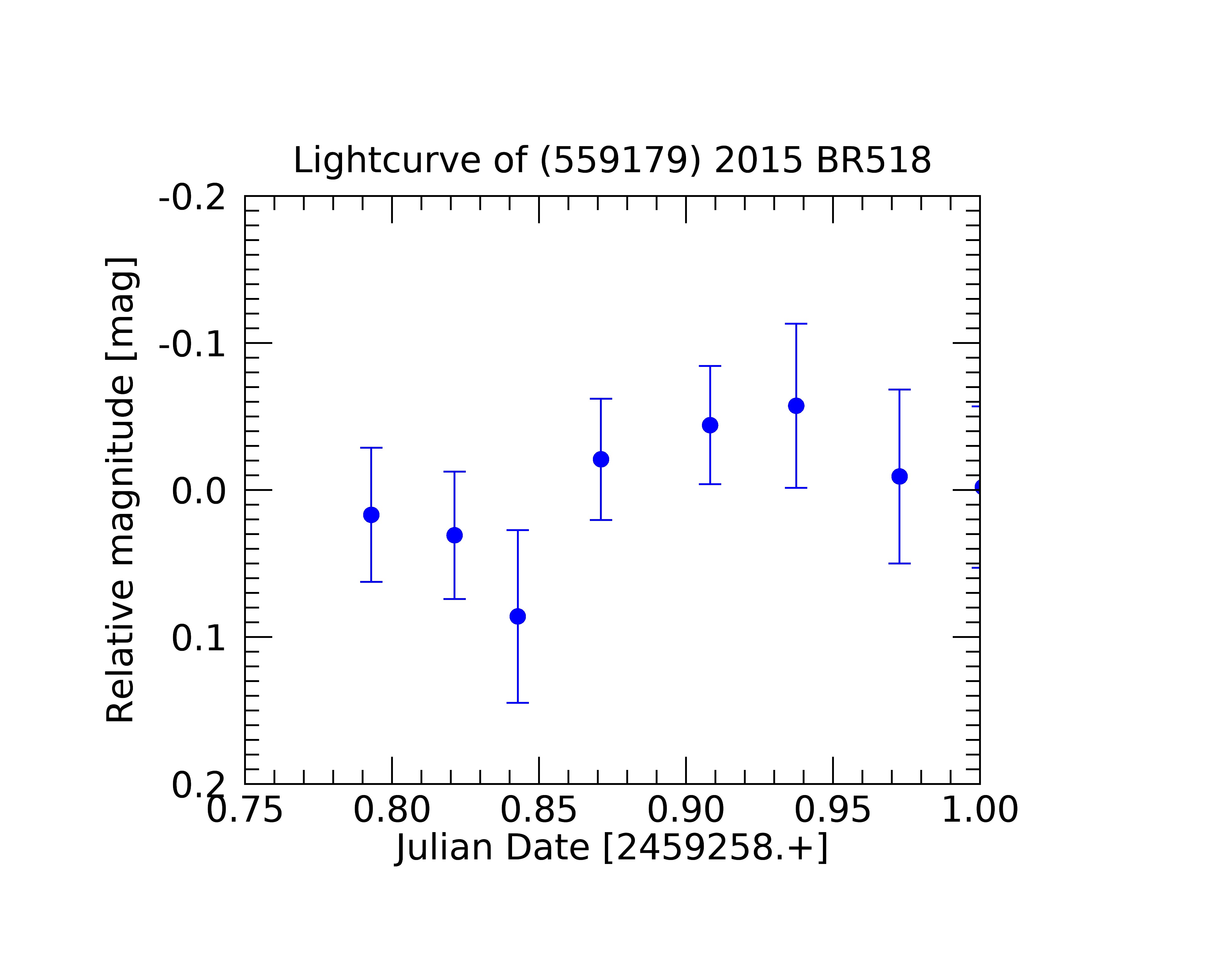} 
\caption{\textit{Objects in the 7:4 mean motion resonance with Neptune }   }
\label{fig:LC74_3}
\end{figure*}

 \clearpage

\section{Appendix B}
\label{sec:appB} 
 The photometry of all targets observed in this paper is available below. No light-time correction applied.    
 \startlongtable
\begin{deluxetable*}{lccc}
 \tablecaption{\label{Tab:Summary_photo2}   }
\tablewidth{0pt}
\tablehead{
Object & Julian Date & Relative Magnitude &  Error  \\
           &                &   [mag]                 &  [mag]\\ }
\startdata 
1999~CX$_{131}$   &   &   &   \\
&	2459258.75547	&	-0.02	&	0.05	\\
&	2459258.78617	&	-0.03	&	0.05	\\
&	2459258.81438	&	0.05	&	0.05	\\
&	2459258.83591	&	0.03	&	0.05	\\
&	2459258.86422	&	-0.06	&	0.05	\\
&	2459258.90054	&	-0.08	&	0.05	\\
&	2459258.96507	&	0.02	&	0.05	\\
&	2459258.99426	&	0.09	&	0.09	\\
\hline
2000~FD$_{8}$   &   &   &   \\
&	2459308.84740	&	0.00	&	0.03	\\
&	2459308.87975	&	0.04	&	0.04	\\
&	2459308.90537	&	0.02	&	0.04	\\
&	2459308.93086	&	-0.05	&	0.04	\\
&	2459348.71001	&	-0.01	&	0.03	\\
&	2459348.73323	&	0.01	&	0.03	\\
&	2459348.76254	&	0.02	&	0.03	\\
&	2459348.79140	&	-0.02	&	0.03	\\
\hline
2001~QF$_{331}$   &   &   &   \\
&	2457656.76101	&	-0.11	&	0.05	\\
&	2457656.78746	&	-0.25	&	0.05	\\
&	2457656.84237	&	-0.09	&	0.05	\\
&	2457656.86055	&	-0.05	&	0.05	\\
&	2457656.87852	&	0.21	&	0.07	\\
&	2457656.89662	&	0.26	&	0.07	\\
&	2457656.90491	&	0.19	&	0.06	\\
&	2457656.91321	&	0.15	&	0.06	\\
&	2457656.93038	&	0.04	&	0.06	\\
&	2457656.96512	&	-0.01	&	0.07	\\
&	2457656.97597	&	-0.18	&	0.07	\\
&	2459847.71913	&	0.16	&	0.07	\\
&	2459847.74252	&	0.11	&	0.06	\\
&	2459847.77491	&	-0.06	&	0.05	\\
&	2459847.80063	&	-0.14	&	0.05	\\
&	2459847.84183	&	-0.11	&	0.05	\\
&	2459847.88430	&	0.14	&	0.05	\\
&	2459847.92376	&	0.33	&	0.06	\\
&	2459847.96533	&	-0.06	&	0.04	\\
&	2459848.00671	&	-0.09	&	0.06	\\
&	2459848.72010	&	-0.14	&	0.06	\\
&	2459848.74542	&	-0.05	&	0.06	\\
&	2459848.76934	&	0.14	&	0.06	\\
&	2459848.85164	&	-0.09	&	0.05	\\
&	2459848.95143	&	0.10	&	0.06	\\
&	2459852.68026	&	-0.26	&	0.08	\\
&	2459852.68736	&	-0.23	&	0.08	\\
&	2459852.71033	&	0.02	&	0.08	\\
&	2459852.73322	&	0.18	&	0.08	\\
&	2459852.76496	&	0.13	&	0.08	\\
&	2459852.80785	&	-0.24	&	0.07	\\
&	2459852.85243	&	-0.09	&	0.08	\\
&	2459852.89485	&	0.00	&	0.08	\\
&	2459852.94115	&	0.15	&	0.08	\\
&	2459852.98253	&	0.09	&	0.08	\\
&	2459853.00909	&	-0.11	&	0.08	\\
\hline
2001~XP$_{254}$ &&&\\
&	2459232.84797	&	-0.07	&	0.04	\\
&	2459232.86636	&	-0.06	&	0.04	\\
&	2459232.94616	&	-0.01	&	0.04	\\
&	2459232.97261	&	0.02	&	0.05	\\
&	2459232.99891	&	0.06	&	0.05	\\
&	2459233.02520	&	0.06	&	0.05	\\
\hline 
2002~VA$_{131}$ &&&\\
&	2459253.62881	&	-0.01	&	0.03	\\
&	2459253.64773	&	0.01	&	0.04	\\
&	2459253.66670	&	0.04	&	0.04	\\
&	2459253.69228	&	0.00	&	0.04	\\
&	2459253.71239	&	-0.02	&	0.04	\\
&	2459253.74618	&	-0.02	&	0.04	\\
&	2459606.66398	&	0.04	&	0.05	\\
&	2459606.68525	&	0.00	&	0.05	\\
&	2459606.72043	&	-0.06	&	0.05	\\
&	2459606.74187	&	-0.01	&	0.06	\\
&	2459606.76962	&	0.03	&	0.07	\\
\hline
2002~VV$_{130}$   &   &   &   \\
&	2459253.61561	&	-0.08	&	0.07	\\
&	2459253.63487	&	0.05	&	0.07	\\
&	2459253.65380	&	0.03	&	0.07	\\
&	2459253.67944	&	-0.03	&	0.07	\\
&	2459253.73233	&	0.03	&	0.08	\\
&	2459545.61607	&	-0.01	&	0.07	\\
&	2459545.69119	&	-0.04	&	0.06	\\
&	2459545.71317	&	-0.03	&	0.06	\\
&	2459545.74415	&	0.06	&	0.06	\\
&	2459545.86380	&	-0.08	&	0.07	\\
&	2459545.88621	&	0.05	&	0.08	\\
&	2459545.90896	&	0.02	&	0.09	\\
&	2459545.93203	&	0.03	&	0.10	\\
&	2459545.94703	&	0.01	&	0.10	\\
&	2459932.70900	&	0.05	&	0.06	\\
&	2459932.74510	&	0.04	&	0.06	\\
&	2459932.81636	&	-0.01	&	0.06	\\
&	2459932.84829	&	-0.08	&	0.06	\\
\hline
2003~YW$_{179}$   &   &   &   \\
&	2459232.79800	&	0.26	&	0.07	\\
&	2459232.81691	&	0.32	&	0.07	\\
&	2459232.84207	&	0.34	&	0.07	\\
&	2459232.86052	&	0.32	&	0.07	\\
&	2459232.96678	&	-0.16	&	0.05	\\
&	2459232.99306	&	-0.21	&	0.05	\\
&	2459233.01931	&	-0.13	&	0.05	\\
&	2459233.03917	&	-0.21	&	0.05	\\
&	2459606.70582	&	-0.11	&	0.07	\\
&	2459606.72739	&	0.12	&	0.08	\\
&	2459606.77674	&	0.43	&	0.09	\\
&	2459606.79696	&	0.22	&	0.08	\\
&	2459606.83672	&	0.20	&	0.08	\\
&	2459606.87105	&	-0.05	&	0.07	\\
&	2459606.90525	&	-0.31	&	0.06	\\
&	2459606.93189	&	-0.32	&	0.07	\\
&	2459606.95221	&	-0.29	&	0.07	\\
&	2459606.97225	&	-0.35	&	0.06	\\
&	2459606.99235	&	-0.31	&	0.07	\\
&	2459607.01227	&	0.15	&	0.10	\\
&	2459648.63000	&	-0.23	&	0.06	\\
&	2459648.70837	&	-0.28	&	0.04	\\
&	2459648.73503	&	-0.18	&	0.05	\\
&	2459648.76981	&	-0.11	&	0.05	\\
&	2459648.84596	&	0.35	&	0.08	\\
&	2459648.87254	&	0.22	&	0.05	\\
&	2459664.64074	&	0.00	&	0.05	\\
&	2459664.66732	&	-0.05	&	0.05	\\
&	2459664.71242	&	-0.25	&	0.03	\\
&	2459664.73712	&	-0.23	&	0.04	\\
&	2459664.77819	&	-0.08	&	0.04	\\
&	2459664.80776	&	0.02	&	0.04	\\
&	2459664.84965	&	0.18	&	0.05	\\
&	2459664.87230	&	0.19	&	0.06	\\
&	2459672.62241	&	0.31	&	0.05	\\
&	2459672.65478	&	0.18	&	0.07	\\
&	2459672.78325	&	-0.16	&	0.04	\\
&	2459672.82205	&	-0.07	&	0.05	\\
&	2459672.84860	&	0.27	&	0.08	\\
\hline
2004~SC$_{60}$   &   &   &   \\
&	2458759.88797	&	-0.11	&	0.04	\\
&	2458759.91251	&	-0.08	&	0.04	\\
&	2458759.93704	&	-0.03	&	0.04	\\
&	2458759.96107	&	-0.01	&	0.04	\\
&	2458759.98532	&	-0.01	&	0.04	\\
&	2458762.66657	&	0.15	&	0.07	\\
&	2458762.70094	&	0.16	&	0.06	\\
&	2458762.75454	&	0.09	&	0.05	\\
&	2458762.82694	&	0.06	&	0.05	\\
&	2458762.86575	&	0.05	&	0.04	\\
&	2458762.90433	&	0.06	&	0.05	\\
&	2458763.00534	&	-0.08	&	0.05	\\
&	2458818.62093	&	0.05	&	0.08	\\
&	2458818.67022	&	-0.02	&	0.07	\\
&	2458818.70312	&	-0.03	&	0.08	\\
&	2458819.56317	&	0.22	&	0.04	\\
&	2458819.59278	&	0.27	&	0.04	\\
&	2458819.61914	&	0.30	&	0.04	\\
&	2458819.64497	&	0.20	&	0.05	\\
&	2459116.69810	&	-0.15	&	0.06	\\
&	2459116.72943	&	-0.13	&	0.05	\\
&	2459116.76875	&	-0.12	&	0.04	\\
&	2459116.79524	&	-0.13	&	0.04	\\
&	2459116.82539	&	-0.13	&	0.04	\\
&	2459116.85376	&	-0.13	&	0.04	\\
&	2459116.88123	&	-0.11	&	0.04	\\
&	2459116.90871	&	-0.12	&	0.04	\\
&	2459116.93653	&	-0.09	&	0.04	\\
&	2459116.96452	&	-0.10	&	0.04	\\
&	2459116.99221	&	-0.04	&	0.04	\\
&	2459117.01083	&	-0.02	&	0.07	\\
&	2459470.73798	&	0.11	&	0.05	\\
&	2459470.76584	&	0.09	&	0.04	\\
&	2459470.79325	&	0.16	&	0.05	\\
&	2459470.82815	&	0.10	&	0.04	\\
&	2459470.85571	&	0.03	&	0.04	\\
&	2459470.88328	&	-0.01	&	0.04	\\
&	2459470.91152	&	0.01	&	0.04	\\
&	2459470.93964	&	-0.04	&	0.04	\\
&	2459470.96774	&	0.01	&	0.04	\\
&	2459470.99560	&	-0.06	&	0.04	\\
&	2459852.67290	&	-0.06	&	0.05	\\
&	2459852.70297	&	-0.06	&	0.04	\\
&	2459852.72581	&	-0.01	&	0.04	\\
&	2459852.88718	&	0.00	&	0.04	\\
&	2459852.93407	&	0.02	&	0.04	\\
&	2459852.97569	&	0.03	&	0.04	\\
\hline
2004~VE$_{131}$   &   &   &   \\
&	2459232.60454	&	0.10	&	0.06	\\
&	2459232.63495	&	0.06	&	0.06	\\
&	2459232.68799	&	0.02	&	0.06	\\
&	2459232.73410	&	-0.04	&	0.05	\\
&	2459232.75959	&	-0.02	&	0.05	\\
&	2459232.78510	&	-0.08	&	0.05	\\
&	2459232.81024	&	-0.05	&	0.06	\\
&	2459232.83547	&	0.02	&	0.07	\\
&	2459253.62254	&	0.12	&	0.07	\\
&	2459253.66057	&	0.06	&	0.07	\\
&	2459253.68615	&	0.08	&	0.07	\\
&	2459253.70599	&	0.03	&	0.06	\\
&	2459253.73970	&	-0.13	&	0.06	\\
&	2459253.78423	&	-0.16	&	0.08	\\
&	2459545.67471	&	-0.17	&	0.05	\\
&	2459545.70283	&	-0.14	&	0.05	\\
&	2459545.72471	&	-0.04	&	0.05	\\
&	2459545.75566	&	-0.01	&	0.05	\\
&	2459545.81021	&	0.13	&	0.06	\\
&	2459545.85633	&	0.16	&	0.07	\\
&	2459545.87882	&	0.17	&	0.07	\\
&	2459545.92407	&	0.04	&	0.07	\\
&	2459545.96264	&	-0.14	&	0.07	\\
&	2459551.65501	&	0.14	&	0.07	\\
&	2459551.68352	&	0.08	&	0.06	\\
&	2459551.71927	&	-0.06	&	0.05	\\
&	2459551.75052	&	-0.12	&	0.05	\\
&	2459551.80790	&	-0.22	&	0.05	\\
&	2459551.91223	&	-0.14	&	0.06	\\
&	2459551.94738	&	-0.04	&	0.06	\\
&	2459551.96620	&	0.19	&	0.08	\\
&	2459551.99083	&	0.17	&	0.09	\\
&	2459553.61560	&	0.26	&	0.09	\\
&	2459553.64415	&	0.09	&	0.07	\\
&	2459553.66463	&	0.07	&	0.07	\\
&	2459553.72879	&	0.06	&	0.06	\\
&	2459553.75968	&	-0.14	&	0.05	\\
&	2459553.79627	&	-0.14	&	0.05	\\
&	2459553.82344	&	-0.07	&	0.05	\\
&	2459553.85051	&	-0.13	&	0.05	\\
&	2459553.90546	&	-0.02	&	0.06	\\
&	2459553.92079	&	0.02	&	0.07	\\
\hline
2006~CJ$_{69}$   &   &   &   \\
&	2459253.86438	&	0.04	&	0.08	\\
&	2459253.88975	&	0.08	&	0.08	\\
&	2459253.91677	&	0.13	&	0.08	\\
&	2459253.97069	&	0.18	&	0.08	\\
&	2459254.01888	&	0.04	&	0.08	\\
&	2459254.04042	&	0.01	&	0.08	\\
&	2459258.71757	&	0.02	&	0.08	\\
&	2459258.77880	&	0.11	&	0.08	\\
&	2459258.80710	&	0.18	&	0.08	\\
&	2459258.82858	&	0.24	&	0.08	\\
&	2459258.85692	&	0.14	&	0.08	\\
&	2459258.89292	&	0.05	&	0.08	\\
&	2459258.92240	&	0.03	&	0.08	\\
&	2459258.95777	&	-0.05	&	0.08	\\
&	2459258.98689	&	-0.09	&	0.08	\\
&	2459259.01587	&	-0.15	&	0.10	\\
&	2459317.64558	&	0.08	&	0.08	\\
&	2459317.66679	&	0.13	&	0.08	\\
&	2459317.68752	&	0.09	&	0.08	\\
&	2459317.70842	&	0.19	&	0.08	\\
&	2459317.84544	&	0.10	&	0.09	\\
&	2459317.88818	&	0.04	&	0.08	\\
&	2459606.79019	&	0.18	&	0.06	\\
&	2459606.81076	&	0.17	&	0.06	\\
&	2459606.85048	&	0.09	&	0.05	\\
&	2459606.88462	&	0.06	&	0.05	\\
&	2459606.91224	&	0.05	&	0.05	\\
&	2459606.93895	&	-0.08	&	0.05	\\
&	2459606.95912	&	-0.14	&	0.05	\\
&	2459606.97919	&	-0.09	&	0.05	\\
&	2459606.99918	&	-0.15	&	0.05	\\
&	2459607.01908	&	-0.14	&	0.05	\\
&	2459607.03897	&	-0.18	&	0.05	\\
&	2459670.74009	&	-0.01	&	0.05	\\
&	2459670.77309	&	-0.05	&	0.04	\\
&	2459670.80824	&	-0.05	&	0.04	\\
&	2459670.84083	&	-0.12	&	0.05	\\
&	2459670.86080	&	-0.17	&	0.06	\\
&	2459670.90779	&	-0.17	&	0.05	\\
&	2459672.71069	&	0.05	&	0.06	\\
&	2459672.74316	&	-0.03	&	0.05	\\
&	2459672.79006	&	-0.13	&	0.04	\\
&	2459672.82883	&	-0.17	&	0.04	\\
&	2459672.86956	&	-0.09	&	0.07	\\
\hline
2008~CS$_{190}$ &&&\\
&	2458542.64090	&	0.03	&	0.02	\\
&	2458542.67882	&	-0.07	&	0.02	\\
&	2458542.75091	&	-0.02	&	0.02	\\
&	2458542.82427	&	0.00	&	0.02	\\
&	2458543.61234	&	0.04	&	0.02	\\
&	2458543.67213	&	0.05	&	0.02	\\
&	2458543.80908	&	-0.03	&	0.02	\\
&	2458544.68763	&	0.00	&	0.02	\\
&	2458544.76958	&	-0.01	&	0.02	\\
\hline
2012~BY$_{154}$ &&&\\
&	2458542.62528	&	-0.05	&	0.03	\\
&	2458542.64546	&	0.01	&	0.03	\\
&	2458542.68343	&	0.03	&	0.03	\\
&	2458542.75597	&	-0.07	&	0.04	\\
&	2458542.82921	&	0.03	&	0.04	\\
&	2458544.69163	&	0.00	&	0.04	\\
&	2458544.77345	&	0.04	&	0.04	\\
\hline
2014~DK$_{143}$ &&& \\
&	2459348.86423	&	0.01	&	0.03	\\
&	2459348.87195	&	0.02	&	0.03	\\
&	2459348.89304	&	-0.06	&	0.03	\\
&	2459348.89980	&	-0.04	&	0.02	\\
&	2460055.90235	&	0.02	&	0.03	\\
&	2460055.90946	&	-0.05	&	0.03	\\
&	2460055.93151	&	-0.07	&	0.04	\\
&	2460055.93921	&	-0.03	&	0.03	\\
&	2460055.94632	&	-0.02	&	0.03	\\
&	2460055.95342	&	-0.04	&	0.03	\\
&	2460055.96060	&	0.00	&	0.03	\\
&	2460055.96770	&	-0.01	&	0.03	\\
&	2460055.97481	&	-0.01	&	0.04	\\
&	2460055.98221	&	0.03	&	0.04	\\
&	2460056.89858	&	-0.04	&	0.03	\\
&	2460056.90568	&	-0.01	&	0.03	\\
&	2460056.91279	&	0.01	&	0.03	\\
&	2460056.93450	&	0.07	&	0.03	\\
&	2460056.94161	&	0.10	&	0.03	\\
&	2460056.95582	&	0.11	&	0.04	\\
&	2460056.97079	&	0.14	&	0.04	\\
&	2460056.97790	&	0.15	&	0.04	\\
&	2460056.98500	&	0.11	&	0.04	\\
&	2460059.79363	&	0.14	&	0.04	\\
&	2460059.81846	&	0.04	&	0.04	\\
&	2460059.84265	&	-0.02	&	0.04	\\
&	2460059.88511	&	-0.03	&	0.03	\\
&	2460059.91946	&	0.03	&	0.03	\\
&	2460059.93221	&	0.06	&	0.03	\\
&	2460059.95006	&	0.09	&	0.04	\\
&	2460059.96313	&	0.11	&	0.04	\\
&	2460059.97621	&	0.10	&	0.04	\\
&	2460061.86269	&	0.05	&	0.03	\\
&	2460061.88983	&	0.02	&	0.03	\\
&	2460061.90393	&	-0.05	&	0.03	\\
&	2460061.90874	&	-0.08	&	0.03	\\
&	2460061.91356	&	-0.08	&	0.03	\\
&	2460061.92736	&	-0.09	&	0.03	\\
&	2460061.95164	&	-0.03	&	0.03	\\
&	2460061.97603	&	0.02	&	0.03	\\
&	2460061.98082	&	0.01	&	0.04	\\
\hline
2014~OL$_{394}$ &&& \\
&	2457656.69773	&	-0.14	&	0.06	\\
&	2457656.73362	&	-0.14	&	0.05	\\
&	2457656.75237	&	-0.08	&	0.05	\\
&	2457656.77861	&	-0.02	&	0.05	\\
&	2457656.82548	&	0.08	&	0.06	\\
&	2457656.83321	&	0.13	&	0.06	\\
&	2457656.85112	&	0.14	&	0.06	\\
&	2457656.86963	&	0.03	&	0.06	\\
&	2458054.63685	&	0.00	&	0.07	\\
&	2458054.65506	&	0.03	&	0.07	\\
&	2458054.66667	&	0.05	&	0.07	\\
&	2458054.69588	&	0.16	&	0.08	\\
&	2458054.71288	&	0.12	&	0.07	\\
&	2458054.73058	&	-0.01	&	0.07	\\
&	2458054.74840	&	0.02	&	0.06	\\
&	2458054.76615	&	-0.03	&	0.06	\\
&	2458054.78400	&	-0.11	&	0.05	\\
&	2458054.80200	&	-0.25	&	0.05	\\
\hline
1999~HG$_{12}$ &&& \\
&	2460056.68876	&	0.01	&	0.06	\\
&	2460056.71146	&	0.06	&	0.07	\\
&	2460056.73007	&	0.06	&	0.06	\\
&	2460056.74211	&	0.05	&	0.06	\\
&	2460056.76182	&	0.01	&	0.06	\\
&	2460056.79360	&	-0.09	&	0.06	\\
&	2460056.91359	&	-0.12	&	0.07	\\
\hline
1999~HR$_{11}$ &&& \\
&	2458254.50918	&	-0.15	&	0.10	\\
&	2458254.53348	&	-0.06	&	0.07	\\
&	2458254.58262	&	0.01	&	0.06	\\
&	2458254.60499	&	0.04	&	0.07	\\
&	2458254.62638	&	-0.06	&	0.07	\\
&	2458254.68414	&	0.06	&	0.06	\\
&	2458254.73908	&	0.01	&	0.07	\\
&	2458255.48621	&	0.16	&	0.07	\\
\hline
1999~HT$_{11}$ &&& \\
&	2460059.80206	&	-0.26	&	0.12	\\
&	2460059.85116	&	-0.08	&	0.12	\\
&	2460059.87654	&	-0.07	&	0.10	\\
&	2460059.89377	&	0.16	&	0.12	\\
&	2460059.91092	&	0.26	&	0.12	\\
&	2460059.94115 	&	0.18	&	0.13	\\
&	2460061.82222 	&	0.01	&	0.10	\\
&	2460061.85583	&	-0.16	&	0.07	\\
&	2460061.92049	&	0.24	&	0.10	\\
&	2460061.93482	&	0.06	&	0.08	\\
&	2460061.94424	&	-0.07	&	0.08	\\
&	2460061.95911 	&	-0.08	&	0.09	\\
&	2460061.96856	&	-0.20	&	0.08	\\
\hline
2004~OK$_{14}$ &&&\\
&	2457937.83476	&	-0.03	&	0.13	\\
&	2457937.86010	&	-0.05	&	0.10	\\
&	2457937.88543	&	-0.03	&	0.09	\\
&	2457937.90002	&	-0.08	&	0.09	\\
&	2457937.91466	&	0.08	&	0.12	\\
&	2457937.92934	&	0.11	&	0.11	\\
&	2457937.94396	&	0.11	&	0.11	\\
&	2457937.95858	&	-0.11	&	0.12	\\
\hline
2004~OQ$_{15}$ &&& \\
&	2457936.88223	&	0.04	&	0.09	\\
&	2457936.88936	&	0.06	&	0.09	\\
&	2457936.89649	&	-0.01	&	0.09	\\
&	2457936.90361	&	-0.12	&	0.08	\\
&	2457936.91071	&	-0.04	&	0.08	\\
&	2457936.91784	&	0.07	&	0.09	\\
&	2457936.93207	&	0.15	&	0.10	\\
&	2457936.93917	&	0.07	&	0.10	\\
&	2457936.94629	&	0.00	&	0.09	\\
&	2457936.95339	&	-0.22	&	0.08	\\
\hline
2004~SC$_{60}$ &&& \\
&	2458759.88797	&	-0.13	&	0.04	\\
&	2458759.91251	&	-0.10	&	0.04	\\
&	2458759.93704	&	-0.05	&	0.04	\\
&	2458759.96107	&	-0.03	&	0.04	\\
&	2458759.98532	&	-0.03	&	0.04	\\
&	2458762.66657	&	0.15	&	0.08	\\
&	2458762.70094	&	0.18	&	0.07	\\
&	2458762.75454	&	0.02	&	0.05	\\
&	2458762.82694	&	0.05	&	0.05	\\
&	2458762.86575	&	0.06	&	0.05	\\
&	2458762.90433	&	0.03	&	0.05	\\
&	2458763.00534	&	-0.13	&	0.05	\\
&	2458818.53528	&	0.11	&	0.07	\\
&	2458818.58641	&	0.11	&	0.07	\\
&	2458818.62093	&	0.03	&	0.08	\\
&	2458818.67022	&	-0.04	&	0.07	\\
&	2458818.70312	&	-0.05	&	0.08	\\
&	2458819.56317	&	0.20	&	0.04	\\
&	2458819.59278	&	0.25	&	0.04	\\
&	2458819.61914	&	0.28	&	0.04	\\
&	2458819.64497	&	0.18	&	0.05	\\
&	2459116.74908	&	-0.10	&	0.04	\\
&	2459116.76875	&	-0.14	&	0.04	\\
&	2459116.79524	&	-0.15	&	0.04	\\
&	2459116.82539	&	-0.15	&	0.04	\\
&	2459116.85376	&	-0.15	&	0.04	\\
&	2459116.88123	&	-0.13	&	0.04	\\
&	2459116.90871	&	-0.14	&	0.04	\\
&	2459116.93653	&	-0.11	&	0.04	\\
&	2459116.96452	&	-0.12	&	0.04	\\
&	2459116.99221	&	-0.06	&	0.04	\\
&	2459117.01083	&	-0.04	&	0.07	\\
&	2459470.73798	&	0.09	&	0.05	\\
&	2459470.76584	&	0.07	&	0.04	\\
&	2459470.79325	&	0.14	&	0.05	\\
&	2459470.82815	&	0.08	&	0.04	\\
&	2459470.85571	&	0.01	&	0.04	\\
&	2459470.88328	&	-0.03	&	0.04	\\
&	2459470.91152	&	-0.01	&	0.04	\\
&	2459470.93964	&	-0.06	&	0.04	\\
&	2459470.96774	&	0.00	&	0.04	\\
&	2459470.99560	&	-0.08	&	0.04	\\
\hline
2013~BN$_{82}$ &&&\\
&	2459253.67281	&	0.01	&	0.03	\\
&	2459253.72544	&	-0.14	&	0.03	\\
&	2459253.75937	&	-0.16	&	0.03	\\
&	2459253.77747	&	-0.20	&	0.03	\\
&	2459253.83323	&	-0.14	&	0.03	\\
&	2459253.85819	&	-0.10	&	0.03	\\
&	2459258.63404	&	0.04	&	0.04	\\
&	2459258.64969	&	0.03	&	0.04	\\
&	2459258.66754	&	-0.09	&	0.03	\\
&	2459258.67723	&	-0.11	&	0.03	\\
&	2459258.70077	&	-0.14	&	0.03	\\
&	2459258.73174	&	-0.14	&	0.03	\\
&	2459258.87843	&	0.08	&	0.03	\\
&	2459258.95093	&	-0.04	&	0.04	\\
&	2459281.64279	&	0.05	&	0.05	\\
&	2459281.65768	&	0.08	&	0.04	\\
&	2459281.67960	&	0.20	&	0.05	\\
&	2459281.69916	&	0.15	&	0.05	\\
&	2459648.64348	&	-0.08	&	0.04	\\
&	2459648.66986	&	0.07	&	0.04	\\
&	2459648.69506	&	0.18	&	0.04	\\
&	2459648.72165	&	0.14	&	0.04	\\
&	2459648.75651	&	0.08	&	0.04	\\
&	2459664.61984	&	0.10	&	0.05	\\
&	2459664.64693	&	0.08	&	0.03	\\
&	2459664.74270	&	0.02	&	0.03	\\
\hline
2013~FR$_{28}$ &&&\\
&	2458543.72514	&	0.51	&	0.06	\\
&	2458543.79974	&	-0.07	&	0.05	\\
&	2458543.83569	&	-0.15	&	0.05	\\
&	2458544.65482	&	0.00	&	0.08	\\
&	2458544.67808	&	-0.12	&	0.07	\\
&	2458544.69804	&	-0.27	&	0.07	\\
&	2458544.72458	&	-0.31	&	0.07	\\
&	2458544.74718	&	-0.23	&	0.07	\\
&	2458544.77917	&	-0.21	&	0.07	\\
&	2458544.80329	&	-0.14	&	0.07	\\
&	2458544.83153	&	-0.05	&	0.07	\\
&	2458544.85746	&	0.22	&	0.08	\\
&	2458544.88904	&	0.61	&	0.09	\\
&	2458607.65110	&	-0.29	&	0.06	\\
&	2458607.67441	&	-0.24	&	0.09	\\
&	2458607.70575	&	-0.14	&	0.08	\\
&	2458607.73639	&	0.32	&	0.08	\\
&	2458607.76607	&	0.54	&	0.09	\\
&	2458607.78900	&	0.31	&	0.08	\\
&	2458607.81189	&	0.06	&	0.06	\\
&	2458607.83476	&	-0.10	&	0.06	\\
&	2458607.85749	&	-0.20	&	0.05	\\
&	2458607.87260	&	-0.24	&	0.05	\\
&	2458607.88770	&	-0.20	&	0.06	\\
&	2458988.66601	&	-0.20	&	0.06	\\
&	2458988.69248	&	-0.17	&	0.06	\\
&	2458988.72555	&	0.03	&	0.08	\\
&	2458988.75065	&	0.16	&	0.08	\\
&	2458988.78370	&	0.68	&	0.13	\\
&	2458988.81611	&	0.14	&	0.09	\\
&	2458988.84849	&	-0.13	&	0.08	\\
&	2458989.68445	&	0.23	&	0.08	\\
&	2458989.73926	&	-0.10	&	0.06	\\
&	2458989.76650	&	-0.20	&	0.06	\\
&	2458989.79304	&	-0.06	&	0.07	\\
\hline
2013~SJ$_{102}$ &&& \\
&	2459116.70415	&	-0.07	&	0.07	\\
&	2459116.71618	&	0.05	&	0.08	\\
&	2459116.73527	&	0.05	&	0.06	\\
&	2459116.75496	&	-0.03	&	0.05	\\
&	2459116.80112	&	-0.08	&	0.04	\\
&	2459116.83123	&	0.07	&	0.05	\\
&	2459116.85962	&	0.02	&	0.04	\\
&	2459116.88707	&	0.06	&	0.04	\\
&	2459116.91458	&	-0.04	&	0.05	\\
&	2459116.94238	&	-0.01	&	0.05	\\
&	2459116.97048	&	-0.05	&	0.04	\\
&	2459116.99822	&	0.03	&	0.05	\\
\hline
2014~AL$_{55}$ &&&\\
&	2457830.65249	&	-0.05	&	0.04	\\
&	2457830.66413	&	-0.06	&	0.03	\\
&	2457830.67686	&	-0.05	&	0.03	\\
&	2457830.68739	&	-0.01	&	0.03	\\
&	2457830.70409	&	0.02	&	0.04	\\
&	2457830.71513	&	0.02	&	0.04	\\
&	2457830.72608	&	0.05	&	0.04	\\
&	2457830.73726	&	-0.01	&	0.05	\\
&	2457830.74739	&	0.03	&	0.05	\\
&	2457830.75775	&	0.07	&	0.06	\\
&	2459253.77247	&	-0.01	&	0.02	\\
&	2459253.79761	&	0.01	&	0.03	\\
&	2459253.85312	&	-0.03	&	0.02	\\
&	2459253.93774	&	-0.03	&	0.03	\\
&	2459253.96466	&	0.04	&	0.03	\\
&	2459253.99152	&	0.02	&	0.03	\\
\hline
2015~BR$_{518}$ &&&\\
&	2458910.80091	&	0.00	&	0.05	\\
&	2458910.81845	&	-0.02	&	0.05	\\
&	2458910.90074	&	-0.01	&	0.04	\\
&	2458910.92419	&	-0.01	&	0.04	\\
&	2458910.94818	&	0.04	&	0.05	\\
&	2459258.79327	&	0.02	&	0.05	\\
&	2459258.82144	&	0.03	&	0.04	\\
&	2459258.84301	&	0.09	&	0.06	\\
&	2459258.87143	&	-0.02	&	0.04	\\
&	2459258.90781	&	-0.04	&	0.04	\\
&	2459258.93706	&	-0.06	&	0.06	\\
&	2459258.97235	&	-0.01	&	0.06	\\
&	2459259.00142	&	0.00	&	0.05	\\
\hline
2015~FP$_{345}$ &&&\\
&	2458607.69027	&	0.11	&	0.06	\\
&	2458607.72142	&	-0.04	&	0.06	\\
&	2458607.75149	&	-0.21	&	0.03	\\
&	2458607.84233	&	0.18	&	0.05	\\
&	2458910.97822	&	0.08	&	0.04	\\
&	2458911.03204	&	0.26	&	0.06	\\
&	2458988.71933	&	0.09	&	0.05	\\
&	2458988.74440	&	-0.05	&	0.04	\\
&	2458988.80942	&	-0.07	&	0.04	\\
&	2458988.84176	&	0.06	&	0.05	\\
&	2458988.87796	&	0.15	&	0.06	\\
&	2458989.73250	&	0.12	&	0.04	\\
&	2458989.75979	&	0.30	&	0.05	\\
&	2458989.78626	&	0.11	&	0.04	\\
&	2458989.81253	&	-0.03	&	0.04	\\
&	2458989.83804	&	-0.24	&	0.04	\\
&	2458989.86349	&	-0.20	&	0.04	\\
&	2459290.72603	&	-0.13	&	0.06	\\
&	2459290.76258	&	-0.03	&	0.05	\\
&	2459290.81363	&	0.09	&	0.06	\\
&	2459290.85421	&	-0.24	&	0.06	\\
&	2459290.87501	&	-0.22	&	0.05	\\
&	2459290.89154	&	-0.26	&	0.06	\\
&	2459290.91071	&	-0.18	&	0.09	\\
&	2459291.78944	&	-0.12	&	0.06	\\
&	2459291.81847	&	0.02	&	0.06	\\
&	2459291.85850	&	0.16	&	0.06	\\
&	2459291.86395	&	0.20	&	0.07	\\
&	2459291.87006	&	0.09	&	0.06	\\
&	2459291.87559	&	0.09	&	0.06	\\
&	2459291.88116	&	0.02	&	0.07	\\
&	2459291.88654	&	-0.01	&	0.07	\\
&	2459291.89192	&	-0.06	&	0.07	\\
&	2459291.89729	&	-0.08	&	0.07	\\
&	2459291.90266	&	-0.09	&	0.07	\\
&	2459291.90818	&	-0.16	&	0.08	\\
&	2459294.82479	&	0.08	&	0.05	\\
&	2459294.86221	&	0.23	&	0.04	\\
&	2459294.89248	&	0.02	&	0.04	\\
&	2459294.91594	&	-0.04	&	0.05	\\
\hline
\hline
\enddata
\end{deluxetable*}

 \clearpage

\end{document}